\documentclass[aps,prd,groupedaddress,nofootinbib]{revtex4-2}
\usepackage{graphicx}
\usepackage{slashed}
\usepackage{xcolor}

\usepackage{amsmath,amssymb}
\usepackage{verbatim}
\usepackage{hyperref} 
\hypersetup{
    colorlinks,
    citecolor=black,
    filecolor=black,
    linkcolor=black,
    urlcolor=black
}

\newcommand{\be}{\begin{equation}}
\newcommand{\ee}{\end{equation}}
\newcommand{\ba}{\begin{eqnarray}}
\newcommand{\ea}{\end{eqnarray}}
\newcommand{\ban}{\begin{eqnarray*}} 
\newcommand{\ean}{\end{eqnarray*}}
\newcommand \nn {\nonumber}
\def\d{{\partial}} 
\def\P{{\mathbf P}}
\def\b{{\mathbf b}}

\def\k{{\mathbf k}}
\def\x{{\mathbf x}}  

\def\r{{\mathbf r}}
\def\z{{\mathbf z}}
\def\w{{\mathbf w}}
\def\v{{\mathbf v}}

\def\uv{{\underline{v}}}

\begin{document}


\title{Back-to-back dijet production in DIS at next-to-eikonal accuracy\\
and twist-3 gluon TMDs
}


\author{Tolga Altinoluk$^{a}$, Guillaume Beuf$\,^{a}$, Alina Czajka$^{a}$ and Cyrille Marquet$^{b}$}
\affiliation{ $^{a}$Theoretical Physics Division, National Centre for Nuclear Research,
Pasteura 7, Warsaw 02-093, Poland\\
$^{b}$CPHT, CNRS, \'Ecole polytechnique, Institut Polytechnique de Paris, 91120 Palaiseau, France}


\date{\today}

\begin{abstract}
We consider dijet production in deep inelastic scattering at small $x$, on a purely gluonic unpolarized target. Starting from earlier results obtained at next-to-eikonal accuracy in the high-energy limit, we perform the expansion in the back-to-back dijet limit, at next-to-leading power accuracy. We rewrite our results in the language of transverse-momentum-dependent (TMD) factorization, in terms of twist-2 and twist-3 TMD gluon distributions (gluon TMDs). Among the next-to-eikonal corrections, we find in particular twist-2 contributions corresponding to the $x$ dependent phase of the twist-2 gluon TMDs. We also find two types of twist-3 unpolarized gluon TMDs, as well as correlators of three gluon field strength tensors.
\\

\end{abstract}


\maketitle

\tableofcontents



\section{Introduction}

The Color Glass Condensate (CGC) effective theory (see \cite{Gelis:2010nm,Albacete:2014fwa,Blaizot:2016qgz} for reviews and references therein) has been frequently used to describe hadronic collisions in the high energy regime for about three decades now. The main feature of this effective theory is known as the gluon saturation phenomenon and it is expected to be reached with increasing energy in the Regge-Gribov limit. In this limit, the density of the gluons in the colliding hadrons increases rapidly with the increasing energy. At sufficiently high scattering energies, the high gluon density effects become dominant and tame the rapid growth of the density by taking into account the nonlinear interactions between the emitted gluons. In this regime, the nonlinear rapidity evolution is given by the well-known Balitsky-Kovchegov/Jalilian-Marian-Iancu-McLerran-Wiegert-Leonidov-Kovner (BK/JIMLWK) evolution equations \cite{Balitsky:1995ub,Kovchegov:1999yj,Kovchegov:1999ua,Jalilian-Marian:1996mkd,Jalilian-Marian:1997qno,Jalilian-Marian:1997jhx,Jalilian-Marian:1997ubg,Kovner:2000pt,Weigert:2000gi,Iancu:2000hn,Iancu:2001ad,Ferreiro:2001qy}.  

Even though hints of gluon saturation have been observed at the Relativistic Heavy Ion Collider (RHIC) and the Large Hadron Collider (LHC), a definite conclusion can not be drawn. A new generation collider, the Electron-Ion Collider (EIC), that will be built in the USA will provide a clean environment to study gluon saturation effects. However, in order to fully benefit from the existing and future experimental data, it is vital to increase the precision of the theoretical calculations. Within the CGC framework, this can be achieved by either going further in the perturbative expansion and perform the calculations of the observables at next-to-leading order (NLO) in coupling constant or by staying at leading-order (LO) in coupling constant but relaxing the adopted approximations systematically and include the corrections that are discarded previously. Each case probes different kind of corrections, thus improving the theory in different aspects and eventually provides a better quantitative comparison with existing and future experimental data.   

Deep inelastic scattering (DIS) on a dense target has been one of the most frequently used processes to study the gluon saturation effects since it provides a clean environment and also will be at the focus of the future EIC. DIS related observables are studied within the dipole factorization framework \cite{Bjorken:1970ah,Nikolaev:1990ja} where the incoming lepton emits a virtual photon which perturbatively splits into a quark-antiquark pair. The scattering of the pair on the dense target is treated within the CGC framework, thus the multiple interactions of the quark and antiquark with the dense target are taken into account. The effects of the multiple interactions are resummed into path ordered exponentials known as Wilson lines. Over the last decade, NLO corrections to DIS related observables have been computed within the CGC framework.  Computations of inclusive DIS with massless quarks \cite{Balitsky:2010ze,Balitsky:2012bs,Beuf:2011xd,Beuf:2016wdz,Beuf:2017bpd,Ducloue:2017ftk,Hanninen:2017ddy} and its fits to HERA data \cite{Beuf:2020dxl} have been performed at NLO. In \cite{Beuf:2021qqa,Beuf:2021srj,Beuf:2022ndu} inclusive DIS is computed at NLO with massive quarks and its fit to HERA data is analyzed in \cite{Hanninen:2022gje}. Diffractive structure functions  in DIS have been also computed at NLO with massless quarks in \cite{Beuf:2022kyp,Beuf:2024msh}.  
%
Apart from the structure functions, inclusive dijet \cite{Caucal:2021ent,Taels:2022tza,Taels:2023czt} and dihadron \cite{Bergabo:2023wed,Bergabo:2022tcu,Iancu:2022gpw,Caucal:2024nsb} production in DIS, as well as the diffractive dijet \cite{Boussarie:2016ogo,Boussarie:2019ero} and dihadron \cite{Fucilla:2022wcg} production have been also computed at NLO.  The results of these studies have allowed to obtain inclusive (and diffractive) single jet and hadron production in DIS \cite{Caucal:2024cdq,Bergabo:2022zhe,Bergabo:2024ivx,Altinoluk:2024vgg,Fucilla:2023mkl} by integrating over one of the final jets or hadrons. Moreover, inclusive and diffractive single and multi jet/hadron production have been also studied extensively within the CGC framework \cite{Kovchegov:2001ni,Kovchegov:2001sc,Marquet:2004xa,Golec-Biernat:2005prq,Marquet:2009ca,Altinoluk:2015dpi,Hatta:2016dxp,Mantysaari:2019csc,Salazar:2019ncp,Iancu:2020jch,Iancu:2021rup,Hatta:2022lzj,Iancu:2022lcw,Iancu:2023lel,Rodriguez-Aguilar:2023ihz,Kar:2023jkn,Hauksson:2024bvv}.

A remarkable observation regarding dihadron or dijet production has been made in \cite{Dominguez:2011wm} where it is shown that one can probe the high energy limit of the leading twist gluon Transverse Momentum Dependent distribution functions (TMDs) of the target from the CGC calculations when one considers a specific kinematic regime known as correlation limit. 
This observation has triggered many efforts to study the production of two or more final state particles not only in DIS but also in proton-nucleus (pA) collisions at forward rapidity over the last decade. The computation framework for (multi-) particle production in forward pA collisions is known as hybrid factorization
\cite{Dumitru:2005gt}. In this approach one treats the dilute projectile in the spirit of the collinear factorization which can be calculated perturbatively, while the scattering of the projectile partons on the dense target is accounted for via eikonal approximation in the CGC framework. In the correlation limit, the two jets (or hadrons) are produced nearly back-to-back in momentum space (see \cite{Petreska:2018cbf,Boussarie:2023izj} for review and references there in) which allows to perform a small dijet size expansion in coordinate space and gives access to the high energy limit of a whole set of different gluon TMDs. Apart from the correlation limit of the forward dijet production in pA collisions \cite{Marquet:2016cgx,Marquet:2017xwy}, the correlation limit of the production of three final state particles has been also studied \cite{Altinoluk:2018uax,Altinoluk:2018byz,Altinoluk:2020qet}.   

The studies exploring the connection between the CGC and TMD factorization frameworks are extended in two ways up to now. In \cite{Altinoluk:2019fui,Altinoluk:2019wyu,Boussarie:2020vzf}, the equivalence between these two frameworks has been extended beyond the correlation limit for dijet production by performing a resummation of the small transverse size of the produced jets at LO. The results of this study have provided an opportunity to distinguish between the kinematic twist corrections and the genuine saturation effects. It is shown that, the resummation of the kinematic twist corrections leads to the so-called small-x improved TMD (iTMD) factorization framework that was conjectured earlier in \cite{Kotko:2015ura,vanHameren:2016ftb}. The iTMD framework interpolates between the dilute limit of the CGC and the correlation limit of the CGC. In \cite{Fujii:2020bkl,Altinoluk:2021ygv,Boussarie:2021ybe}, the iTMD formulation is studied numerically for massive dijet production both in DIS and forward pA collisions where the unpolarized gluon TMDs and their linearly polarized counterparts are probed. On the other hand, in \cite{Caucal:2022ulg,Caucal:2023nci,Caucal:2023fsf}, it is shown that the equivalence between the CGC and TMD frameworks at leading twist can be extended to NLO. 

The eikonal approximation is one of the most frequently used kinematical approximations when performing calculations within CGC effective theory. This approximation corresponds to discarding the contributions that are suppressed by the energy of the collision in the computation of the observables. While this approximation can be reliable for asymptotic values of the colliding energies, finite energy corrections might be important especially for the kinematics planned for the EIC. For a highly boosted target that is described by a gluon background field ${\cal A}^\mu_a(x)$, the eikonal approximation corresponds to the following three assumptions: (i) the highly boosted background field is localized in the longitudinal direction (around $x^+=0$) due to the Lorentz contraction, (ii) the interaction with the projectile partons occurs only with the enhanced component of the background field which in our setup corresponds to the "$-$" component (the interaction with the suppressed (transverse and "$+$") components are discarded) and (iii) the target fields are assumed to be static (i.e. $x^-$ independent) due to Lorentz time dilation and therefore the dynamics of the target is neglected. Under these three assumptions, the highly boosted background field that describes the target has the form 
\be
{\cal A}^{\mu}_a(x^-,x^+,\x)\simeq\delta^{\mu -}\delta(x^+){\cal A}^-_a(\x)
\ee
which is known as the shockwave approximation. There is yet another source of subeikonal corrections apart from the above listed ones. Within the CGC framework, at eikonal accuracy, interaction between the projectile and the target occurs through t-channel gluon exchanges. At NEik order, t-channel quark exchanges are also allowed which provide another source of subeikonal corrections.  

Over the last decade, there have been many improvements in computing subeikonal corrections in the CGC framework. In \cite{Altinoluk:2014oxa,Altinoluk:2015gia}, a systematic method has been developed to account for the finite longitudinal extent of the target for the gluon propagator, that originates from relaxing the above mentioned assumption (i), at next-to-eikonal (NEik) and next-to-next-to-eikonal (NNEik) orders. These results are used to compute single inclusive gluon production at mid rapidity and  various spin asymmetries. The effects of NEik corrections on particle production and azimuthal asymmetries are studied both in dilute-dilute \cite{Altinoluk:2015xuy,Agostini:2019avp,Agostini:2019hkj} and in dilute-dense \cite{Agostini:2022ctk,Agostini:2022oge} collisions. In \cite{Altinoluk:2020oyd,Agostini:2023cvc}, NEik corrections to the quark propagator that are not only associated with the finite longitudinal width of the target but also with the subleading component of the background field (relaxing the above mentioned assumptions (i) and (ii)) have been computed.  The effects of relaxing the static approximation (above mentioned approximation (iii)) for the quark and scalar propagators are studied in \cite{Altinoluk:2021lvu}. In \cite{Altinoluk:2022jkk}, the full NEik quark propagator in dynamical gluon background field is computed and it is used to compute the DIS dijet production cross section at NEik accuracy. The dilute limit of DIS dijet production cross section is studied numerically in \cite{Agostini:2024xqs}. Finally, the subeikonal corrections that stem from the t-channel quark exchanges have been studied in the context of quark-gluon dijet production in DIS have been studied in \cite{Altinoluk:2023qfr}. Apart from the aforementioned works that focus on the derivation of the NEik corrections to the parton propagators and their applications to observables, in \cite{Kovchegov:2015pbl,Kovchegov:2016zex,Kovchegov:2016weo,Kovchegov:2017jxc,Kovchegov:2017lsr,Kovchegov:2018znm,Kovchegov:2018zeq,Kovchegov:2020hgb,Adamiak:2021ppq,Kovchegov:2021lvz,Cougoulic:2022gbk,Borden:2023ugd,Borden:2024bxa} quark and gluon helicity distributions together with their evolutions have been studied at NEik accuracy. The helicity dependent generalization of the non-linear rapidity evolution that goes beyond eikonal accuracy has been derived in \cite{Cougoulic:2019aja,Cougoulic:2020tbc}. Quark TMD distributions \cite{Kovchegov:2021iyc,Kovchegov:2022kyy} and gluon double-spin asymmetries \cite{Kovchegov:2024aus} are studied beyond eikonal accuracy. In \cite{Chirilli:2018kkw,Chirilli:2021lif}, NEik corrections to both quark and gluon propagators are derived in the high-energy operator product expansion (OPE) formalism. On the other hand, the studies beyond eikonal accuracy have been also performed in the context of rapidity evolution of gluon TMDs \cite{Balitsky:2015qba,Balitsky:2016dgz,Balitsky:2017flc}. A similar idea is pursued in \cite{Boussarie:2020fpb,Boussarie:2021wkn,Boussarie:2023xun} where a new formulation for the unintegrated gluon distributions that interpolate between the moderate and high values of energy is introduced.  A different approach to subeikonal corrections is introduced in \cite{Jalilian-Marian:2017ttv,Jalilian-Marian:2018iui,Jalilian-Marian:2019kaf} by including longitudinal momentum exchange between the projectile and the target during the interaction. In \cite{Li:2023tlw,Li:2024fdb}, corrections to eikonal approximation are studied by adopting an effective Hamiltonian approach. Last but not least, in \cite{Hatta:2016aoc,Kovchegov:2019rrz,Boussarie:2019icw,Kovchegov:2023yzd}, NEik effects are studied in the context of orbital angular momentum. 

Our main goal in this paper is to study the correlation limit of the DIS dijet production at NEik accuracy and extend the studies to comprehend the connection between the CGC and TMD factorization frameworks  beyond leading twist and beyond eikonal order.  A similar idea is recently pursued in \cite{Fu:2023jqv,Fu:2024sba} where the matching between the CGC and high-twist formalism is studied, but in the context of collinear factorization instead of TMD factorization. The outline of the paper is as follows. In Sec. \ref{NEik_DIS_gen_Kin}, after summarizing the results for the NEik DIS dijet production cross section via longitudinal and transverse photons, computed in \cite{Altinoluk:2022jkk}, we rewrite all the decorated Wilson lines in terms of the components of the field strength tensor ${\cal F}_{\mu\nu}$. In Sec. \ref{Corr_limit_amplitude_NEik}, we perform the expansion of the NEik DIS dijet production amplitudes in the correlation limit. Then, we use these results to calculate the back-to-back dijet production in DIS at NEik accuracy via longitudinal photon in Sec. \ref{sec:gamma_L_case} and via transverse photon in Sec. \ref{sec:gamma_T_case}. In Sec. \ref{sec:CGC_vs_TMD}, we rewrite our results in terms of the standard gluon TMD distributions  and compute the corresponding coefficient functions. Finally, in Sec. \ref{sec:Conc} we provide a short discussion of our results and possible future projects. The list of the integrals needed for computing hard factors at amplitude level is given in Appendix \ref{app:int}. The discussion about the extraction of the non-eikonal corrections from the generalized eikonal cross section for longitudinal photon is provided in Appendix \ref{app:geik-eik}. Appendix \ref{app:rewrite_non_fact} is devoted to the technical details concerning how the back-to-back dijet production amplitude for transverse photon is rewritten. In Appendix \ref{app:coeffs}, explicit expressions for the tensor hard factors are provided.


\section{DIS dijet production cross section at next-to-eikonal accuracy in general kinematics}
\label{NEik_DIS_gen_Kin}

In the single photon exchange approximation, DIS processes can be expressed as the product of a leptonic tensor, encoding the virtual photon emission by the incoming lepton, and a hadronic tensor, encoding the interaction of the virtual photon with the target. Integrating over the azimuthal angle of the scattered lepton, the hadronic tensor is projected into two scalar functions, and the cross section is written as a linear combination of these two scalar functions, for example as  
\begin{align}
\frac{d\sigma^{l+\textrm{target}\rightarrow l'+ \textrm{dijet}+X}}{d x_{Bj}\, d Q^2\, d {\rm P.S.}}
=&\, 
\frac{\alpha_{\textrm{em}}}{\pi\, x_{Bj}\, Q^2}
\left[
\left(1-y+\frac{y^2}{2}\right) \frac{d\sigma_{\gamma^{*}_T\rightarrow \textrm{dijet}}}{d {\rm P.S.}}(x_{Bj},Q^2)
+\left(1-y\right) \frac{d\sigma_{\gamma^{*}_L\rightarrow \textrm{dijet}}}{d {\rm P.S.}}(x_{Bj},Q^2)
\right]
\, ,
 \label{lepton_to_photon_DIS} 
\end{align}
for dijet production in DIS.
Let us denote $k_l^{\mu}$ the momentum of the incoming lepton, $q^{\mu}$ the momentum of the exchanged virtual photon (so that $k_l^{\mu}-q^{\mu}$ is the momentum of the scattered lepton), and $P_{\textrm{tar}}^{\mu}$  the momentum of the target. Then, the usual Lorentz invariant variables for DIS are defined as follows. The Mandelstam $s$ variable for the lepton-target collision is $s=(k_l+P_{\textrm{tar}})^2$, the photon virtuality is $Q^2=-q^2>0$, and the Bjorken variable is $x_{Bj}=Q^2/(2P_{\textrm{tar}}\!\cdot\! q)$. Finally, the inelasticity variable $y$ is defined as $y=(2P_{\textrm{tar}}\!\cdot\! q)/s=Q^2/(s x_{Bj})$.
The two scalar functions appearing in Eq.~\eqref{lepton_to_photon_DIS} can be interpreted as cross sections for the virtual photon - target subprocess, in which the photon has either a transverse or longitudinal polarization. 

 In Eq.~\eqref{lepton_to_photon_DIS}, $d {\rm P.S.}$ is the Lorentz invariant phase space measure for the produced hadronic final state, which in the dijet case is defined as 
\begin{align}
d{\rm P.S.}=\frac{d^2\k_1}{(2\pi)^2}\frac{dk_1^+}{(2\pi)2k_1^+}\frac{d^2\k_2}{(2\pi)^2}\frac{dk_2^+}{(2\pi)2k_2^+}\, , 
\label{phase_space_1_2}
\end{align}
where $k_1$ and $k_2$ are the produced jets momenta, identified as the momenta of the produced quark and antiquark  at leading order (LO) in the QCD coupling $\alpha_s$.

In this section, we first recall the results for the DIS dijet production amplitudes,  via either longitudinal or transverse photon exchange, obtained at NEik accuracy in \cite{Altinoluk:2022jkk} at LO in $\alpha_s$, in the sections \ref{sec:gen_kinematics_sigma_L} and \ref{sec:gen_kinematics_sigma_T}. In Ref.~\cite{Altinoluk:2022jkk}, the NEik corrections to the dijet cross section are written in terms of new types of dipole and quadrupole operators. These new operators involve Wilson lines with insertions of covariant derivatives along their longitudinal extent and therefore dubbed as decorated Wilson lines. 
In the section \ref{sec:dec_Wilson_F}, we will show that these covariant derivative insertions can be written as field strength insertions, and provide the expressions for the scattering amplitudes at NEik accuracy calculated in Ref.~\cite{Altinoluk:2022jkk} now rewritten in terms Wilson lines with field strength insertions along their longitudinal extent.


\subsection{NEik DIS dijet production cross section via longitudinal photon}
\label{sec:gen_kinematics_sigma_L}
The cross section at NEik accuracy for DIS dijet for longitudinal photon is written as \cite{Altinoluk:2022jkk}
%
\begin{align}
 \frac{d\sigma_{\gamma^{*}_L\rightarrow q_1\bar q_2}}{d {\rm P.S.}} =  \frac{d\sigma_{\gamma^{*}_L\rightarrow q_1\bar q_2}}{d {\rm P.S.}}\Bigg|_{\rm Gen. \, Eik}+  \frac{d\sigma_{\gamma^{*}_L\rightarrow q_1\bar q_2}}{d {\rm P.S.}}\Bigg|_{\rm NEik \, corr.} + O({\rm NNEik})
 \label{def_L_cross_sec}
 \, .
\end{align}
%

The first term is the generalized eikonal contribution, which goes beyond the strict eikonal approximation by including the weak dependence of the gluon background field on the light-cone coordinate $z^-$ and thus accounts for the dynamics of the target beyond infinite Lorentz time dilation. It is given by  
%
\begin{align}
\label{Cross_Section_GEik}
\frac{d\sigma_{\gamma^{*}_L\rightarrow q_1\bar q_2}}{d {\rm P.S.}}\Bigg|_{\rm Gen. \, Eik}
= 2q^+ \int d (\Delta b^-) e^{i\Delta b^-(k_1^++k_2^+ - q^+)} 
\sum_{\rm hel. \,,\,   col. }
\Big\langle  \Big(\mathbf{M}_{q_1 \bar q_2 \leftarrow \gamma^*_L}^{\rm Gen.\,  Eik} \Big(- \frac{\Delta b^-}{2}\Big)\Big)^\dag \mathbf{M}_{q_1 \bar q_2 \leftarrow \gamma^*_L}^{\rm Gen.\,  Eik} \Big(\frac{\Delta b^-}{2} \Big) \Big\rangle
\end{align}
%
where the summation is performed over the colors and light front helicities of the produced quark and antiquark, $\langle \cdots\rangle$ stands for averaging over the background field of the target and the $b^-$-dependent amplitude reads\footnote{Throughout the manuscript, we use $\int_{\z}\equiv\int d^2\z$ as shorthand notation for the coordinate space integrals over the transverse direction.} 
\begin{align}
i \mathbf{M}_{q_1 \bar q_2 \leftarrow \gamma^*_L}^{\rm Gen.\,  Eik} (b^-)
=&\, 
 -Q\,  \frac{e e_f}{2\pi} \, \bar u(1) \gamma^+ v(2)\, 
 \frac{(q^+\!+\!k_1^+\!-\!k_2^+)(q^+\!+\!k_2^+\!-\!k_1^+)}{4(q^+)^3}\,
 \theta(q^+\!+\!k_1^+\!-\!k_2^+)\,
 \theta(q^+\!+\!k_2^+\!-\!k_1^+)\,
  \nn \\
&
\times
\int_{\v,\w}\,  e^{-i\v \cdot\k_1} \, e^{-i\w\cdot\k_2}\, 
\textrm{K}_0\left(\hat{Q}\, |\v\!-\!\w|\right)
\left[\mathcal{U}_F\left(\frac{L^+}{2},-\frac{L^+}{2},\v,b^- \right)
\mathcal{U}_F\left(\frac{L^+}{2},-\frac{L^+}{2},\w,b^- \right)^{\dag}
\!-\!1\right]
\label{bdep_Ampl-GenEik_L} 
\, .
\end{align}
Here, ${\rm K_\alpha(x)}$ is the modified Bessel function of the second kind and ${\hat Q}$ is defined as 
\begin{align}
\hat Q=\sqrt{m^2+\frac{(q^++k_1^+-k_2^+)(q^+-k_1^++k_2^+)}{4(q^+)^2}Q^2} \,. 
\label{def_hat_Q}
\end{align}
The Wilson line in the gluon background field $\mathcal{A}_a^-(x)$ of the target is defined as\footnote{Here and in the following, when we omit the adjoint color index for the gauge field or field strength, the contraction with the fundamental color generators is implied, as $\mathcal{A}^{\mu}(x)\equiv t^a \mathcal{A}_a^{\mu}(x)$ and $\mathcal{F}^{\mu\nu}(x)\equiv t^a \mathcal{F}_a^{\mu\nu}(x)$.} 
\ba
\label{def:Wilson}
\mathcal{U}_F(x^+,y^+;\v,v^-) = \mathcal{P}_+ \exp \Big[-ig \int_{y^+}^{x^+} dv^+ 
\mathcal{A}^-(v) \Big]
\, .
\ea
In Eq.~\eqref{bdep_Ampl-GenEik_L}, $L^+$ is the typical longitudinal width of the target, which was used in Ref.~\cite{Altinoluk:2022jkk} in order to keep track of power counting of corrections beyond the eikonal approximation, in particular the corrections beyond the shockwave limit for the target. In the rest of this section, we will drop the first two arguments of the Wilson line if they extend from $-L^+/2$ to $L^+/2$, $\mathcal{U}_F(\v,v^-)\equiv \mathcal{U}_F( L^+/2,-L^+/2,\v,v^-)$.

The strict eikonal amplitude can be obtained from Eq.~\eqref{bdep_Ampl-GenEik_L} by neglecting the $b^-$ dependence (which now provides the condition $k_1^++k_2^+=q^+$) and its explicit expression reads  (with the notation  $\mathcal{U}_F(\v)\equiv\mathcal{U}_F(\v,b^-=0)$) 
\begin{align}
i {\cal M}_{q_1 \bar q_2 \leftarrow \gamma^*_L}^{\rm strict~Eik}
 =&\, 
 -Q\,  \frac{e e_f}{2\pi} \, \bar u(1) \gamma^+ v(2)\, 
 \frac{k_1^+k_2^+}{(q^+)^3}\,
 \int_{\v,\w} e^{-i\v \cdot\k_1}\, e^{-i\w\cdot\k_2}\,
\textrm{K}_0\left(\bar{Q}\, |\v\!-\!\w|\right)
\Big[\mathcal{U}_F(\v )
\mathcal{U}_F^{\dag}(\w )
-1\Big]
\label{Ampl-StrictEik_L} 
\, .
\end{align}

The second term in Eq.~\eqref{def_L_cross_sec}, the explicit NEik correction to the cross section, is given by 
\begin{align}
 \frac{d\sigma_{\gamma^{*}_L\rightarrow q_1\bar q_2}}{d {\rm P.S.}}\Bigg|_{\rm NEik \, corr.}
&
= (2q^+)\, 2\pi \delta(k_1^+\!+\!k_2^+\!-\!q^+) \sum_{\rm hel. \,,\,  col. }
\Big[ \Big\langle  \Big({\cal M}_{q_1 \bar q_2 \leftarrow \gamma^*_L}^{\rm strict~Eik} \Big)^\dag  
{\cal M }_{q_1 \bar q_2 \leftarrow \gamma^*_L}^{\rm NEik \, corr.} \Big\rangle
+ \Big \langle \Big({\cal M}_{q_1 \bar q_2 \leftarrow \gamma^*_L}^{\rm NEik \, corr.} \Big)^\dag {\cal M}_{q_1 \bar q_2 \leftarrow \gamma^*_L}^{\rm strict~Eik} \Big\rangle \Big] \nn \\
&
=  (2q^+)\, 2\pi \delta(k_1^+\!+\!k_2^+\!-\!q^+) \sum_{\rm hel. \,,\,  col. }
2 {\rm Re}  \Big\langle  \Big({\cal M}_{q_1 \bar q_2 \leftarrow \gamma^*_L}^{\rm strict~Eik}\Big)^\dag  {\cal M}_{q_1 \bar q_2 \leftarrow \gamma^*_L}^{\rm NEik \, corr.} \Big\rangle
\, .
\label{Cross_Section_long_NEik}
\end{align}
The NEik correction to the amplitude can be organized as the following three different contributions:
\begin{align}
i {\cal M }_{q_1 \bar q_2 \leftarrow \gamma^*_L}^{\rm NEik \, corr.}= 
i{\cal M }_{q_1 \bar q_2 \leftarrow \gamma^*_L}^{\textrm{dec. on }q}
 +
i{\cal M }_{q_1 \bar q_2 \leftarrow \gamma^*_L}^{\textrm{dec. on }\bar{q}}
+
i{\cal M }_{q_1 \bar q_2 \leftarrow \gamma^*_L}^{\textrm{dyn. target}}
\, ,
\label{NEik_corr_Ampl_L}
\end{align}   
with the decorations on the quark Wilson line (stemming from including the effects of the finite width of the target and the interaction with the transverse component of the background field in the quark propagator) 
\begin{align}
i{\cal M }_{q_1 \bar q_2 \leftarrow \gamma^*_L}^{\textrm{dec. on }q}
=&\, 
 -Q\,\frac{e e_f}{2\pi} \,  \frac{k_2^+}{2(q^+)^3}
\int_{\v,\w} e^{-i\v \cdot\k_1}\, e^{-i\w\cdot\k_2}\, 
\textrm{K}_0\left(\bar{Q}\, |\v\!-\!\w|\right)
 \nn \\
& \times
\bar u(1)  \gamma^+ 
\left[\frac{[\gamma^i,\gamma^j]}{4}\, \mathcal{U}^{(3)}_{F; ij} ( \v) 
- i\, \mathcal{U}^{(2)}_F ( \v)\,
+ \mathcal{U}^{(1)}_{F;j} ( \v) \,  \bigg(\frac{(\k_2^j\!-\!\k_1^j)}{2}+\frac{i}{2}\, \partial_{\w^j}\bigg)
 \right]
\mathcal{U}_F(\w)^{\dag}\,
v(2) 
\label{ampl-q_dec_L} 
\, ,
\end{align}
similarly, the decorations on the antiquark Wilson line\footnote{In Eq.~\eqref{ampl-qbar_dec_L} the partial derivative in $\v$ acts only within the bracket, on the Wilson line, and not on the phase factor.} 
\begin{align}
i{\cal M }_{q_1 \bar q_2 \leftarrow \gamma^*_L}^{\textrm{dec. on }\bar{q}}
=&\,
 -Q\,\frac{e e_f}{2\pi} \,  \frac{k_1^+}{2(q^+)^3}
\int_{\v,\w} e^{-i\v \cdot\k_1}\, e^{-i\w\cdot\k_2}\, 
\textrm{K}_0\left(\bar{Q}\, |\v\!-\!\w|\right)
 \nn \\
& \times
\bar u(1)  \gamma^+ 
\Bigg[
\mathcal{U}_F(\v)
\left(\frac{[\gamma^i,\gamma^j]}{4}\,  \mathcal{U}^{(3)}_{F; ij} ( \w)^{\dag} 
\!-\!i\, \mathcal{U}^{(2)}_F ( \w)^{\dag} 
\!+\!\bigg(\frac{i}{2}\, \overleftarrow{\partial_{\v^j}} -\frac{(\k_2^j\!-\!\k_1^j)}{2} \bigg)\mathcal{U}^{(1)}_{F;j} ( \w)^{\dag}
\right)
\Bigg]
v(2) 
\label{ampl-qbar_dec_L}
\, ,
\end{align}
with $\bar Q$ defined as
\begin{align}
\bar Q \equiv \sqrt{m^2+Q^2\, \frac{k_1^+k_2^+}{(q^+)^2}} \; .
\end{align}
The decorated Wilson lines that appear in the amplitudes in Eqs. \eqref{ampl-q_dec_L} and \eqref{ampl-qbar_dec_L} are defined as 
\begin{align}
\label{Wilson_dec_1}
\mathcal{U}^{(1)}_{F;j} ( \v) =&\,  \int_{-\frac{L^+}{2}}^{\frac{L^+}{2}}dv^+\, 
\mathcal{U}_F\Big(\frac{L^+}{2},v^+;\v\Big) 
\overleftrightarrow{\mathcal{D}_{\v^j}}
\mathcal{U}_F\Big(v^+,-\frac{L^+}{2};\v\Big) 
\\
\label{Wilson_dec_2}
\mathcal{U}^{(2)}_F ( \v) =&\,
\int_{-\frac{L^+}{2}}^{\frac{L^+}{2}}dv^+\,  
\mathcal{U}_F\Big(\frac{L^+}{2},v^+;\v\Big) 
\overleftarrow{\mathcal{D}_{\v^j}}\, \overrightarrow{\mathcal{D}_{\v^j}} 
\mathcal{U}_F\Big(v^+,-\frac{L^+}{2};\v\Big) 
\\
\mathcal{U}^{(3)}_{F; ij} ( \v)
 =&\,  
\int_{-\frac{L^+}{2}}^{\frac{L^+}{2}}dv^+\,  
\mathcal{U}_F\Big(\frac{L^+}{2},v^+;\v\Big) 
g \mathcal{F}_{ij}(\uv) 
\mathcal{U}_F\Big(v^+,-\frac{L^+}{2};\v\Big) 
\label{Wilson_dec_3}
\end{align}
Their Hermitian conjugates are
\ba
\label{Wilson_dec_1_hc}
\mathcal{U}^{(1)}_{F;j} ( \v)^{\dag} &=&  
-\int_{-\frac{L^+}{2}}^{\frac{L^+}{2}}dv^+\, 
\mathcal{U}_F\Big(v^+,-\frac{L^+}{2};\v\Big)^{\dag} 
\overleftrightarrow{\mathcal{D}_{\v^j}}
\mathcal{U}_F\Big(\frac{L^+}{2},v^+;\v\Big)^{\dag}  
\\
\label{Wilson_dec_2_hc}
\mathcal{U}^{(2)}_F ( \v)^{\dag} &=& 
\int_{-\frac{L^+}{2}}^{\frac{L^+}{2}}dv^+\,  
\mathcal{U}_F\Big(v^+,-\frac{L^+}{2};\v\Big)^{\dag} 
\overleftarrow{\mathcal{D}_{\v^j}}\, \overrightarrow{\mathcal{D}_{\v^j}} 
\mathcal{U}_F\Big(\frac{L^+}{2},v^+;\v\Big)^{\dag}
\\
\mathcal{U}^{(3)}_{F; ij} ( \v)^{\dag} &=&  
\int_{-\frac{L^+}{2}}^{\frac{L^+}{2}}dv^+\, 
\mathcal{U}_F\Big(v^+,-\frac{L^+}{2};\v\Big)^{\dag} 
g\mathcal{F}_{ij}(\uv)
\mathcal{U}_F\Big(\frac{L^+}{2},v^+;\v\Big)^{\dag} 
\label{Wilson_dec_3_hc}
\, .
\ea
Here, the background covariant derivative and field strength are defined with the following convention
\ba
\overleftrightarrow{\mathcal{D}_{\mu}} &\equiv& \overrightarrow{\mathcal{D}_{\mu}} - \overleftarrow{\mathcal{D}_{\mu}} \\
\overrightarrow{\mathcal{D}_{\mu}} &\equiv& \overrightarrow{\partial_\mu} + ig  \mathcal{A}_\mu \\
\overleftarrow{\mathcal{D}_{\mu}} &\equiv& \overleftarrow{\partial_\mu} - ig \mathcal{A}_\mu \\
\label{F}
\mathcal{F}_{\mu\nu} &\equiv& \partial_\mu \mathcal{A}_\nu - \partial_\nu \mathcal{A}_\mu
+ig [\mathcal{A}_\mu, \mathcal{A}_\nu]
\label{def_F_mu_nu}
\, .
\ea

The third term in Eq.~\eqref{NEik_corr_Ampl_L} is a correction beyond the static approximation for the target (or infinite Lorentz time dilation). More precisely, it is the first NEik power correction associated with the relative separation along the $x^-$ direction that can exist between the quark and antiquark Wilson lines in the amplitude. 
It reads
\begin{align}
i{\cal M }_{q_1 \bar q_2 \leftarrow \gamma^*_L}^{\textrm{dyn. target}}
 = &\,
 i Q\, \frac{e e_f}{2\pi} \,  \bar u(1) \gamma^+ v(2) \,
 \frac{(k_1^+\!-\!k_2^+)}{2(q^+)^3}\,
 \int_{\v,\w} e^{-i\v \cdot\k_1}\, e^{-i\w\cdot\k_2}\, 
 \nn \\
&
\times\,
\left[
\textrm{K}_0\left(\bar{Q}\, |\v\!-\!\w|\right)
-\frac{\left(\bar{Q}^2\!-\!m^2\right)}{2\bar{Q}}\, |\v\!-\!\w|\, 
\textrm{K}_1\left(\bar{Q}\, |\v\!-\!\w|\right)
\right]
\bigg[\mathcal{U}_F\Big(\v,b^- \Big)\overleftrightarrow{\partial_{b^-}}
\mathcal{U}_F\Big(\w,b^- \Big)^{\dag}
\bigg]\bigg|_{b^-=0}
 \, .
\label{ampl-dyn_L} 
\end{align}

\subsection{NEik DIS dijet production cross section via transverse photon}
\label{sec:gen_kinematics_sigma_T}

In the transverse photon case as well, the dijet cross section at NEik accuracy can be written as a sum of a generalized eikonal contribution and an explicit NEik correction: 
\begin{align}
 \frac{d\sigma_{\gamma^{*}_T\rightarrow q_1\bar q_2}}{d {\rm P.S.}} =  \frac{d\sigma_{\gamma^{*}_T\rightarrow q_1\bar q_2}}{d {\rm P.S.}}\Bigg|_{\rm Gen. \, Eik}+  \frac{d\sigma_{\gamma^{*}_T\rightarrow q_1\bar q_2}}{d {\rm P.S.}}\Bigg|_{\rm NEik \, corr.} + O({\rm NNEik})
 \label{def_T_cross_sec}
 \, .
\end{align}
 The first term is obtained as 
\begin{align}
\label{Cross_Section_GEik_T}
\frac{d\sigma_{\gamma^{*}_T\rightarrow q_1\bar q_2}}{d {\rm P.S.}}\Bigg|_{\rm Gen. \, Eik}
= 2q^+ \int d (\Delta b^-) e^{i\Delta b^-(k_1^++k_2^+ - q^+)} 
\frac{1}{2}\sum_{\lambda}\sum_{\rm hel. \,,\,   col. }
\Big\langle  \Big(\mathbf{M}_{q_1 \bar q_2 \leftarrow \gamma^*_T}^{\rm Gen.\,  Eik} \Big(- \frac{\Delta b^-}{2}\Big)\Big)^\dag \mathbf{M}_{q_1 \bar q_2 \leftarrow \gamma^*_T}^{\rm Gen.\,  Eik} \Big(\frac{\Delta b^-}{2} \Big) \Big\rangle
\, ,
\end{align}
%
which includes an averaging over the polarization $\lambda$ of the incoming transverse photon. The $b^-$ dependent amplitude involved in Eq.~\eqref{Cross_Section_GEik_T} is given by 
\begin{align}
i \mathbf{M}_{q_1 \bar q_2 \leftarrow \gamma^*_T}^{\rm Gen.\,  Eik} (b^-)
 =&\, 
 \frac{e e_f}{2\pi} \, 
  \varepsilon_{\lambda}^i \; 
  \frac{1}{2q^+}\;
 \theta(q^+\!+\!k_1^+\!-\!k_2^+)\,
 \theta(q^+\!+\!k_2^+\!-\!k_1^+)\,
\int_{\v,\w} e^{-i\v \cdot\k_1}\, e^{-i\w\cdot\k_2}
  \nn \\
&
\times
\bigg\{
i\, \frac{(\v^j\!-\!\w^j)}{|\v\!-\!\w|}\, \hat{Q}\, \textrm{K}_1\left(\hat{Q}\, |\v\!-\!\w|\right)\, 
 \bar u(1) \gamma^+ 
 \! \left[\frac{(k_2^+\!-\!k_1^+)}{q^+}\, \delta^{ij}
 +\frac{[\gamma^i,\gamma^j]}{2}\right] v(2)\, 
 \nn \\
&\,
+\textrm{K}_0\left(\hat{Q}\, |\v\!-\!\w|\right)\, 
m\, \bar u(1) \gamma^+  \gamma^i v(2)\, 
\bigg\}
\bigg[\mathcal{U}_F\Big(\v,b^- \Big)
\mathcal{U}_F\Big(\w,b^- \Big)^{\dag}
-1\bigg]
\label{Ampl-GenEik_T} 
\, .
\end{align}

As discussed in \cite{Altinoluk:2022jkk}, in the strict eikonal approximation, Eq.~\eqref{Ampl-GenEik_T} reduces to
\begin{align}
& i {\cal M}_{q_1 \bar q_2 \leftarrow \gamma^*_T}^{\rm strict~Eik}
 = 
 \frac{e e_f}{2\pi} \, 
  \varepsilon_{\lambda}^i \, \frac{1}{2q^+}
 \int_{\v,\w} \, e^{-i\v \cdot\k_1}\, e^{-i\w\cdot\k_2}\,
 \Big[\mathcal{U}_F(\v )
\mathcal{U}_F(\w )^{\dag}
-1\Big]
  \nn \\
&
\times
\bigg\{
i\, \frac{(\v^j\!-\!\w^j)}{|\v\!-\!\w|}\, \bar{Q}\, \textrm{K}_1\left(\bar{Q}\, |\v\!-\!\w|\right)\, 
 \bar u(1) \gamma^+ \! \left[\frac{(k_2^+\!-\!k_1^+)}{q^+}\, \delta^{ij}
 +\frac{[\gamma^i,\gamma^j]}{2}\right] v(2)\, 
+\textrm{K}_0\left(\bar{Q}\, |\v\!-\!\w|\right)\, 
m\, \bar u(1) \gamma^+  \gamma^i v(2)\, 
\bigg\}
\label{Ampl-StrictEik_T} 
\, .
\end{align}

The second term in Eq.~\eqref{def_T_cross_sec}  is calculated in the same way as in the longitudinal case, up to the averaging over $\lambda$, as
\begin{align}
 \frac{d\sigma_{\gamma^{*}_T\rightarrow q_1\bar q_2}}{d {\rm P.S.}}\Bigg|_{\rm NEik \, corr.}
&
=  (2q^+)\, 2\pi \delta(k_1^+\!+\!k_2^+\!-\!q^+)\, \frac{1}{2}\sum_{\lambda}\, \sum_{\rm hel. \,,\,  col. }
2 {\rm Re}  \Big\langle  \Big({\cal M}_{q_1 \bar q_2 \leftarrow \gamma^*_T}^{\rm strict~Eik}\Big)^\dag  {\cal M}_{q_1 \bar q_2 \leftarrow \gamma^*_T}^{\rm NEik \, corr.} \Big\rangle
\, .
\label{Cross_Section_T_NEik}
\end{align}
By contrast, the NEik correction to the amplitude is the sum of five contributions in the transverse photon case, as
\begin{align}
i {\cal M }_{q_1 \bar q_2 \leftarrow \gamma^*_T}^{\rm NEik \, corr.}= 
i{\cal M }_{q_1 \bar q_2 \leftarrow \gamma^*_T}^{\textrm{dec. on }q}
 +
i{\cal M }_{q_1 \bar q_2 \leftarrow \gamma^*_T}^{\textrm{dec. on }\bar{q}}
+
i{\cal M }_{q_1 \bar q_2 \leftarrow \gamma^*_T}^{\textrm{dyn. target}}
+
i{\cal M }_{q_1 \bar q_2 \leftarrow \gamma^*_T}^{L^+\textrm{ phase}}
+
i{\cal M }_{q_1 \bar q_2 \leftarrow \gamma^*_T}^{\textrm{in}}
\, .
\label{NEik_corr_Ampl_T}
\end{align}   
The first three terms are the direct analogs of the ones appearing in the longitudinal photon case. They read
\begin{align}
i{\cal M }_{q_1 \bar q_2 \leftarrow \gamma^*_T}^{\textrm{dec. on }q}
 = &\,
 \frac{e e_f}{2\pi} \,  \varepsilon_{\lambda}^l \, \frac{1}{2 q^+}\,\frac{1}{2 k_1^+}\,
\int_{\v,\w} e^{-i\v \cdot\k_1}\, e^{-i\w\cdot\k_2} 
 \nn \\
& \times
\bar u(1) \gamma^+   \bigg[
\left(\frac{[\gamma^i,\gamma^j]}{4}\,     \mathcal{U}^{(3)}_{F; ij} ( \v)
- i\,  \mathcal{U}^{(2)}_F ( \v) 
+  \mathcal{U}^{(1)}_{F;j} ( \v)\, \bigg( \frac{(\k_2^j\!-\!\k_1^j)}{2}  +\frac{i}{2}\, \overrightarrow{\partial_{\w^j}}\bigg)
\right)
\mathcal{U}_F(\w)^{\dag}
\bigg]  
\nn \\
& \times
\bigg\{
i\, \frac{(\v^m\!-\!\w^m)}{|\v\!-\!\w|}\, \bar{Q}\, \textrm{K}_1\left(\bar{Q}\, |\v\!-\!\w|\right)\, 
 \left[\frac{(k_2^+\!-\!k_1^+)}{q^+}\, \delta^{lm}+\frac{[\gamma^l,\gamma^m]}{2}\right]
 +\textrm{K}_0\left(\bar{Q}\, |\v\!-\!\w|\right)\; m\, \gamma^l
 \bigg\}  v(2)
 \, ,
\label{Ampl-q_dec_T}
\end{align}

\begin{align}
i{\cal M }_{q_1 \bar q_2 \leftarrow \gamma^*_T}^{\textrm{dec. on }\bar{q}}
= &\, 
 \frac{e e_f}{2\pi} \, \varepsilon_{\lambda}^l \, \frac{1}{2 q^+}\, \frac{1}{2 k_2^+}\, 
\int_{\v,\w} e^{-i\v \cdot\k_1}\, e^{-i\w\cdot\k_2}\,
 \nn \\
&
\times\;
\bar u(1)
 \gamma^+
\bigg\{
i\, \frac{(\v^m\!-\!\w^m)}{|\v\!-\!\w|}\, \bar{Q}\, \textrm{K}_1\left(\bar{Q}\, |\v\!-\!\w|\right)\, 
 \left[\frac{(k_2^+\!-\!k_1^+)}{q^+}\, \delta^{lm}+\frac{[\gamma^l,\gamma^m]}{2}\right]
 +\textrm{K}_0\left(\bar{Q}\, |\v\!-\!\w|\right)\; m\, \gamma^l
 \bigg\}
 \nn \\
&
\times 
\bigg[
\mathcal{U}_F(\v)
\left(\frac{[\gamma^i,\gamma^j]}{4} \mathcal{U}^{(3)}_{F; ij}(\w)^{\dag} 
\!-\!i\, \mathcal{U}^{(2)}_F(\w)^{\dag}
+\bigg(\frac{i}{2}\,  \overleftarrow{\partial_{\v^j}}\!-\!\frac{(\k_2^j\!-\!\k_1^j)}{2}\bigg)\, \mathcal{U}^{(1)}_{F;j} ( \w)^{\dag} 
\right)
\bigg]
 v(2) 
\label{Ampl-qbar_dec_T}
\, ,
\end{align}
and
\begin{align}
i{\cal M }_{q_1 \bar q_2 \leftarrow \gamma^*_T}^{\textrm{dyn. target}}
= &\,
 \frac{e e_f}{2\pi}\, \, \varepsilon_{\lambda}^i \, \frac{1}{2(q^+)^2}
\int_{\v,\w} e^{-i\v \cdot\k_1}\, e^{-i\w\cdot\k_2}\,
\bigg[\mathcal{U}_F\Big(\v,b^- \Big)\overleftrightarrow{\partial_{b^-}}
\mathcal{U}_F\Big(\w,b^- \Big)^{\dag}
\bigg]\bigg|_{b^-=0}\;  
 \nn \\
&
\times
\bar u(1) \gamma^+\Bigg\{ 
 \frac{(\v^i\!-\!\w^i)\, \bar{Q}}{|\v\!-\!\w|}\, \textrm{K}_1\left(\bar{Q}\, |\v\!-\!\w|\right)\,
-\frac{i(k_2^+\!-\!k_1^+)\, Q^2\, |\v\!-\!\w|}{4q^+\, \bar{Q}}\: 
 \textrm{K}_1\left(\bar{Q}\, |\v\!-\!\w|\right)\, m\, \gamma^i
 \nn\\
&\;\;\;\;
+\frac{(k_2^+\!-\!k_1^+)\, Q^2}{4q^+}\, (\v^j\!-\!\w^j) \textrm{K}_0\left(\bar{Q}\, |\v\!-\!\w|\right)  \left[\frac{(k_2^+\!-\!k_1^+)}{q^+}\, \delta^{ij}+\frac{[\gamma^i,\gamma^j]}{2}\right]
\Bigg\}v(2)
\label{Ampl-dyn_T} 
\,  .
\end{align}  
The last two terms  in Eq.~\eqref{NEik_corr_Ampl_T} correspond to further  contributions which vanish for longitudinal photon but survive for transverse photon.
The fourth term comes from the expansion of a $L^+$ dependent phase factor appearing when integrating over the $x^+$ coordinate of the photon splitting vertex before the target. It is obtained as \cite{Altinoluk:2022jkk}
\begin{align}
i{\cal M }_{q_1 \bar q_2 \leftarrow \gamma^*_T}^{L^+\textrm{ phase}}
 =  &\, 
  e e_f \, \varepsilon_{\lambda}^i \,
  \frac{(-1)L^+}{16k_1^+k_2^+}\;
  \bar u(1) \gamma^+\!
  \left[\frac{(k_2^+\!-\!k_1^+)}{q^+}\, \delta^{ij}+\frac{[\gamma^i,\gamma^j]}{2}\right]  
  v(2) \: 
 \int_{\z} e^{-i\z \cdot(\k_1+\k_2)}\:
\Big[\mathcal{U}_F(\z)\overleftrightarrow{\partial_{\z^j}}
\mathcal{U}_F(\z)^{\dag}
\Big]
\, .
\label{Ampl-Lplus_T} 
\end{align}
 The fifth term in Eq.~\eqref{NEik_corr_Ampl_T} is the contribution from the diagram in which the photon splitting happens within the target instead of before. It is a contribution beyond the standard dipole picture of DIS, which is allowed beyond the eikonal approximation. It is found to be
\begin{align}
i{\cal M }_{q_1 \bar q_2 \leftarrow \gamma^*_T}^{\textrm{in}} 
=&\,
  e e_f \, \varepsilon_{\lambda}^i\; \frac{1}{8k_1^+k_2^+}\;
\bar u(1) \gamma^+\!\left[\frac{(k_2^+\!-\!k_1^+)}{q^+}\, \delta^{i j}
+\frac{1}{2}\, [ \gamma^i,  \gamma^j] 
\right] v(2)
\nonumber\\
&\, \times \, 
\int_{\z}
e^{-i(\k_1+\k_2) \cdot \z}\: \int_{-L^+/2}^{L^+/2} dz^+\, 
\bigg[ \mathcal{U}_F\Big(\frac{L^+}{2},z^+;\z\Big) \overleftrightarrow{ {\cal D}_{\z^j}}  \mathcal{U}_F\Big(\frac{L^+}{2},z^+;\z\Big)^{\dag}\bigg]
\label{ampl-in_T} 
\, .
\end{align}

\subsection{Local decorations for the Wilson lines}
\label{sec:dec_Wilson_F}

For later convenience, we aim to rewrite all decorated Wilson lines in terms of the components of the strength tensor $\mathcal{F}_{\mu\nu}$. Field strength insertions correspond to local decorations, by contrast with the covariant or ordinary derivatives acting on whole gauge links.
$\mathcal{U}^{(3)}_{F;ij}$ and $\mathcal{U}_{F;ij}^{(3)\dag}$ are already expressed in terms of $\mathcal{F}_{ij}$ insertions as desired.
 In order to convert the other types of decorated Wilson lines, let us first derive a general formula for $\partial_{\tilde \mu} \mathcal{U}_F$, where $\tilde \mu \in \{-,j\}$. 
The derivative of the Wilson line Eq. \eqref{def:Wilson} is
\ba
\label{deriv_Wilson_1}
\partial_{\tilde \mu} \mathcal{U}_F(x^+,y^+;\v,v^-) = -ig \int_{y^+}^{x^+} dv^+ 
\mathcal{U}_F(x^+,v^+;\v,v^-) \partial_{\tilde \mu} \mathcal{A}^-(v) 
\mathcal{U}_F (v^+,y^+;\v,v^-).
\ea
Using the formula (\ref{F}) we can rewrite Eq. (\ref{deriv_Wilson_1}) as
\ba
\partial_{\tilde \mu} \mathcal{U}_F(x^+,y^+;\v,v^-) 
&=& 
-ig \int_{y^+}^{x^+} dv^+ 
\mathcal{U}_F(x^+,v^+;\v,v^-) \Big( \mathcal{F}_{\tilde \mu}^{\;-} (v)
+  \partial^- \mathcal{A}_{\tilde \mu}(v) 
-ig [ \mathcal{A}_{\tilde \mu}(v), \mathcal{A}^-(v)] \Big)
\mathcal{U}_F (v^+,y^+;\v,v^-) 
\nn \\
&=& 
-ig \int_{y^+}^{x^+} dv^+ 
\partial_{v^+}\bigg\{\mathcal{U}_F(x^+,v^+;\v,v^-)  \mathcal{A}_{\tilde \mu}(v)
\mathcal{U}_F (v^+,y^+;\v,v^-) \bigg\}
\nn \\
&& 
-ig \int_{y^+}^{x^+} dv^+ 
\bigg\{\mathcal{U}_F(x^+,v^+;\v,v^-) 
\Big( \mathcal{F}_{\tilde \mu}^{\;-} (v)
-\overleftarrow{\mathcal{D}_{v^+}}\mathcal{A}_{\tilde \mu}(v)
-\mathcal{A}_{\tilde \mu}(v)\overrightarrow{\mathcal{D}_{v^+}}  \Big)
\mathcal{U}_F (v^+,y^+;\v,v^-) 
\bigg\}
\label{deriv_Wilson_2}
\, ,
\ea
integrating by parts in $v^+$, and grouping the terms into covariant derivatives. Note that in the last line of Eq.\eqref{deriv_Wilson_2}, the covariant derivatives act only within the bracket, on the Wilson lines. By definition of the Wilson line 
one has
\ba
\label{cov_conserv_Wilson_1}
\overrightarrow{\mathcal{D}_{v^+}} \mathcal{U}_F (v^+,y^+;\v,v^-) &=& 
0
\, , \\
\label{cov_conserv_Wilson_2}
\mathcal{U}_F (x^+,v^+;\v,v^-) \overleftarrow{\mathcal{D}_{v^+}}  &=& 0 
\, .
\ea
Using these two relations, one can rewrite Eq.\eqref{deriv_Wilson_2} as
\ba
\label{deriv_Wilson}
\partial_{\tilde \mu} \mathcal{U}_F(x^+,y^+;\v,v^-)
&=&
-ig \int_{y^+}^{x^+} dv^+ 
\mathcal{U}_F(x^+,v^+;\v,v^-)  \mathcal{F}_{\tilde \mu}^{\;-} (v)
\mathcal{U}_F (v^+,y^+;\v,v^-) \nn \\
&&
-ig   \mathcal{A}_{\tilde \mu}(x^+,\v,v^-) \mathcal{U}_F (x^+,y^+;\v,v^-) 
+ig \mathcal{U}_F (x^+,y^+;\v,v^-) 
 \mathcal{A}_{\tilde \mu}(y^+,\v,v^-)
 \, .
\ea

Using the identity \eqref{deriv_Wilson}, the decorated Wilson lines appearing in the NEik DIS dijet amplitudes, presented in this section, can all be rewritten with local insertions of the background field strength. 
Reminding that we are using a gauge in which the gauge field vanishes outside of the target, we have in particular
%
\ba
\label{gauge_field_boundary_cond}
\mathcal{A}_{\tilde \mu}(L^+/2,\v,v^-) = \mathcal{A}_{\tilde \mu}(-L^+/2,\v,v^-)=0 
\, .
\ea
Combining Eqs.~\eqref{deriv_Wilson} and \eqref{gauge_field_boundary_cond}, 
the decorated Wilson lines of type 1 and 2 that appear in both longitudinal and transverse photon cases can be converted into
\ba  
\label{dec-D}
\mathcal{U}_{F;j}^{(1)} (\v)
&=&
\int_{-\frac{L^+}{2}}^{\frac{L^+}{2}} dz^+
(2z^+) \, \mathcal{U}_F\Big(\frac{L^+}{2},z^+;\v\Big) 
\Big(ig \mathcal{F}_{j}^{\;-} (z^+,\v)\Big)
\mathcal{U}_F \Big(z^+,-\frac{L^+}{2};\v\Big) \, ,\\
\label{dec-DD}
\mathcal{U}_F^{(2)} (\v)
&=&
\int_{-\frac{L^+}{2}}^{\frac{L^+}{2}} dz^+
\int^{\frac{L^+}{2}}_{-\frac{L^+}{2}}  dz'^+ 
\theta(z^+-z'^+) (z^+-z'^+)\nn \\
&&
\times\; 
\mathcal{U}_F\Big(\frac{L^+}{2},z^+;\v\Big) 
\Big(ig \mathcal{F}_{j}^{\;-} (z^+,\v)\Big)
\mathcal{U}_F \Big(z^+,z'^+;\v\Big)
\Big(ig \mathcal{F}_{j}^{\;-} (z'^+,\v)\Big)
\mathcal{U}_F \Big(z'^+,-\frac{L^+}{2};\v\Big)\, .
\ea
In order to write the contributions \eqref{ampl-q_dec_L}, \eqref{ampl-qbar_dec_L}, \eqref{Ampl-q_dec_T} and \eqref{Ampl-qbar_dec_T} with local decorations only, one also needs 
\ba
\label{partial_deriv_Wilson_full}
\partial_j \mathcal{U}_F(\v,v^-)
&=&
- \int_{-\frac{L^+}{2}}^{\frac{L^+}{2}}  dv^+ 
\mathcal{U}_F\left(\frac{L^+}{2},v^+;\v,v^-\right)  \Big(ig\mathcal{F}_{j}^{\;-} (v) \Big)\; 
\mathcal{U}_F \left(v^+,-\frac{L^+}{2};\v,v^-\right)
 \, .
\ea
Moreover, the structure that appears in the contributions \eqref{ampl-dyn_L} and \eqref{Ampl-dyn_T}  can be rewritten as
\ba
\label{dec-partial-min-both}
\mathcal{U}_F(\v,b^-) 
\overleftrightarrow{\partial_{-}} 
\mathcal{U}_F ( \w,b^-)^{\dag} \Big|_{b^-=0}
&=& 
\int_{-\frac{L^+}{2}}^{\frac{L^+}{2}}  dw^+ 
\mathcal{U}_F(\v)
\mathcal{U}_F \Big(w^+,-\frac{L^+}{2};\w\Big)^{\dag}
\Big( ig  \mathcal{F}^{+-} (w^+,\w)\Big)
\mathcal{U}_F\Big(\frac{L^+}{2},w^+;\w\Big)^{\dag} 
\nn \\
&&
-\int_{-\frac{L^+}{2}}^{\frac{L^+}{2}}  dv^+ 
\mathcal{U}_F\Big(\frac{L^+}{2},v^+;\v\Big) 
\Big( -ig  \mathcal{F}^{+-} (v^+,\v)\Big)
\mathcal{U}_F \Big(v^+,-\frac{L^+}{2};\v\Big) 
\mathcal{U}_F(\w)^{\dag}\, . 
\ea
Finally, in the transverse photon case the two extra structures that appear in Eqs. \eqref{Ampl-Lplus_T} and \eqref{ampl-in_T} can be transformed into   
\ba 
\label{loc_dec_Lplus}
\mathcal{U}_F(\z) \, 
\overleftrightarrow{\partial_j} \, 
\mathcal{U}_F ( \z)^{\dag} 
&=&
2 \int_{-\frac{L^+}{2}}^{\frac{L^+}{2}}  dz^+ 
\mathcal{U}_F \Big(\frac{L^+}{2}, z^+;\z\Big)
\Big( ig \mathcal{F}_{j}^{\;-} (z^+,\z)\Big)
\mathcal{U}_F \Big(\frac{L^+}{2},z^+;\z\Big)^{\dag} 
\ea
and 
\ba
&&
\label{loc_dec_in}
\int_{-\frac{L^+}{2}}^{\frac{L^+}{2}}  dz^+ 
\mathcal{U}_F\Big(\frac{L^+}{2},z^+;\z\Big) 
\overleftrightarrow{\mathcal{D}_j} \,  
\mathcal{U}_F \Big(\frac{L^+}{2},z^+;\z\Big)^{\dag}\nn \\
&&
=
2\int_{-\frac{L^+}{2}}^{\frac{L^+}{2}}  dz^+ \,
\Big(z^+ + \frac{L^+}{2}\Big)
\mathcal{U}_F \Big(\frac{L^+}{2},z^+;\z\Big) 
\Big(ig \mathcal{F}_{j}^{\;-} (z^+,\z)\Big)
\mathcal{U}_F \Big(\frac{L^+}{2},z^+;\z\Big)^{\dag} 
\, .
\ea
Remarkably, inserting the expressions \eqref{loc_dec_Lplus} and \eqref{loc_dec_in} into Eqs.\eqref{Ampl-Lplus_T}  and \eqref{ampl-in_T},  one finds a cancellation of the terms proportional to $L^+$, and one is left with
\begin{align}
i{\cal M }_{q_1 \bar q_2 \leftarrow \gamma^*_T}^{L^+\textrm{ phase}}
+ i{\cal M }_{q_1 \bar q_2 \leftarrow \gamma^*_T}^{\textrm{in}} 
 =  &\, 
  e e_f \, \varepsilon_{\lambda}^i \,
  \frac{1}{8k_1^+k_2^+}\;
  \bar u(1) \gamma^+\!
  \left[\frac{(k_2^+\!-\!k_1^+)}{q^+}\, \delta^{ij}+\frac{[\gamma^i,\gamma^j]}{2}\right]  
  v(2) \: 
 \int_{\z}\, e^{-i\z \cdot(\k_1+\k_2)}\:
 \nn \\
& \times\,
 \int_{-\frac{L^+}{2}}^{\frac{L^+}{2}}  dz^+\, (2z^+)\,  
\mathcal{U}_F \Big(\frac{L^+}{2}, z^+;\z\Big)
\Big( ig \mathcal{F}_{j}^{\;-} (z^+,\z)\Big)
\mathcal{U}_F \Big(\frac{L^+}{2},z^+;\z\Big)^{\dag}
\label{Ampl-Lplus-in_combined} 
\, .
\end{align}
At this stage, in each of the contributions to the amplitude, the target width $L^+$ appears only in two places: as endpoint of the Wilson lines, and as boundary of the integration over the position of the decorations.
However, our calculations have been done in a generic gauge assuming only that the background gauge field vanishes outside of the target. Hence, it is safe to extend all the background Wilson lines beyond the extent of the target, from $-\infty$ to $+\infty$ along the $x^+$ direction.
Moreover, after rewriting the decorations as local insertions of the background field strength, all the integration along the $x^+$ directions are now integrations over the position of a field strength insertion. Due to confinement, the background field strength of the target has to decay faster than a power in the longitudinal direction, and thus along $x^+$. Hence it is safe to extend all these integrations 
from $-\infty$ to $+\infty$ as well.
Hence, from now on, we will systematically drop $L^+$ in the boundary of both Wilson lines and integrations.

\section{Correlation limit of the next-to-eikonal DIS dijet production amplitude}
\label{Corr_limit_amplitude_NEik}


\subsection{Change of variables and definition of the limit}
\label{sec:cv_and_b2b_limit}
In order to study the correlation limit, it is convenient to change variables for the jet momenta as $(\k_1,\k_2) \to (\P,\k) $
with
\ba
\k &=& \k_1+\k_2,
\label{def_k}
 \\
\P &=& z_2\k_1 - z_1 \k_2
\label{def_P}
\, ,
\ea
so that
\ba
\k_1 &=& \P + z_1 \k, \\
\k_2 &=& -\P + z_2 \k
\, .
\ea
Here, we have used the notation
\ba
z_{1,2} &\equiv& \frac{k^+_{1,2}}{k_1^+\!+\!k_2^+} , 
\label{def_z1_z2}
 \ea
for the fraction of the lightcone $+$ momentum of the dijet system carried by each of the two jets.\footnote{Note that with the definition \eqref{def_z1_z2} which does not involve the photon momentum $q^+$, one has the relation $z_1+z_{2}=1$ automatically, even if a nonzero $+$ momentum is exchanged with the target.}
With these definitions, $\k$ is the total transverse momentum of the dijet system, whereas $\P$ is the relative transverse momentum between the two jets. For completeness, we also introduce the notation 
\ba
k^+ &\equiv& k_1^+\!+\!k_2^+ 
\ea
for the total $+$ momentum of the dijet system.
If one changes variables from $(k^+_1,\k_1,k^+_2,\k_2)$ to $(k^+,\k,z_1,\P)$ in order to describe the dijet final state, the two-particle phase space measure \eqref{phase_space_1_2} becomes
\begin{align}
d{\rm P.S.}=\frac{d^2\k}{(2\pi)^2}\frac{dk^+}{(2\pi)2k^+}\frac{d^2\P}{(2\pi)^2}\frac{dz_1}{(2\pi)2z_1(1\!-\!z_1)}\, . 
\label{phase_space_k_P_z}
\end{align}
The correlation limit (or back-to-back jets limit) is defined as $|\k_1+\k_2| \ll |\k_1|,|\k_2| $
or equivalently 
$|\k|\ll |\P|$ after the change of variables. 

In most of the contributions to the amplitude (apart from the simpler \eqref{Ampl-Lplus-in_combined}), the phase factors become 
\ba
e^{-i\v \cdot\k_1}\; e^{-i\w \cdot\k_2} = e^{-i\P \cdot(\v-\w)}\; e^{-i\k\cdot(z_1\v+z_2\w)}
\ea
under the change of variables~\eqref{def_k} and~\eqref{def_P}. Hence, it is convenient to change variables for the transverse positions as well, as  $(\v,\w) \to (\r,\b)$, where
\ba
\r &=& \v-\w
\label{def_r_from_v_w} \\
\b &=& z_1 \v + z_2 \w \, , 
\label{def_b_from_v_w}
\ea
and thus
\ba
\v &=& \b +z_2 \r  
\label{def_v_from_r_b}\\
\w&=& \b-z_1 \r
\label{def_w_from_r_b}
\, .
\ea
In these new variables, the dipole size $\r$ is then the Fourier conjugate of the relative momentum $\P$, and the dipole impact paramater $\b$ is the Fourier conjugate of the total transverse momentum $\k$ of the dijet, as
\ba
e^{-i\v \cdot\k_1}\; e^{-i\w \cdot\k_2} = e^{-i\P \cdot\r}\; e^{-i\k\cdot\b}
\label{phase_fact_change_var}
\, .
\ea
Hence, enforcing the back-to-back jets kinematics $|\k|\ll |\P|$ on each contribution to the amplitude amounts to impose  $|\r| \ll |\b| $ in the rest of the integrand multiplying the phase factors~\eqref{phase_fact_change_var}. In particular, the condition $|\r| \ll |\b|$ allows one to perform the expansion of the color operators involved in the amplitude order by order as powers of $\r$, that we will perform in Sec.~\ref{sec:color_struct_exp}. But first, let us discuss power counting. 

%

\subsection{Power counting: high-energy vs correlation limits}
\label{sec:powcount}    

In this article, we are studying the overlap of the high-energy regime and of the back-to-back jets regime of the cross section, each associated with an expansion in powers. The energy or momentum  scales relevant to define these regimes for DIS dijet production are the total transverse momentum of the dijet system $\k$ (see Eq.~\eqref{def_k}), the relative transverse momentum between the jets $\P$ (see Eq.~\eqref{def_P}) and the center-of-mass energy of the photon-target collision $W$.

The high energy regime, aka the Regge-Gribov limit, is when $W$ is much larger than any transverse momentum (or virtuality or mass) scale, all considered of the same order:
\begin{align}
\k^2 \sim   &\, \P^2 \ll W^2   \, . 
\label{Regge_regime_def}
\end{align}
The eikonal approximation corresponds to the leading power in this regime, and subleading power corrections in this regime are suppressed as
\begin{align}
 &\, \left(\frac{\P^2}{W^2}\right)^n\, ,\left(\frac{\k^2}{W^2}\right)^n   \, , 
\label{NnEik_def}
\end{align}
where $n=1$ corresponds to next-to-eikonal (NEik) corrections, $n=2$ to next-to-next-to-eikonal (NNEik) corrections and so on.

By contrast, the back-to-back jets regime, also called correlation limit, corresponds to $|\P|$ much larger than $|\k|$:
\begin{align}
|\k| \ll    &\, |\P|  \, . 
\label{b2b_regime_def}
\end{align}
In that regime, subleading power corrections are suppressed as 
\begin{align}
 &\, \left(\frac{|\k|}{|\P|}\right)^n   \, , 
\label{NnLP_def}
\end{align}
with respect to the leading power contribution, with $n=0$ corresponding to the leading power (LP), $n=1$ to next-to-leading power correction (NLP), $n=2$ to next-to-next-to-leading power correction (NNLP), and so on.

In the Bjorken limit, where $|\P|\sim W \to \infty$, next-to-eikonal corrections of the type $(\k^2/W^2)^n$ will look like next-to-leading power corrections. In order to understand the interplay between these two expansions in powers, we are led to consider the overlap region between the Regge-Gribov limit and the back-to-back jets regimes, meaning that 
\begin{align}
|\k| \ll    &\, |\P| \ll  W   \, . 
\label{overlap_regime_def}
\end{align}
In that case both types of power corrections \eqref{NnEik_def} and \eqref{NnLP_def} coexist and can unambiguously be distinguished.
 
As extra scales in the problem, we also have the photon virtuality $Q$ and the quark mass $m$. Provided that they are of the order of $|\P|$ at most, 
\begin{align}
0 \leq    &\, Q \lesssim  |\P|   \quad \textrm{and} \quad 0 \leq   m \lesssim  |\P|\, , 
\label{Q_m_less_or_order_P}
\end{align}
they will not affect the power counting. In particular, their relative size compared to $|\k|$ is irrelevant. The only effect of $Q$, under the conditions \eqref{Q_m_less_or_order_P}, is that the cross section via longitudinal photon exchange is suppressed by an overall factor $Q^2/\P^2$ compared to the cross section via transverse photon exchange, in order to ensure the cancellation of the former in the real photon limit $Q^2\rightarrow 0$. Hence, we will provide our results with full dependence on $Q$ and $m$, assuming only the conditions \eqref{Q_m_less_or_order_P}.

The center-of-mass energy of the photon-target collision $W$ is defined as $W^2 \equiv (P_{\textrm{tar}}\!+\! q)^2$. In the regime of interest, $W^2 \gg Q^2 \gtrsim M_{\textrm{tar}}^2$ (where $M_{\textrm{tar}}$ is the target mass), one finds
\begin{align}
2q^+P_{{tar}}^- =    &\, W^2 + O(Q^2) \, . 
\label{W^2_approx}
\end{align}
Since the eikonal contribution to cross sections is independent of $W^2$ (possibly up to logarithms when resumming higher orders in $\alpha_s$), the NEik correction can be equivalently defined as suppressed by one power of $W^2$ (like in Eq.~\eqref{NnEik_def}) or as suppressed by one power of  $2q^+P_{{tar}}^-$, with the same coefficient.

It remains now to specify the power counting assigned to quantities associated with the target, such as background field and covariant derivatives.
First, gauge links are irrelevant for the power counting (once no derivative or covariant derivative is acting on them). Second, for each direction $\mu$, we assign the same power counting to the background covariant derivative $\mathcal{D}_{\mu}$, the partial derivative $\partial_{\mu}$ and the background gauge field $\mathcal{A}_{\mu}$, provided that the derivatives are acting on a background field (gauge field, field strength or gauge link).

In the regime $|\k|\ll |\P|$ and thus $|\r| \ll |\b| $, it is clear from Eqs.~\eqref{def_v_from_r_b} and~\eqref{def_w_from_r_b} that all transverse derivatives acting on a background field tend to derivatives with respect to $\b$, so that 
\begin{align}
\mathcal{D}_{i}=    &\, O\left(|\b|^{-1}\right)=O(|\k|) \, . 
\label{perp_deriv_power_count}
\end{align}

As discussed in Refs.~\cite{Altinoluk:2020oyd,Altinoluk:2021lvu,Altinoluk:2022jkk}, the high energy limit can be understood as the limit of infinite boost of the target, and thus the power counting in the 
high energy regime is determined by covariance properties under longitudinal Lorentz boosts. 
For example, under such as boost, the component $\mathcal{A}^{-}=\mathcal{A}_{+}$ of the background field is enhanced by one power of the boost factor. Moreover, due to Lorentz length contraction, the dependence of the background fields on $x^+$ become faster, so that $\partial_+$ and $\mathcal{D}_+$ should be counted as enhanced by one power of the boost factor as well. Since derivatives have a dimension of momentum, and that $P_{{tar}}^-$ is the component of the momentum of the target which is enhanced by one power of the boost factor of the target, we assign the power counting rule
\begin{align}
\mathcal{D}_{+}=    &\, O(P_{{tar}}^-) \, . 
\label{+_deriv_power_count}
\end{align}

By contrast, the component $\mathcal{A}^{+}=\mathcal{A}_{-}$ of the background gauge field is suppressed by one power of the boost factor, under a longitudinal Lorentz boost of the target. It is also the case for derivatives $\partial_-$ and $\mathcal{D}_-$ acting on a background field, due to Lorentz time dilation. In order to match the physical dimension and the behavior under boosts, 
we assign 
\begin{align}
\mathcal{D}_{-}=    &\, O\left(\frac{\k^2}{P_{{tar}}^-}\right) \, . 
\label{-_deriv_power_count}
\end{align}

Remembering the definition of the background field strength \eqref{def_F_mu_nu}, the power counting for $\mathcal{F}_{\mu\nu}$ is obtained from the ones for the background derivatives $\mathcal{D}_{\mu}$ and  $\mathcal{D}_{\nu}$ as
\begin{align}
O\left(\mathcal{F}_{\mu\nu}\right)=    &\, O\left(\mathcal{D}_{\mu}\right)\; O\left(\mathcal{D}_{\nu}\right) \, , 
\label{F_mu_nu_power_count_from_Dmu_Dnu}
\end{align}
so that in particular
\begin{align}
\mathcal{F}_{i+}
=    &\, 
{\mathcal{F}_{i}}^{-} 
=O\left(|\k|\, {P_{{tar}}^-}\right)
\\
\mathcal{F}_{ij}
=    &\, 
O\left(\k^2\right)
\\
\mathcal{F}^{+-}
=    &\, 
O\left(\k^2\right)
\\
\mathcal{F}_{i-}
=    &\, 
{\mathcal{F}_{i}}^{+} 
=O\left(\frac{|\k|^3}{ {P_{{tar}}^-}}\right)
\, .
\end{align}
Hence, the components $\mathcal{F}_{ij}$ and $\mathcal{F}^{+-}$ of the background field strength appear power suppressed with respect to the leading components ${\mathcal{F}_{i}}^{-} $ both in the high-energy regime \eqref{Regge_regime_def} and in the back-to-back jets regime \eqref{b2b_regime_def}. And the components ${\mathcal{F}_{i}}^{+}$ are further power suppressed in both regimes. Note that the hierarchy between the components of the background field strength, or between the components of the background gauge field or covariant derivative, is the same in the high-energy regime~\eqref{Regge_regime_def} and in the back-to-back jets regime~\eqref{b2b_regime_def}.

Finally, let us remind that the background field strength $\mathcal{F}_{\mu\nu}(x)$ has a fast decay outside of an interval of width $L^+$ (centered around $x^+=0$ by convention) representing the target, as discussed in the paragraph after Eq.~\eqref{Ampl-Lplus-in_combined}. Due to Lorentz length contraction, this width $L^+$ is suppressed by one power of the boost factor, under a longitudinal Lorentz boost of the target, leading to the shockwave approximation at the eikonal order. Hence, the power counting for the integration over the $x^+$ coordinate of  $\mathcal{F}_{\mu\nu}(x)$ is
\begin{align}
\int dx^+\, \mathcal{F}_{\mu\nu}(x) =    &\, O\left(L^+\, \mathcal{F}_{\mu\nu}\right)
= O\left(\frac{\mathcal{F}_{\mu\nu}}{P_{{tar}}^-}\right)
\\
\int dx^+\, x^+\, \mathcal{F}_{\mu\nu}(x) =    &\, O\left((L^+)^2\, \mathcal{F}_{\mu\nu}\right)
= O\left(\frac{\mathcal{F}_{\mu\nu}}{(P_{{tar}}^-)^2}\right)
 \, . 
\label{int_+_F_mu_nu_power_count}
\end{align}

With all the power counting rules introduced in this section, one can estimate the order of any contribution to the amplitude or cross section for DIS dijet production in the regime \eqref{overlap_regime_def}, and thus identify leading power terms and powers corrections of the type \eqref{NnEik_def} or \eqref{NnLP_def}. In particular,  for $|\k|\sim|\P|$, these power counting rules reduce to the ones of the high-energy regime (see for example Refs.~\cite{Altinoluk:2020oyd,Altinoluk:2021lvu,Altinoluk:2022jkk}). And for $|\P|\sim W$ these power counting rules reduce to the ones used for example in Ref.~\cite{Mulders:2000sh} in the context of TMD factorization.

As an example, with these power counting rules, one finds for the decorated Wilson lines \eqref{Wilson_dec_1},  \eqref{Wilson_dec_2}, and  \eqref{Wilson_dec_3}, 
\begin{align}
\mathcal{U}^{(1)}_{F;j} ( \v) 
=&\,   O\left(\frac{|\k|}{P_{{tar}}^-}\right) 
\label{dec_Wilson_line_1_power_count}\\
\mathcal{U}^{(2)}_F ( \v) 
=&\, O\left(\frac{\k^2}{P_{{tar}}^-}\right) 
\label{dec_Wilson_line_2_power_count}\\
\mathcal{U}^{(3)}_{F; ij} ( \v)
 =&\,  
O\left(\frac{\k^2}{P_{{tar}}^-}\right)
\label{dec_Wilson_line_3_power_count}\, .
\end{align}


\subsection{Expansion of the color structures in the correlation limit}
\label{sec:color_struct_exp}

\subsubsection{Color structure in generalized eikonal amplitudes}

Both in the  longitudinal~\eqref{bdep_Ampl-GenEik_L} and transverse~\eqref{Ampl-GenEik_T}   photon cases, the generalized eikonal amplitude is of the form
%
\begin{align}
{\cal O}^{\rm Gen.\,  Eik} [f]
=&\, 
\int_{\v,\w}\,  e^{-i\v \cdot\k_1} \, e^{-i\w\cdot\k_2}\, 
f\left(\v\!-\!\w\right)
\Big[\mathcal{U}_F(\v,b^- )
\mathcal{U}_F(\w,b^- )^{\dag}
\!-\!1\Big]
\, ,
\end{align}
with some numerical function $f(\r)$. Performing the change of variables introduced in Sec.~\ref{sec:cv_and_b2b_limit} (and dropping $L^+$ as discussed in the section \ref{sec:dec_Wilson_F}), one finds
\begin{align}
{\cal O}^{\rm Gen.\,  Eik} [f]
=&\, 
\int_{\r,\b}\,  e^{-i\r \cdot\P} \, e^{-i\b\cdot\k}\, 
f\left(\r\right)
\Big[\mathcal{U}_F(\b +z_2 \r ,b^- )
\mathcal{U}_F(\b-z_1 \r,b^- )^{\dag}
\!-\!1\Big]
\, .
\end{align}
At zeroth order in $|\r|$, one obtains a trivial result $\mathcal{U}_F(\b  ,b^- )
\mathcal{U}_F(\b,b^- )^{\dag}
\!-\!1=0$. By contrast, the terms of first order in $|\r|$ provide the LP contribution in the back-to-back limit~\cite{Dominguez:2011wm}. Since we are interested in the interplay between NEik corrections and NLP terms in the back-to-back limit, we should expand the open dipole operator to second order in $|\r|$, as
\begin{align}
{\cal O}^{\rm Gen.\,  Eik} [f]
=&\, 
\int_{\r,\b}\,  e^{-i\r \cdot\P} \, e^{-i\b\cdot\k}\, 
f\left(\r\right)\bigg[ z_2\r^i\big(\d_i\mathcal{U}_F(\b,b^-)\big)\mathcal{U}_F(\b,b^-)^\dagger
- z_1\r^i\mathcal{U}_F(\b,b^-)\big(\d_i\mathcal{U}_F(\b,b^-)^\dagger\big)\nonumber\\
&
+\frac{z^2_2}{2}\r^i\r^j\big(\d_i\d_j\mathcal{U}_F(\b,b^-)\big)\mathcal{U}_F(\b,b^-)^\dagger
+\frac{z^2_1}{2}\r^i\r^j\mathcal{U}_F(\b,b^-)\big(\d_i\d_j\mathcal{U}_F(\b,b^-)^\dagger\big)\nonumber\\
&
-z_1z_2\r^i\r^j\big(\d_i\mathcal{U}_F(\b,b^-)\big)\big(\d_j\mathcal{U}_F(\b,b^-)^\dagger\big)
+O\left(\frac{|\r|^3}{|\b|^3}\right)\bigg] \nonumber\\
=&\, 
\int_{\r,\b}\,  e^{-i\r \cdot\P} \, e^{-i\b\cdot\k}\, 
f\left(\r\right)
\bigg[ \Big(1+i\frac{(z_2-z_1)}{2}\r\cdot \k \Big)\r^i\big(\d_i\mathcal{U}_F(\b,b^-)\big)\mathcal{U}_F(\b,b^-)^\dagger
\nonumber\\
&
-\frac{1}{2}\r^i\r^j\big(\d_i\mathcal{U}_F(\b,b^-)\big)\big(\d_j\mathcal{U}_F(\b,b^-)^\dagger\big)
+O\left(\frac{|\r|^3}{|\b|^3}\right)\bigg]\, ,
\end{align}
where we performed integration by parts and used $z_1\!+\!z_2=1$ in order to combine the terms. Using Eq.~\eqref{partial_deriv_Wilson_full}, the expanded open dipole operator can be rewritten as 
\begin{align}
{\cal O}^{\rm Gen.\,  Eik} [f]
=&\, 
\int_{\r,\b}\,  e^{-i\r \cdot\P} \, e^{-i\b\cdot\k}\, 
f\left(\r\right)\nonumber\\
&\times\, 
\bigg[ \Big(1+\frac{i(z_2\!-\!z_1)}{2}\r\!\cdot\! \k \Big)\r^i
 \int  dv^+ 
\mathcal{U}_F\left(+\infty,v^+;\b,b^-\right) \big[ -ig\mathcal{F}_{i}^{\;-} (v^+,\b,b^-) \big]\; 
\mathcal{U}_F\left(+\infty,v^+;\b,b^-\right)^\dagger
\nonumber\\
&\hspace{0.5cm}
-\frac{1}{2}\r^i\r^j
\int  dv^+ \int  dw^+ 
\mathcal{U}_F\left(+\infty,v^+;\b,b^-\right) \big[ -ig\mathcal{F}_{i}^{\;-} (v^+,\b,b^-) \big]\; \mathcal{U}_F\left(v^+,-\infty;\b,b^-\right) 
\nonumber\\
&\hspace{1cm}\times\, 
\mathcal{U}_F\left(w^+,-\infty;\b,b^-\right)^\dagger
\big[ +ig\mathcal{F}_{j}^{\;-} (w^+,\b,b^-) \big]\; \mathcal{U}_F\left(+\infty,w^+;\b,b^-\right)^\dagger
+O\left(\frac{|\r|^3}{|\b|^3}\right)\bigg]\, ,
\label{expand_GEik_op_1}
\end{align}
%
From the definition of the adjoint representation, one has the relation between fundamental and adjoint Wilson lines on the same path
\begin{align}
\mathcal{U}_F\left(x^+,y^+;\b,b^-\right) t^a \; \mathcal{U}_F\left(x^+,y^+;\b,b^-\right)^\dagger
=&\,   t^b \, \mathcal{U}_A\left(x^+,y^+;\b,b^-\right)_{ba}
\label{rel_adj_fund}
\, .
\end{align}
   
     Using the relation \eqref{rel_adj_fund}, the expression \eqref{expand_GEik_op_1} can be simplified as
\begin{align}
{\cal O}^{\rm Gen.\,  Eik} [f]
=&\, 
\int_{\r,\b}\,  e^{-i\r \cdot\P} \, e^{-i\b\cdot\k}\, 
f\left(\r\right)\nonumber\\
&\times\, 
\bigg[ \Big(1+\frac{i(z_2\!-\!z_1)}{2}\r\!\cdot\! \k \Big)\r^i 
 (-i)\int  dv^+ 
t^{a'}\, \mathcal{U}_A\left(+\infty,v^+;\b,b^-\right)_{a'a} g{\mathcal{F}_{i}^{\;-}}_a (v^+,\b,b^-)\; 
\nonumber\\
&\hspace{0.5cm}
-\frac{1}{2}\r^i\r^j
\int  dv^+ \int  dw^+ 
t^{a'} t^{b'}\, \mathcal{U}_A\left(+\infty,v^+;\b,b^-\right)_{a'a} g{\mathcal{F}_{i}^{\;-}}_a (v^+,\b,b^-)\; 
\nonumber\\
&\hspace{1cm}\times
\mathcal{U}_A\left(+\infty,w^+;\b,b^-\right)_{b'b} g{\mathcal{F}_{j}^{\;-}}_b (w^+,\b,b^-)\; 
+O\left(\frac{|\k|^3}{|\P|^3}\right)\bigg]
\, ,
\label{expand_GEik_op_2}
\end{align}
using that $O(|\b|)=O(|\k|^{-1}) $ and  $O(|\r|)=O(|\P|^{-1})$.  
In this small $\r$ expansion, note that at each order, the color structure becomes independent of $\r$, and $\r$ appears only in the numerical factor.

\subsubsection{Color structures of the contributions with decorated Wilson lines $\mathcal{U}^{(1)}_{F;i}$ and $\mathcal{U}^{(2)}_F$}    
  
Let us now consider the contributions to the amplitude with decorated Wilson lines $\mathcal{U}^{(1)}_{F;i}$ and $\mathcal{U}^{(2)}_F$ on the quark line. From Eqs.~\eqref{ampl-q_dec_L}  and~\eqref{Ampl-q_dec_T} in the longitudinal and transverse photon cases respectively, one finds these contributions to be of the form 
\begin{align}
{\cal O}^{\textrm{dec. on }q}_{(1)+(2)}[f]
=&\, 
\int_{\v,\w} e^{-i\v \cdot\k_1}\, e^{-i\w\cdot\k_2}\, \frac{f( \v\!-\!\w)}{2k_1^+} 
\left[
- i\, \mathcal{U}^{(2)}_F ( \v)\,
+ \mathcal{U}^{(1)}_{F;i} ( \v) \,  \bigg(\frac{(\k_2^i\!-\!\k_1^i)}{2}+\frac{i}{2}\, \partial_{\w^i}\bigg)
 \right]
\mathcal{U}_F(\w)^{\dag}\,
\nonumber
\\
=&\, 
\int_{\r,\b} e^{-i\r \cdot\P}\, e^{-i\b\cdot\k}\, \,\frac{f( \r)}{2k_1^+}
\bigg[
- i\, \mathcal{U}^{(2)}_F ( \b +z_2 \r)\, \mathcal{U}_F(\b-z_1 \r)^{\dag}
\nonumber\\
&
-\Big(\P^i+\frac{(z_1-z_2)}{2}\k^i\Big) \, \mathcal{U}^{(1)}_{F;i} ( \b +z_2 \r) \, \mathcal{U}_F(\b-z_1 \r)^{\dag}
+\frac{i}{2}  \, \mathcal{U}^{(1)}_{F;i} ( \b +z_2 \r)\, \partial_{i}  \mathcal{U}_F(\b-z_1 \r)^{\dag}\bigg]
\label{O_dec_q_1plus2_1}
\, ,
\end{align}
 again with a numerical function $f(\r)$, and an explicit factor $1/2k_1^+$ included for later convenience.
 
 This time, all terms in Eq.~\eqref{O_dec_q_1plus2_1} provide a non-zero result at zeroth order in $\r$. Note that one term in the bracket comes with a $\P$ prefactor. Since the small $\r$ expansion is equivalent to a large $\P$ expansion, as discussed in sec.~\ref{sec:cv_and_b2b_limit}, the term multiplied by $\P$ should be expanded to one more order in $\r$ compared to the others in order to compensate the large $\P$. Hence, in order to obtain the first two non-trivial contributions in the small $\r$ and large $\P$ limit, one should expand the operator multiplied by $\P$ to linear order in  $\r$ and all the other terms to zeroth order in $\r$.
\begin{align}
{\cal O}^{\textrm{dec. on }q}_{(1)+(2)}[f]
=&\, 
\int_{\r,\b} e^{-i\r \cdot\P}\, e^{-i\b\cdot\k}\, \,\frac{f( \r)}{2k_1^+}
\Bigg\{
- i\, \mathcal{U}^{(2)}_F ( \b )\, \mathcal{U}_F(\b)^{\dag}
+\frac{(z_2\!-\!z_1)}{2}\k^i\, \mathcal{U}^{(1)}_{F;i} ( \b ) \, \mathcal{U}_F(\b)^{\dag}
+\frac{i}{2}  \, \mathcal{U}^{(1)}_{F;i} ( \b )\, \partial_{i}  \mathcal{U}_F(\b)^{\dag}
\nonumber\\
&
-\P^i \, \bigg[\mathcal{U}^{(1)}_{F;i} ( \b ) \, \mathcal{U}_F(\b)^{\dag}
-z_1 \r^j\, \mathcal{U}^{(1)}_{F;i} ( \b ) \, \partial_{j}\mathcal{U}_F(\b)^{\dag}
+z_2 \r^j\Big(\partial_j\mathcal{U}^{(1)}_{F;i} ( \b) \Big)\, \mathcal{U}_F(\b)^{\dag}
\bigg]
 +O\left(\frac{|\r|}{|\b|^3\, P_{{tar}}^-}\right)
\Bigg\}
\nonumber\\
=&\, 
\int_{\r,\b} e^{-i\r \cdot\P}\, e^{-i\b\cdot\k}\, \frac{f( \r)}{2k_1^+}
\Bigg\{
\bigg(-\P^i +\frac{(z_2\!-\!z_1)}{2}\k^i\, -i z_2\P^i (\r\!\cdot\!\k)
\bigg)\, \mathcal{U}^{(1)}_{F;i} ( \b ) \, \mathcal{U}_F(\b)^{\dag}
\nonumber\\
&
- i\, \mathcal{U}^{(2)}_F ( \b )\, \mathcal{U}_F(\b)^{\dag}
+\bigg(\frac{i}{2}\, \delta^{ij} +\P^i \r^j \bigg)\, \mathcal{U}^{(1)}_{F;i} ( \b )\, \partial_{j}  \mathcal{U}_F(\b)^{\dag}
+O\left(\frac{|\k|^3}{|\P|\, P_{{tar}}^-}\right)
\Bigg\}\, ,
\label{O_dec_q_1plus2_2}
\end{align}
 after integrating by parts in $\b$ in the term with $\partial_j\mathcal{U}^{(1)}_{F;i} ( \b)$. In Eq.~\eqref{O_dec_q_1plus2_2}, the first term in the bracket, proportional to $-\P^i$, is of order  $|\P||\k|/P_{{tar}}^-$ (see Eq.~\eqref{dec_Wilson_line_1_power_count}). By contrast, all five other terms explicitly written in the bracket are of order $|\k|^2/P_{{tar}}^-$.

 Using the identity~\eqref{rel_adj_fund} together with the expressions \eqref{dec-D}, \eqref{dec-DD} and \eqref{partial_deriv_Wilson_full}, the color structures appearing in Eq.~\eqref{O_dec_q_1plus2_2} can be rewritten as 
\begin{align}
 \mathcal{U}^{(1)}_{F;i} ( \b )\,   \mathcal{U}_F(\b)^{\dag}
 =&
\int dz^+
(2z^+) \, \mathcal{U}_F(+\infty,z^+;\b) 
\Big(ig \mathcal{F}_{i}^{\;-} (z^+,\b)\Big)
\mathcal{U}_F(+\infty,z^+;\b)^\dagger
 \nonumber\\
 =&\,
i \int dz^+ \, (2z^+) \,  
t^{a'} \, 
\mathcal{U}_A(+\infty,z^+;\b)_{a'a}\;  g {\mathcal{F}_{i}^{\;-}}_a (z^+,\b)
\label{U1_Udag}
\, ,
\end{align}
\begin{align}
\mathcal{U}^{(2)}_F ( \b)\, \mathcal{U}_F(\b)^{\dag}
&=
\int dz^+ \int dz'^+ \theta(z^+\!-\!z'^+) (z^+\!-\!z'^+)
\mathcal{U}_F(+\infty,z^+;\b) \Big(ig \mathcal{F}_{j}^{\;-} (z^+,\b)\Big) \mathcal{U}_F(+\infty,z^+;\b)^\dagger
\nn \\
&
\times\; 
\mathcal{U}_F(+\infty,z'^+;\b) \Big(ig \mathcal{F}_{j}^{\;-} (z'^+,\b)\Big) \mathcal{U}_F(+\infty,z'^+;\b)^\dagger
 \nonumber\\
&=-
\int dz^+ \int dz'^+ \theta(z^+\!-\!z'^+) (z^+\!-\!z'^+) \, 
t^{a'}t^{b'} \, 
\mathcal{U}_A(+\infty,z^+;\b)_{a'a}\;  g {\mathcal{F}_{j}^{\;-}}_a (z^+,\b)\nonumber\\
&\hspace{2.5cm}
\times\, 
\mathcal{U}_A(+\infty,z'^+;\b)_{b'b} \; g {\mathcal{F}_{j}^{\;-}}_b (z'^+,\b)
\label{U2_Udag}
\end{align}
and
\begin{align}
 \mathcal{U}^{(1)}_{F;i} ( \b )\, \partial_{j}  \mathcal{U}_F(\b)^{\dag}
 =&
 \int dz^+
(2z^+) \, \mathcal{U}_F(+\infty,z^+;\b) \, 
(ig) \mathcal{F}_{i}^{\;-} (z^+,\b)
\mathcal{U}_F(z^+,-\infty;\b) \nonumber\\
&\times
\int dz'^+ 
\mathcal{U}_F \left(z'^+,-\infty;\b\right)^\dagger\,  (ig)\mathcal{F}_{j}^{\;-} (z'^+,\b) \; 
 \mathcal{U}_F\left(+\infty,z'^+;\b\right)^\dagger \nonumber\\
 =&
 -\int dz^+ \int dz'^+ \, (2z^+) \,  
t^{a'}t^{b'} \, 
\mathcal{U}_A(+\infty,z^+;\b)_{a'a}\;  g {\mathcal{F}_{i}^{\;-}}_a (z^+,\b)
\mathcal{U}_A(+\infty,z'^+;\b)_{b'b} \; g {\mathcal{F}_{j}^{\;-}}_b (z'^+,\b)
\label{U1_dUdag}
\, .
\end{align}

Using  the expressions \eqref{U2_Udag} and \eqref{U1_dUdag}, one finds for the terms in the last line of Eq.~\eqref{O_dec_q_1plus2_2} 
\begin{align}
&\,
 - i\, \mathcal{U}^{(2)}_F ( \b )\, \mathcal{U}_F(\b)^{\dag}
+\bigg[\frac{i}{2}\, \delta^{ij} +\P^i \r^j \bigg]\, \mathcal{U}^{(1)}_{F;i} ( \b )\, \partial_{j}  \mathcal{U}_F(\b)^{\dag}
\nonumber\\
=&\,
i\, \int dz^+ \int dz'^+ \theta(z^+\!-\!z'^+) (z^+\!-\!z'^+) \, t^{a'}t^{b'} \,
\mathcal{U}_A(+\infty,z^+;\b)_{a'a}\;  g {\mathcal{F}_{j}^{\;-}}_a (z^+,\b)
\mathcal{U}_A(+\infty,z'^+;\b)_{b'b} \; g {\mathcal{F}_{j}^{\;-}}_b (z'^+,\b)
\nonumber\\
&\, -\bigg[\frac{i}{2}\, \delta^{ij} +\P^i \r^j \bigg]\,
 \int dz^+ \int dz'^+ \, (2z^+) \,  
t^{a'}t^{b'} \, 
\mathcal{U}_A(+\infty,z^+;\b)_{a'a}\;  g {\mathcal{F}_{i}^{\;-}}_a (z^+,\b)
\mathcal{U}_A(+\infty,z'^+;\b)_{b'b} \; g {\mathcal{F}_{j}^{\;-}}_b (z'^+,\b)
\nonumber\\
=&\,
\, t^{a'}t^{b'} \,\int dz^+ \int dz'^+  
\mathcal{U}_A(+\infty,z^+;\b)_{a'a}\;  g {\mathcal{F}_{i}^{\;-}}_a (z^+,\b)
\mathcal{U}_A(+\infty,z'^+;\b)_{b'b} \; g {\mathcal{F}_{j}^{\;-}}_b (z'^+,\b)
\nonumber\\
&\,
\times\, 
\left\{
i\theta(z^+\!-\!z'^+) (z^+\!-\!z'^+) \, \delta^{ij}
 -\bigg[\frac{i}{2}\, \delta^{ij} +\P^i \r^j \bigg]\, (2z^+) \,
\right\}
\nonumber\\
=&\,
\, t^{a'}t^{b'} \,\int dz^+ \int dz'^+  
\mathcal{U}_A(+\infty,z^+;\b)_{a'a}\;  g {\mathcal{F}_{i}^{\;-}}_a (z^+,\b)
\mathcal{U}_A(+\infty,z'^+;\b)_{b'b} \; g {\mathcal{F}_{j}^{\;-}}_b (z'^+,\b)
\nonumber\\
&\,
\times\, 
\left\{
-i\delta^{ij}\Big[\theta(z^+\!-\!z'^+) z'^+  + \theta(z'^+\!-\!z^+) z^+ \Big]
 -\P^i \r^j\, (2z^+) \,
\right\}
\, ,
\end{align}
so that  Eq.~\eqref{O_dec_q_1plus2_2} becomes (using the expression \eqref{U1_Udag} as well)
\begin{align}
{\cal O}^{\textrm{dec. on }q}_{(1)+(2)}[f]
=&\, 
\int_{\r,\b} e^{-i\r \cdot\P}\, e^{-i\b\cdot\k}\, \frac{f( \r)}{2k_1^+}
\nonumber\\
&\,\times
\Bigg\{
 \bigg[-\P^i +\frac{(z_2\!-\!z_1)}{2}\k^i\, -i z_2\P^i (\r\!\cdot\!\k)
\bigg]\, 
i t^{a'} \, 
 \int dz^+ \, (2z^+) \,  
\mathcal{U}_A(+\infty,z^+;\b)_{a'a}\;  g {\mathcal{F}_{i}^{\;-}}_a (z^+,\b)
\nonumber\\
&\,
+t^{a'}t^{b'} \,\int dz^+ \int dz'^+  
\mathcal{U}_A(+\infty,z^+;\b)_{a'a}\;  g {\mathcal{F}_{i}^{\;-}}_a (z^+,\b)
\mathcal{U}_A(+\infty,z'^+;\b)_{b'b} \; g {\mathcal{F}_{j}^{\;-}}_b (z'^+,\b)
\nonumber\\
&\,
\hspace{1cm}
\times\, 
\Big\{
-i\delta^{ij} \min(z^+,z'^+)
 -\P^i \r^j\, (2z^+) \,
\Big\}
+ O\left(\frac{|\k|^3}{|\P|\, P_{{tar}}^-}\right) 
\Bigg\}\, .
\label{O_dec_q_1plus2_3_bis}
\end{align}

The contributions to the amplitude with decorated Wilson lines $\mathcal{U}^{(1)}_{F;i}$ and $\mathcal{U}^{(2)}_F$ on the antiquark line can be read off from Eqs.~\eqref{ampl-qbar_dec_L}  and~\eqref{Ampl-qbar_dec_T} in the longitudinal and transverse photon cases respectively. One finds
\begin{align}
{\cal O}^{\textrm{dec. on }\bar{q}}_{(1)+(2)}[f]
=&\, 
\int_{\v,\w} e^{-i\v \cdot\k_1}\, e^{-i\w\cdot\k_2}\, \frac{f( \v\!-\!\w)}{2k_2^+}\; 
\Bigg\{ \mathcal{U}_F ( \v)\,\left[
- i\, \mathcal{U}^{(2)}_F (\w)^{\dag}\,
+   \bigg(-\frac{(\k_2^i\!-\!\k_1^i)}{2}+\frac{i}{2}\, \overleftarrow{\partial_{\v^i}}\bigg)\, \mathcal{U}^{(1)}_{F;i} (\w)^{\dag}
 \right]\Bigg\}
\nonumber
\\
=&\, 
\int_{\r,\b} e^{-i\r \cdot\P}\, e^{-i\b\cdot\k}\, \,\frac{f(\r)}{2k_2^+}\;
\bigg[
- i\, \mathcal{U}_F ( \b +z_2 \r)\, \mathcal{U}_F^{(2)}(\b-z_1 \r)^{\dag}
\nonumber\\
&
+\Big(\P^i-\frac{(z_2-z_1)}{2}\k^i\Big) \, \mathcal{U}_F ( \b +z_2 \r) \, \mathcal{U}^{(1)}_{F;i}(\b-z_1 \r)^{\dag}
+\frac{i}{2}  \,   \big(\partial_{i}\mathcal{U}_F ( \b +z_2 \r)\big)\,  \mathcal{U}^{(1)}_{F;i}(\b-z_1 \r)^{\dag}\bigg]
\nonumber\\
=&\, 
\int_{\r,\b} e^{-i\r \cdot\P}\, e^{-i\b\cdot\k}\, \frac{f( \r)}{2k_2^+}
\Bigg\{
\bigg[\P^i -\frac{(z_2\!-\!z_1)}{2}\k^i\, -i z_1\P^i (\r\!\cdot\!\k)
\bigg]\, \mathcal{U}_F(\b)\, \mathcal{U}^{(1)}_{F;i} ( \b )^{\dag} \, 
\nonumber\\
&
- i\, \mathcal{U}_F ( \b )\, \mathcal{U}_F^{(2)}(\b)^{\dag}
+\bigg[\frac{i}{2}\, \delta^{ij} +\P^j \r^i \bigg]\, \big(\partial_{i}  \mathcal{U}_F(\b)\big)\,\mathcal{U}^{(1)}_{F;j} ( \b )^{\dag} 
+ O\left(\frac{|\k|^3}{|\P|\, P_{{tar}}^-}\right) 
\Bigg\}\, ,
\label{O_dec_qbar_1plus2_1}
\end{align}
in the small $\r$ (and large $\P$) limit. In the last term in Eq.~\eqref{O_dec_qbar_1plus2_1}, for further convenience, we have interchanged the labels $i$ and $j$ which are summed over. Following the same steps as in the case of decorations on the quark line, one obtains
\begin{align}
{\cal O}^{\textrm{dec. on }\bar{q}}_{(1)+(2)}[f]
=&\, 
\int_{\r,\b} e^{-i\r \cdot\P}\, e^{-i\b\cdot\k}\, \frac{f( \r)}{2k_2^+}
\nonumber\\
&\,\times
\Bigg\{
 \bigg[-\P^i +\frac{(z_2\!-\!z_1)}{2}\k^i\, +i z_1\P^i (\r\!\cdot\!\k)
\bigg]\, 
 i t^{a'} \,  \int dz^+ \, (2z^+) \,  
\mathcal{U}_A(+\infty,z^+;\b)_{a'a}\;  g {\mathcal{F}_{i}^{\;-}}_a (z^+,\b)
\nonumber\\
&
+t^{a'}t^{b'} \int dz^+ \int dz'^+ \, 
\Big[
-i\delta^{ij} \min(z^+,z'^+)
 -(2z'^+) \,\P^j \r^i\, 
\Big]
\nonumber\\
&\, \hspace{2cm}\times
\mathcal{U}_A(+\infty,z^+;\b)_{a'a}\;  g {\mathcal{F}_{i}^{\;-}}_a (z^+,\b)\,
\mathcal{U}_A(+\infty,z'^+;\b)_{b'b} \; g {\mathcal{F}_{j}^{\;-}}_b (z'^+,\b)
+ O\left(\frac{|\k|^3}{|\P|\, P_{{tar}}^-}\right) 
\Bigg\}\, .
\label{O_dec_qbar_1plus2_3}
\end{align}

By comparing Eqs.~\eqref{ampl-q_dec_L} and ~\eqref{ampl-qbar_dec_L}, it is clear that the same  function $f(\r)$ appears as a prefactor in Eqs.~\eqref{O_dec_q_1plus2_1} for the decorations on the quark line and in  Eqs.~\eqref{O_dec_qbar_1plus2_1} for the decorations on the antiquark line, thanks to the denominators $2k_1^+$ and $2k_2^+$ that we have introduced explicitly. The situation is the same in the transverse photon case, but with a different  function $f(\r)$ than in the longitudinal photon case.
Hence, in both cases, the total contribution with decorated Wilson lines of type $\mathcal{U}^{(1)}_{F;i}$ and $\mathcal{U}^{(2)}_F$  on either the quark or the antiquark line writes
\begin{align}
{\cal O}^{\textrm{dec. on }q+\bar{q}}_{(1)+(2)}[f]
=&\, 
\int_{\r,\b} e^{-i\r \cdot\P}\, e^{-i\b\cdot\k}\, \frac{f( \r)}{(2k^+)z_1z_2}
\nonumber\\
&\,\times
\Bigg\{(2i)t^{a'} \,
 \bigg[-\P^i +(z_2\!-\!z_1)\Big(\frac{\k^i}{2} -i \P^i (\r\!\cdot\!\k)\Big)
\bigg]\, 
 \int dv^+ \, v^+ \,   
\mathcal{U}_A(+\infty,v^+;\b)_{a'a}\;  g {\mathcal{F}_{i}^{\;-}}_a (v^+,\b)
\nonumber\\
&
+t^{a'}t^{b'} \int dv^+ \int dw^+ \, 
\mathcal{U}_A(+\infty,v^+;\b)_{a'a}\;  g {\mathcal{F}_{i}^{\;-}}_a (v^+,\b)\,
\mathcal{U}_A(+\infty,w^+;\b)_{b'b} \; g {\mathcal{F}_{j}^{\;-}}_b (w^+,\b)
\nonumber\\
&\,
\times
\Big\{
-i\delta^{ij} \min(v^+,w^+)
-(\P^j \r^i+\r^j \P^i)( z_2 v^+\!+\!z_1 w^+)
 +(\P^j \r^i-\r^j \P^i)( z_2 v^+\!-\!z_1 w^+)
\Big\}
+ O\left(\frac{|\k|^3}{|\P|\, P_{{tar}}^-}\right) 
\Bigg\}\, ,
\label{O_dec_q_qbar_1plus2}
\end{align}

\subsubsection{Color structures of the contributions with decorated Wilson line $\mathcal{U}^{(3)}_{F; ij}$}    

We have for the moment excluded the contributions to the amplitude with decorated Wilson lines  $ \mathcal{U}^{(3)}_{F; ij}$.
For such decoration on the quark line, one finds from Eqs.~\eqref{ampl-q_dec_L}  and~\eqref{Ampl-q_dec_T} these contributions to be of the form 
\begin{align}
{\cal O}^{\textrm{dec. on }q}_{(3)}[f]
=&\, 
\int_{\v,\w} e^{-i\v \cdot\k_1}\, e^{-i\w\cdot\k_2}\, f( \v\!-\!\w) \,
\mathcal{U}^{(3)}_{F; ij}( \v)\,
\mathcal{U}_F(\w)^{\dag}\,
\nonumber
\\
=&\, 
\int_{\r,\b} e^{-i\r \cdot\P}\, e^{-i\b\cdot\k}\, \,f( \r)\,
\mathcal{U}^{(3)}_{F; ij} ( \b +z_2 \r)\, \mathcal{U}_F(\b-z_1 \r)^{\dag}
\label{O_dec_q_3_def}
\end{align}
with some function $f(\r)$.
For our purposes, we will need to expand that contribution only at zeroth order in $\r$. Hence, we obtain 
\begin{align}
{\cal O}^{\textrm{dec. on }q}_{(3)}[f]
=&\, 
\int_{\r,\b} e^{-i\r \cdot\P}\, e^{-i\b\cdot\k}\, \,f( \r)\,
\bigg\{
\mathcal{U}^{(3)}_{F; ij} ( \b )\, \mathcal{U}_F(\b)^{\dag}
+O\left(\frac{|\r|}{|\b|^3\, P_{{tar}}^-}\right)
\bigg\}
\nonumber
\\
=&\, 
\int_{\r,\b} e^{-i\r \cdot\P}\, e^{-i\b\cdot\k}\, \,f( \r)\,
\bigg\{
  \int dv^+\,  
\mathcal{U}_F(+\infty,v^+;\b) \, 
g \mathcal{F}_{ij}(v^+,\b) \, 
\mathcal{U}_F(+\infty,v^+;\b)^{\dag}
+ O\left(\frac{|\k|^3}{|\P|\, P_{{tar}}^-}\right) 
\bigg\}
\nonumber
\\
=&\, 
\int_{\r,\b} e^{-i\r \cdot\P}\, e^{-i\b\cdot\k}\, \,f( \r)\,
\bigg\{
  \int dv^+ 
\,  t^{a'}
\mathcal{U}_A(+\infty,v^+;\b)_{a'a} \;
g \mathcal{F}_{ij}^a(v^+,\b)
+ O\left(\frac{|\k|^3}{|\P|\, P_{{tar}}^-}\right) 
\bigg\}
\, ,
\label{O_dec_q_3}
\end{align}
 where the term explicitly written is of order $\k^2/P_{{tar}}^-$.

 Similarly, for the contribution of the  decorated Wilson line $ \mathcal{U}^{(3)\dagger}_{F; ij}$ on the antiquark line, one finds from 
  Eqs.~\eqref{ampl-qbar_dec_L}  and~\eqref{Ampl-qbar_dec_T} an expression of the form
\begin{align}
{\cal O}^{\textrm{dec. on }\bar{q}}_{(3)}[f]
=&\, 
\int_{\v,\w} e^{-i\v \cdot\k_1}\, e^{-i\w\cdot\k_2}\, f( \v\!-\!\w) \,
\mathcal{U}_F( \v)\,
\mathcal{U}^{(3)}_{F; ij}(\w)^{\dag}\,
\nonumber
\\
=&\, 
\int_{\r,\b} e^{-i\r \cdot\P}\, e^{-i\b\cdot\k}\, \,f( \r)\,
\bigg\{
\mathcal{U}_F ( \b )\,\mathcal{U}^{(3)}_{F; ij}(\b)^{\dag}
+ O\left(\frac{|\k|^3}{|\P|\, P_{{tar}}^-}\right) 
\bigg\}
\nonumber
\\
=&\, 
\int_{\r,\b} e^{-i\r \cdot\P}\, e^{-i\b\cdot\k}\, \,f( \r)\,
\bigg\{
  \int dv^+ 
\,  t^{a'}
\mathcal{U}_A(+\infty,v^+;\b)_{a'a} \;
g \mathcal{F}_{ij}^a(v^+,\b)
+ O\left(\frac{|\k|^3}{|\P|\, P_{{tar}}^-}\right) 
\bigg\}
\label{O_dec_qbar_3}
\, .
\end{align}
This is the same formal expression as in the quark case~\eqref{O_dec_q_3}, but the relation between the prefactors $f( \r)$ involved in the quark and anti-quark cases is not as direct (in the transverse photon case) as with ${\cal O}^{\textrm{dec. on }q(\bar{q})}_{(1)+(2)}$.
Hence, in order to get the full contribution of the decorated Wilson line $ \mathcal{U}^{(3)}_{F; ij}$ both on the quark and antiquark lines, one simply needs to sum their corresponding factors $f( \r)$.

\subsubsection{Color structures of the contribution from the dynamics of the target}    
 
The explicit NEik corrections to the amplitudes beyond the static approximation for the target, respectively given in Eqs.~\eqref{ampl-dyn_L} and ~\eqref{Ampl-dyn_T}, are of the form 
\begin{align}
{\cal O}^{\textrm{dyn. tar.}}[f]
=&\, 
\int_{\v,\w} e^{-i\v \cdot\k_1}\, e^{-i\w\cdot\k_2}\, f( \v\!-\!\w) \,
\bigg[\mathcal{U}_F(\v,b^- )\overleftrightarrow{\partial_{b^-}}
\mathcal{U}_F(\w,b^- )^{\dag}
\bigg]\bigg|_{b^-=0}
\nonumber
\\
=&\, 
\int_{\r,\b} e^{-i\r \cdot\P}\, e^{-i\b\cdot\k}\, \,f( \r)\,
\bigg[\mathcal{U}_F( \b +z_2 \r,b^- )\overleftrightarrow{\partial_{b^-}}
\mathcal{U}_F(\b-z_1 \r,b^-)^{\dag}
\bigg]\bigg|_{b^-=0}
\, .
\label{O_dyn_tar_def}
\end{align}
In this study, we will need to keep only the zeroth order of that expression in the small $\r$ expansion. Then, using Eqs.~\eqref{dec-partial-min-both} and~\eqref{rel_adj_fund}, one obtains
\begin{align}
{\cal O}^{\textrm{dyn. tar.}}[f]
=&\, 
\int_{\r,\b} e^{-i\r \cdot\P}\, e^{-i\b\cdot\k}\, \,f( \r)\,
\bigg\{
\bigg[\mathcal{U}_F( \b,b^- )\overleftrightarrow{\partial_{b^-}}
\mathcal{U}_F(\b,b^-)^{\dag}
\bigg]\bigg|_{b^-=0}\,
+ O\left(\frac{|\k|^3}{|\P|\, P_{{tar}}^-}\right)  
\bigg\}
\nonumber
\\
=&\, 
\int_{\r,\b} e^{-i\r \cdot\P}\, e^{-i\b\cdot\k}\, \,f( \r)\,
\bigg\{
(2i)
  \int dv^+\,  
\mathcal{U}_F(+\infty,v^+;\b) \, 
g \mathcal{F}^{+-}(v^+,\b) \, 
\mathcal{U}_F(+\infty,v^+;\b)^{\dag}
+ O\left(\frac{|\k|^3}{|\P|\, P_{{tar}}^-}\right) 
\bigg\}
\nonumber
\\
=&\, 
\int_{\r,\b} e^{-i\r \cdot\P}\, e^{-i\b\cdot\k}\, \,f( \r)\,
\bigg\{
(2i)
  \int dv^+ 
\,  t^{a'}
\mathcal{U}_A(+\infty,v^+;\b)_{a'a} \;
g \mathcal{F}^{+-}_a(v^+,\b)
+ O\left(\frac{|\k|^3}{|\P|\, P_{{tar}}^-}\right) 
\bigg\}
\, 
\label{O_dyn_tar}
\end{align}
 where the term explicitly written is of order $\k^2/P_{{tar}}^-$.

\subsubsection{Color structure of the extra contributions to the amplitude for transverse photon}    

In the last contribution~\eqref{Ampl-Lplus-in_combined}, present in the amplitude for transverse photon, the color structure  involves a single transverse coordinate. Hence, no small $\r$ expansion can be performed in that case. Using the identity~\eqref{rel_adj_fund}, one finds
\begin{align}
{\cal O}_j^{L^+\, \textrm{phase}\, +\, \textrm{in}} =  &\, \int_{\z}\, e^{-i\z \cdot(\k_1+\k_2)}\:
 \int  dz^+\, (2z^+)\,  
\mathcal{U}_F(+\infty, z^+;\z)
\Big( ig \mathcal{F}_{j}^{\;-} (z^+,\z)\Big)
\mathcal{U}_F (+\infty,z^+;\z)^\dag
 \nn \\
=  &\, 
2i\,  t^{a'} \, \int_{\b}\, e^{-i\b \cdot \k}\:
 \int  dv^+\, v^+\,  
\mathcal{U}_A(+\infty,v^+;\b)_{a'a}\;  g {\mathcal{F}_{j}^{\;-}}_a (v^+,\b)
\label{O_-Lplus-in_combined} 
\, .
\end{align}


\section{Back-to-back dijet production in DIS via longitudinal photon\label{sec:gamma_L_case}}

\subsection{Back-to-back limit for the longitudinal photon amplitude}

Having discussed the small $\r$ expansion of the different color structures appearing in the amplitude, the next step is to insert back the obtained expressions in the amplitude, and perform the integration over $\r$, since the $\r$ dependence has now been extracted out of the color operators. 
Inserting the expansion \eqref{expand_GEik_op_2} into the generalized eikonal amplitude \eqref{bdep_Ampl-GenEik_L} in the longitudinal photon case, and using the identities \eqref{r_integrals} to perform the integration over $\r$, one finds
\begin{align}
&i \mathbf{M}_{q_1 \bar q_2 \leftarrow \gamma^*_L}^{\rm Gen.\,  Eik} (b^-)
=\, 
 -Q\,  e e_f\, \bar u(1) \gamma^+ v(2)\, 
 \frac{(q^+\!-\!(z_2\!-\!z_1)k^+)(q^+\!+\!(z_2\!-\!z_1)k^+)}{4(q^+)^3}\,
 \theta(q^+\!-\!(z_2\!-\!z_1)k^+)\,
 \theta(q^+\!+\!(z_2\!-\!z_1)k^+)\,
  \nn \\
&
\times
\int_{\b}\,   e^{-i\b\cdot\k}\,
\Bigg\{
\bigg[
-\frac{2\P^j}{[\P^2+\hat Q^2]^2}
+(z_2\!-\!z_1)\bigg(\frac{\k^j}{[\P^2+\hat Q^2]^2}
-\frac{4(\k\!\cdot\!\P)\P^j}{[\P^2+\hat Q^2]^3}
\bigg)
\bigg]
 \int  dv^+ 
t^{a'}\, \mathcal{U}_A\left(+\infty,v^+;\b,b^-\right)_{a'a} g{\mathcal{F}_{j}^{\;-}}_a (v^+,\b,b^-)\; 
\nonumber\\
&\hspace{0.5cm}
-\bigg[
\frac{\delta^{ij}}{[\P^2+\hat Q^2]^2}
-\frac{4\P^i\P^j}{[\P^2+\hat Q^2]^3}
\bigg]\, t^{a'} t^{b'}\,
\int  dv^+ \int  dw^+ 
 \mathcal{U}_A\left(+\infty,v^+;\b,b^-\right)_{a'a} g{\mathcal{F}_{i}^{\;-}}_a (v^+,\b,b^-)\; 
\nonumber\\
&\hspace{1cm}\times
\mathcal{U}_A\left(+\infty,w^+;\b,b^-\right)_{b'b} g{\mathcal{F}_{j}^{\;-}}_b (w^+,\b,b^-)
\Bigg\}+O\left(\frac{Q|\k|}{|\P|^5}\right)
\label{Ampl-GenEik_L_b2b} 
\, .
\end{align}
In the static approximation for the target, one obtains instead from Eq.~\eqref{Ampl-StrictEik_L}  the strict eikonal amplitude as 
\begin{align}
& i{\cal M }_{q_1 \bar q_2 \leftarrow \gamma^*_L}^{\rm   Eik} 
=\, 
 - \frac{Q\,  e e_f}{q^+}\, \bar u(1) \gamma^+ v(2)\, 
z_1 z_2\, 
\int_{\b}\,   e^{-i\b\cdot\k}\,
  \nn \\
&
\times
\Bigg\{
\bigg[
-\frac{2\P^j}{[\P^2\!+\!\bar Q^2]^2}
+(z_2\!-\!z_1)\bigg(\frac{\k^j}{[\P^2\!+\!\bar Q^2]^2}
-\frac{4(\k\!\cdot\!\P)\P^j}{[\P^2\!+\!\bar Q^2]^3}
\bigg)
\bigg]\,
t^{a'} 
 \int  dv^+ 
\mathcal{U}_A\left(+\infty,v^+;\b\right)_{a'a} g{\mathcal{F}_{j}^{\;-}}_a (v^+,\b)\; 
\nonumber\\
&\hspace{0.5cm}
+\bigg[
-\frac{\delta^{ij}}{[\P^2\!+\!\bar Q^2]^2}
+\frac{4\P^i\P^j}{[\P^2\!+\!\bar Q^2]^3}
\bigg]\, t^{a'} t^{b'}
\int  dv^+ \int  dw^+ 
\, \mathcal{U}_A\left(+\infty,v^+;\b\right)_{a'a} g{\mathcal{F}_{i}^{\;-}}_a (v^+,\b)\; 
\nonumber\\
&\hspace{1cm}\times
\mathcal{U}_A\left(+\infty,w^+;\b\right)_{b'b} g{\mathcal{F}_{j}^{\;-}}_b (w^+,\b)
\Bigg\}
+O\left(\frac{Q|\k|}{|\P|^5}\right)
\label{Ampl-Eik_L_b2b} 
\, .
\end{align}
In both Eqs.~\eqref{Ampl-GenEik_L_b2b} and \eqref{Ampl-Eik_L_b2b}, the first term, proportional to $-2\P^j$, is the LP contribution in the correlation limit, and it is overall of order $Q/(|\k|\, |\P|^3)$ at amplitude level. The other terms explicitly written are NLP corrections, of order $Q/|\P|^4$ overall.

Similarly, using the expansion \eqref{O_dec_q_qbar_1plus2} and the integrals \eqref{r_integrals}, one can calculate the total contribution of the decorated Wilson lines $\mathcal{U}^{(1)}_{F;j}$ and $\mathcal{U}^{(2)}_F$ in the 
 \eqref{ampl-q_dec_L} and \eqref{ampl-qbar_dec_L} terms in the amplitude in the back-to-back regime, and find
\begin{align}
& i{\cal M }_{q_1 \bar q_2 \leftarrow \gamma^*_L}^{(1)+(2)}
=
 -  \frac{Q\,e e_f}{q^+}\, \bar u(1)  \gamma^+ v(2) \int_{\b}\,   e^{-i\b\cdot\k}\,
 \nn \\
& \times
\Bigg\{
\frac{i}{2q^+}\bigg[
-\frac{2\P^j}{[\P^2\!+\!\bar Q^2]}
+(z_2\!-\!z_1)\bigg(\frac{\k^j}{[\P^2\!+\!\bar Q^2]}
-\frac{4(\k\!\cdot\!\P)\P^j}{[\P^2\!+\!\bar Q^2]^2}
\bigg)
\bigg]\,
t^{a'} \int  dv^+\, v^+\, 
 \mathcal{U}_A\left(+\infty,v^+;\b\right)_{a'a} g{\mathcal{F}_{j}^{\;-}}_a (v^+,\b)\; 
\nonumber\\
&\hspace{0.5cm}
+\frac{i}{2q^+}\, t^{a'} t^{b'}\int  dv^+ \int  dw^+
\mathcal{U}_A\left(+\infty,v^+;\b\right)_{a'a} g{\mathcal{F}_{i}^{\;-}}_a (v^+,\b)\; 
\mathcal{U}_A\left(+\infty,w^+;\b\right)_{b'b} g{\mathcal{F}_{j}^{\;-}}_b (w^+,\b)
\nonumber\\
&\hspace{1cm}\times
\Bigg[
-\frac{\delta^{ij}}{[\P^2\!+\!\bar Q^2]}\, \min(v^+,w^+)
+\frac{4\P^i\P^j}{[\P^2\!+\!\bar Q^2]^2}\, ( z_2 v^+\!+\!z_1 w^+)
\Bigg]
\Bigg\}+O\left(\frac{Q |\k|}{W^2\, |\P|^3}\right)
\label{ampl-1_2_L_b2b_bis} 
\, .
\end{align}

The total contribution of the decorated Wilson line $\mathcal{U}^{(3)}_{F; ij}$ in the  terms
 \eqref{ampl-q_dec_L} and \eqref{ampl-qbar_dec_L} in the amplitude in the back-to-back jets regime is found thanks to the expansions  \eqref{O_dec_q_3} and \eqref{O_dec_qbar_3} as
\begin{align}
i{\cal M }_{q_1 \bar q_2 \leftarrow \gamma^*_L}^{(3)}
=&\,
 -\frac{Q\, e e_f}{2(q^+)^2}\, \bar u(1)  \gamma^+ \frac{[\gamma^i,\gamma^j]}{4} v(2)
\,
\frac{1}{[\P^2\!+\!\bar Q^2]}  
\int_{\b}\  e^{-i\b\cdot\k}  
\nonumber\\
&\, \times\,
\int dz^+ 
\,  t^{a'}
\mathcal{U}_A(+\infty,z^+;\b)_{a'a} \,
g \mathcal{F}_{ij}^a(z^+,\b)\:
+O\left(\frac{Q|\k|}{W^2\, |\P|^3}\right)
\label{ampl-3_L_b2b}
\, .
\end{align}
Finally, using Eq.~\eqref{O_dyn_tar}, the last NEik contribution to the longitudinal photon amplitude, written in Eq.~\eqref{ampl-dyn_L}, becomes
\begin{align}
i{\cal M }_{q_1 \bar q_2 \leftarrow \gamma^*_L}^{\textrm{dyn. tar.}}
=&\,
Q\, e e_f\,
 \frac{(z_2\!-\!z_1)}{(q^+)^2}\, \bar u(1)  \gamma^+ v(2)
\,
\frac{[\P^2+m^2]}{[\P^2\!+\!\bar Q^2]^2}  
\int_{\b}\  e^{-i\b\cdot\k}  
\nonumber\\
&\, \times\,
\int dz^+ 
\,  t^{a'}
\mathcal{U}_A(+\infty,z^+;\b)_{a'a} \,
g \mathcal{F}^{+-}_a(z^+,\b)\:
+O\left(\frac{Q|\k|}{W^2\, |\P|^3}\right)
\label{ampl-dyn_tar_L_b2b}
\, .
\end{align}
Hence, in the back-to-back jets regime, we have written the explicit NEik correction to the amplitude~\eqref{NEik_corr_Ampl_L}
as
\begin{align}
i {\cal M }_{q_1 \bar q_2 \leftarrow \gamma^*_L}^{\rm NEik \, corr.}= 
i{\cal M }_{q_1 \bar q_2 \leftarrow \gamma^*_L}^{(1)+(2)}
 +
i{\cal M }_{q_1 \bar q_2 \leftarrow \gamma^*_L}^{(3)}
+
i{\cal M }_{q_1 \bar q_2 \leftarrow \gamma^*_L}^{\textrm{dyn. target}}
\, ,
\label{NEik_corr_Ampl_L_b2b}
\end{align}   
with the three contributions given in Eqs.~\eqref{ampl-1_2_L_b2b_bis}, \eqref{ampl-3_L_b2b} and \eqref{ampl-dyn_tar_L_b2b} respectively.
In Eq.~\eqref{ampl-1_2_L_b2b_bis}, the first term, proportional to $-2\P^j$, is the LP contribution in the correlation limit, and it is overall of order $Q/(|\k|\, |\P|\, W^2)$ at amplitude level. The other terms explicitly written in Eqs.~\eqref{ampl-1_2_L_b2b_bis}, \eqref{ampl-3_L_b2b} and \eqref{ampl-dyn_tar_L_b2b} are NLP corrections, of order $Q/(|\P|^2\, W^2)$.



\subsection{Cross section via longitudinal photon}


The generalized eikonal cross section was defined in Eq. \eqref{Cross_Section_GEik} in terms of the $b^-$-dependent amplitudes, which is given in the correlation limit in Eq. \eqref{Ampl-GenEik_L_b2b}. The generalized eikonal cross section for longitudinal photon in the correlation limit is then found as
\begin{align}
&
\frac{d\sigma_{\gamma^{*}_L\rightarrow q_1\bar q_2}}{d {\rm P.S.}}\Bigg|_{\rm Gen. \, Eik}
= 2q^+ \int d (\Delta b^-) e^{i\Delta b^-(k^+ - q^+)} 
\,  Q^2\,  e^2 e_f^2\, \frac{z_1\, z_2 (k^+)^2}{4(q^+)^6}\Big[ (q^+)^2-(z_2-z_1)^2(k^+)^2\Big]^2
 \int_{\b,\b'}\,   e^{i(\b'-\b)\cdot\k}\,
\nonumber\\
&\times
 \theta(q^+\!-\!(z_2\!-\!z_1)k^+)\,
 \theta(q^+\!+\!(z_2\!-\!z_1)k^+)\,
 \Bigg\{
\Bigg[
\frac{4\P^i\P^j}{[\P^2+\hat Q^2]^4}
-2(z_2\!-\!z_1)\frac{(\P^i\k^j+\k^i\P^j)}{[\P^2+\hat Q^2]^4}
+16(z_2\!-\!z_1)\frac{(\k\cdot\P)\P^i\P^j}{[\P^2+\hat Q^2]^5}
\Bigg]
  \nn \\
&
\times
 \int  dv^+ d{v'}^+\Big\langle 
 g{\mathcal{F}_{j}^{\;-}}_b ({v'}^+,\b',0)
 \Big[  \mathcal{U}_A\left(+\infty,{v'}^+;\b',0\right)^{\dag}  \mathcal{U}_A\left(+\infty,v^+;\b,\Delta b^-\right)\Big]_{ba}
  g{\mathcal{F}_{i}^{\;-}}_a (v^+,\b,\Delta b^-)\Big\rangle
\nonumber\\
&+
\frac{4\P^l}{[\P^2+\hat Q^2]^4}\bigg[ \delta^{ij}-\frac{4\, \P^i\P^j}{[\P^2+\hat Q^2]}\bigg]\,
 {\rm tr}_F\big( t^{a'} t^{b'} t^{c'}\big)
 \int  dv^+ dw^+ d{v'}^+
  \Big\langle    \mathcal{U}_A\left(+\infty,{v'}^+;\b',0\right)_{c'c}  g{\mathcal{F}_{l}^{\;-}}_c ({v'}^+,\b',0)
  \nonumber\\
 &
 \hspace{1.5cm}
 \times
 \mathcal{U}_A\left(+\infty,{v}^+;\b,\Delta b^-\right)_{a'a}  g{\mathcal{F}_{i}^{\;-}}_a ({v}^+,\b,\Delta b^-)
  \mathcal{U}_A\left(+\infty,{w}^+;\b,\Delta b^-\right)_{b'b}  g{\mathcal{F}_{j}^{\;-}}_b ({w}^+,\b,\Delta b^-)
  \Big\rangle
 \nonumber\\
 &+
\frac{4\P^l}{[\P^2+\hat Q^2]^4}\bigg[ \delta^{ij}-\frac{4\, \P^i\P^j}{[\P^2+\hat Q^2]}\bigg]\,
 {\rm tr}_F\big( t^{a'} t^{b'} t^{c'}\big)
 \int  dv^+ d{w'}^+ d{v'}^+
  \Big\langle  \mathcal{U}_A\left(+\infty,{v'}^+;\b',0\right)_{a'a}  g{\mathcal{F}_{i}^{\;-}}_a ({v'}^+,\b',0)
 \nonumber\\  
 &
 \hspace{1.5cm}
 \times
  \mathcal{U}_A\left(+\infty,{w'}^+;\b',0\right)_{b'b}  g{\mathcal{F}_{j}^{\;-}}_b ({w'}^+,\b',0)
  \mathcal{U}_A\left(+\infty,{v}^+;\b,\Delta b^-\right)_{c'c}  g{\mathcal{F}_{l}^{\;-}}_c ({v}^+,\b,\Delta b^-)
  \Big\rangle
  \Bigg\}+O\left(\frac{Q^2}{\P^8}\right)
  \, .
  \label{Gen_Eik_X_sec_L}
\end{align}
%
Here, we have used the relation 
%
\begin{align}
\sum_{h_1,h_2}\, \bar v(2) \gamma^+ u(1)\,\bar u(1) \gamma^+ v(2)= 8\, k_1^+\, k_2^+ =8\, (k^+)^2\, z_1z_2
\, .
\label{Dirac_struct_1}
\end{align}
%
In the strict eikonal approximation, one finds the cross section in the back-to-back regime to be
%
\begin{align}
&
\frac{d\sigma_{\gamma^{*}_L\rightarrow q_1\bar q_2}}{d {\rm P.S.}}\Bigg|_{\rm Eik}
= 2q^+ \, 2\pi\, \delta(q^+-k^+) 
\,  4\, Q^2\,  e^2 e_f^2\, z_1^3z_2^3
 %
 \int_{\b,\b'}\,   e^{i(\b'-\b)\cdot\k}\,
 \Bigg\{
\Bigg[
\frac{4\P^i\P^j}{[\P^2\!+\!\bar Q^2]^4}
-2(z_2\!-\!z_1)\frac{(\P^i\k^j+\k^i\P^j)}{[\P^2\!+\!\bar Q^2]^4}
\nonumber\\
&
+16(z_2\!-\!z_1)\frac{(\k\cdot\P)\P^i\P^j}{[\P^2\!+\!\bar Q^2]^5}
\Bigg]
 \int  dv^+ d{v'}^+\Big\langle 
 g{\mathcal{F}_{j}^{\;-}}_b ({v'}^+,\b')
 \Big[  \mathcal{U}_A\left(+\infty,{v'}^+;\b'\right)^{\dag}  \mathcal{U}_A\left(+\infty,v^+;\b\right)\Big]_{ba}
  g{\mathcal{F}_{i}^{\;-}}_a (v^+,\b)\Big\rangle
\nonumber\\
&+
\frac{4\P^l}{[\P^2\!+\!\bar Q^2]^4}\bigg[ \delta^{ij}-\frac{4\, \P^i\P^j}{[\P^2\!+\!\bar Q^2]}\bigg]\,
 {\rm tr}_F\big( t^{a'} t^{b'} t^{c'}\big)
 \int  dv^+ dw^+ d{v'}^+
  \Big\langle    \mathcal{U}_A\left(+\infty,{v'}^+;\b'\right)_{c'c}  g{\mathcal{F}_{l}^{\;-}}_c ({v'}^+,\b')
  \nonumber\\
 &
 \hspace{5.5cm}
 \times
 \mathcal{U}_A\left(+\infty,{v}^+;\b\right)_{a'a}  g{\mathcal{F}_{i}^{\;-}}_a ({v}^+,\b)
  \mathcal{U}_A\left(+\infty,{w}^+;\b\right)_{b'b}  g{\mathcal{F}_{j}^{\;-}}_b ({w}^+,\b)
  \Big\rangle
 \nonumber\\
 &+
\frac{4\P^l}{[\P^2\!+\!\bar Q^2]^4}\bigg[ \delta^{ij}-\frac{4\, \P^i\P^j}{[\P^2\!+\!\bar Q^2]}\bigg]\,
 {\rm tr}_F\big( t^{a'} t^{b'} t^{c'}\big)
 \int  dv^+ d{w'}^+ d{v'}^+
  \Big\langle   \mathcal{U}_A\left(+\infty,{v}^+;\b\right)_{c'c}  g{\mathcal{F}_{l}^{\;-}}_c ({v}^+,\b)
  \nonumber\\
 &
 \hspace{1.5cm}
 \times
 \mathcal{U}_A\left(+\infty,{v'}^+;\b'\right)_{a'a}  g{\mathcal{F}_{i}^{\;-}}_a ({v'}^+,\b')
  \mathcal{U}_A\left(+\infty,{w'}^+;\b'\right)_{b'b}  g{\mathcal{F}_{j}^{\;-}}_b ({w'}^+,\b')
  \Big\rangle
  \Bigg\}+O\left(\frac{Q^2}{\P^8}\right)
  \, .
  \label{str_Eik_X_sec_L}
\end{align}
This expression can be obtained either by squaring the strict eikonal amplitude~\eqref{Ampl-Eik_L_b2b} or by neglecting the dependence of the color operators on $\Delta b^-$ in the generalized eikonal cross section~\eqref{Gen_Eik_X_sec_L}.
In Eq.~\eqref{str_Eik_X_sec_L}, the very first term, of order $Q^2/(\k^2\, \P^6)$, is the LP contribution, whereas all the other ones written explicitly are NLP corrections, of order $Q^2/(|\k|\, |\P|^7)$. 

As explained in appendix~\ref{app:geik-eik}, the generalized eikonal cross section~\eqref{Gen_Eik_X_sec_L} and the strict eikonal cross section~\eqref{str_Eik_X_sec_L} differ by contributions which are NNLP in the back-to-back dijet limit, which is beyond the accuracy of the present study. Hence, in the following (and in particular in the transverse photon case), we will not discuss further the generalized eikonal contribution, and focus instead on the strict eikonal one and on the explicit NEik corrections to it.

The next step is to calculate the explicit NEik correction to the cross section, as defined in~\eqref{Cross_Section_long_NEik}, as the interference between the amplitudes~\eqref{Ampl-Eik_L_b2b}  and~\eqref{NEik_corr_Ampl_L_b2b} in the correlation limit. The result (including the eikonal contribution), can be written in the form 
%
\begin{align}
&
\frac{d\sigma_{\gamma^{*}_L\rightarrow q_1\bar q_2}}{d {\rm P.S.}}\Bigg|_{{\rm Eik }+{\rm NEik}}
= \frac{d\sigma_{\gamma^{*}_L\rightarrow q_1\bar q_2}}{d {\rm P.S.}}\Bigg|_{{\mathcal{F}^{\perp -}} {\mathcal{F}^{\perp -}}}
+\frac{d\sigma_{\gamma^{*}_L\rightarrow q_1\bar q_2}}{d {\rm P.S.}}\Bigg|_{{\mathcal{F}^{\perp -}} {\mathcal{F}^{\perp -}} {\mathcal{F}^{\perp -}}}
+ \frac{d\sigma_{\gamma^{*}_L\rightarrow q_1\bar q_2}}{d {\rm P.S.}}\Bigg|_{{\mathcal{F}^{+ -}} {\mathcal{F}^{\perp -}}}
\, ,
\label{X_sec_L_gen_form}
\end{align}
%
grouping the terms according to the field strength correlators they involve.

As can be seen from Eq.~\eqref{str_Eik_X_sec_L}, the eikonal cross section contributes to the first two terms on the right hand side of Eq.~\eqref{X_sec_L_gen_form}. 
The interference between the term \eqref{ampl-1_2_L_b2b_bis} in the amplitude (involving decorated Wilson lines of the type $\mathcal{U}^{(1)}_{F;i}$ and $\mathcal{U}^{(2)}_F$) and the strict eikonal amplitude~\eqref{Ampl-Eik_L_b2b} also contributes to the first two terms in Eq.~\eqref{X_sec_L_gen_form}. 
All in all, one obtains
%
%
\begin{align}
&\, \frac{d\sigma_{\gamma^{*}_L\rightarrow q_1\bar q_2}}{d {\rm P.S.}}\Bigg|_{{\mathcal{F}^{\perp -}} {\mathcal{F}^{\perp -}}}
= 2q^+ 2\pi\,  \delta(q^+\!-\!k^+) \, 
\,  4\, Q^2\,  e^2 e_f^2\, z_1^3 z_2^3
\Bigg[
\frac{4\P^i\P^j}{[\P^2\!+\!\bar Q^2]^4}
-2(z_2\!-\!z_1)\frac{(\P^i\k^j+\k^i\P^j)}{[\P^2\!+\!\bar Q^2]^4}
\nonumber\\
 &
 +16(z_2\!-\!z_1)\frac{(\k\!\cdot\!\P)\P^i\P^j}{[\P^2\!+\!\bar Q^2]^5}
\Bigg]
  \int_{\b,\b'}\,   e^{i(\b'-\b)\cdot\k}\, 
   \int  dv^+ d{v'}^+\,
 \bigg[ 1-
 i\frac{[\P^2\!+\!\bar Q^2]}{2q^+z_1 z_2} \, 
 ({v'}^+\!-\!v^+)
 \bigg]
\nonumber\\
&
\times\, 
 \Big\langle 
 g{\mathcal{F}_{j}^{\;-}}_b ({v'}^+,\b')
 \Big[  \mathcal{U}_A\left(+\infty,{v'}^+;\b'\right)^{\dag}  \mathcal{U}_A\left(+\infty,v^+;\b\right)\Big]_{ba}
  g{\mathcal{F}_{i}^{\;-}}_a (v^+,\b)\Big\rangle
 +O\left(\frac{Q^2}{\P^8}\right)
 +O\left(\frac{Q^2}{\k^2 \P^2 W^4}\right)
\label{X_sec_L_Fperpmin_Fperpmin}
\end{align}
%
and
%
%
\begin{align}
&
\frac{d\sigma_{\gamma^{*}_L\rightarrow q_1\bar q_2}}{d {\rm P.S.}}\Bigg|_{{\mathcal{F}^{\perp -}} {\mathcal{F}^{\perp -}} {\mathcal{F}^{\perp -}}}
= 2q^+ 2\pi\,  \delta(q^+-k^+) \, 16\, Q^2\,  e^2 e_f^2\, z_1^3 z_2^3\, \frac{\P^l}{[\P^2\!+\!\bar Q^2]^4}\,     
2\, {\textrm{Re}}\;    
\,  
 \int_{\b,\b'}\,   e^{i(\b'-\b)\cdot\k}\, 
  \int  dv^+ dw^+ d{v'}^+
 \nonumber\\
 &\times\, 
 {\rm tr}_F\big( t^{a'} t^{b'} t^{c'}\big)\;
 \Big\langle   \mathcal{U}_A\left(+\infty,{v'}^+;\b'\right)_{c'c}  g{\mathcal{F}_{l}^{\;-}}_c ({v'}^+,\b')
 \mathcal{U}_A\left(+\infty,{v}^+;\b\right)_{a'a}  g{\mathcal{F}_{i}^{\;-}}_a ({v}^+,\b)
  \mathcal{U}_A\left(+\infty,{w}^+;\b\right)_{b'b}  g{\mathcal{F}_{j}^{\;-}}_b ({w}^+,\b)
  \Big\rangle
   \nonumber\\
 &
 \times
 \Bigg\{
\delta^{ij}\bigg[1-  i\frac{[\P^2\!+\!\bar Q^2]}{2q^+z_1 z_2} \, 
 \left({v'}^+\!-\!\min(v^+,w^+)\right)\bigg]
-\frac{4\, \P^i\P^j}{[\P^2\!+\!\bar Q^2]}
\bigg[1-  i\frac{[\P^2\!+\!\bar Q^2]}{2q^+z_1 z_2} \, 
 \left({v'}^+\!-\!z_2 v^+\!-\!z_1 w^+)\right)\bigg]
   \Bigg\}
  \nonumber\\
 &  
   +O\left(\frac{Q^2}{\P^8}\right)
 +O\left(\frac{Q^2}{\k^2 \P^2 W^4}\right)
   \, .
\label{X_sec_L_Fperpmin_Fperpmin_Fperpmin}   
\end{align}
The third term in Eq.~\eqref{X_sec_L_gen_form} is obtained as the interference between the term~\eqref{ampl-dyn_tar_L_b2b} in the amplitude and the strict eikonal amplitude~\eqref{Ampl-Eik_L_b2b}. One finds
\begin{align}
&\, \frac{d\sigma_{\gamma^{*}_L\rightarrow q_1\bar q_2}}{d {\rm P.S.}}\Bigg|_{{\mathcal{F}^{+ -}} {\mathcal{F}^{\perp -}}}
=  2q^+ \, 2\pi\, \delta(q^+-k^+) 
\,  8\, Q^2\,  e^2 e_f^2\, \frac{z_1^2z_2^2 (z_2\!-\!z_1)}{q^+} \; \frac{\P^j[\P^2+m^2]}{[\P^2\!+\!\bar Q^2]^4}\,
 2\, {\textrm{Re}}\,  \int_{\b,\b'}\,   e^{i(\b'-\b)\cdot\k}\,
 \int  dv^+ d{v'}^+\nonumber\\
 &\times
 \Big\langle 
 g{\mathcal{F}_{j}^{\;-}}_b ({v'}^+,\b')
 \Big[  \mathcal{U}_A\left(+\infty,{v'}^+;\b'\right)^{\dag}  \mathcal{U}_A\left(+\infty,v^+;\b\right)\Big]_{ba}
  g\mathcal{F}^{\, +-}_a (v^+,\b)\Big\rangle
    +O\left(\frac{Q^2}{\P^6 W^2}\right)
 +O\left(\frac{Q^2}{|\k|\, |\P|^3 W^4}\right)
  \, .
\label{X_sec_L_Fplusmin_Fperpmin}   
\end{align}
%
As a remark, the term~\eqref{ampl-3_L_b2b} in the amplitude (including the decorated Wilson line $ \mathcal{U}^{(3)}_{F; ij}$) does not contribute to the cross section at NEik accuracy in the longitudinal photon case. Indeed, its interference with the eikonal amplitude is proportional to the Dirac structure
%
%
\begin{align}
\sum_{h_1,h_2}\, \bar v(2) \gamma^+ u(1)\,\bar u(1) \gamma^+ [\gamma^i,\gamma^j] v(2)= 4k_1^+k_2^+\, {\rm tr}\Big(\frac{\gamma^-\gamma^+}{2}[\gamma^i,\gamma^j]\Big)=0
\, .
\end{align}
%

A crucial feature of our results is that, in Eq.~\eqref{X_sec_L_Fperpmin_Fperpmin}, the NLP and the NEik power corrections factorize from each other. Indeed, in that expression, NLP corrections are purely kinematical corrections to the hard factor, whereas the NEik correction, involving the light-cone time interval ${v'}^+\!-\!v^+$ between the two field strength insertions, is instead a correction to the operator. We will further comment on this issue in Sec.~\ref{sec:CGC_vs_TMD}. Finally, we note that the $3\, \mathcal{F}$ term Eq.~\eqref{X_sec_L_Fperpmin_Fperpmin_Fperpmin} contains the 3-body genuine twists corrections discussed in \cite{Altinoluk:2019wyu} that are power suppressed but eikonal (the 1's inside the two brackets of the third line), as well as additional terms that are both NLP and NEik.


\section{Back-to-back dijet production in DIS via transverse photon\label{sec:gamma_T_case}}


After having calculated the contribution of longitudinal photon exchange to the DIS dijet cross section in the back-to-back regime, we calculate now the contribution of transverse photon exchange in this section.


\subsection{Back-to-back limit for the transverse photon amplitude}

For the same reason as in the case of longitudinal photon exchange, the difference between generalized eikonal and strict eikonal cross section is a NNLP correction in the correlation limit, beyond our accuracy in the present study (see Appendix~\ref{app:geik-eik}). Hence, in the transverse photon case, we focus on the strict eikonal amplitude~\eqref{Ampl-StrictEik_T} and on explicit NEik corrections to it~\eqref{Ampl-q_dec_T}, \eqref{Ampl-qbar_dec_T}, \eqref{Ampl-dyn_T} and~\eqref{Ampl-Lplus-in_combined}. 
The corresponding expressions in the back-to-back dijet limit are obtained by inserting the small $\r$ expansions of the color operators derived in Sec.~\ref{sec:color_struct_exp}, and performing the integration over $\r$ thanks to the identities \eqref{r_integrals}. 
In such a way, the strict eikonal amplitude is found thanks to Eq.~\eqref{expand_GEik_op_2} as
\begin{align}
&i{\cal M }_{q_1 \bar q_2 \leftarrow \gamma^*_T}^{\rm  Eik}
=\, 
  \frac{e e_f\,}{(2q^+)}\, \varepsilon_{\lambda}^l\,  
   \bar u(1) \gamma^+\int_{\b}\,   e^{-i\b\cdot\k}\, 
  \nn \\
&
\times
\Bigg\{
\bigg[
\bigg(
\frac{\delta^{jm}}{[\P^2\!+\!\bar Q^2]}
-\frac{2\P^j\P^m}{[\P^2\!+\!\bar Q^2]^2}
+\frac{(z_2\!-\!z_1)}{[\P^2\!+\!\bar Q^2]^2}\left(\k^j\P^m+\P^j\k^m+(\k\!\cdot\!\P)\delta^{jm}
-4\frac{(\k\!\cdot\!\P)\P^j\P^m}{[\P^2\!+\!\bar Q^2]}\right)
\bigg)
\bigg(
(z_2\!-\!z_1)\, \delta^{lm}+\frac{[\gamma^l,\gamma^m]}{2}
\bigg) 
  \nn \\
&
\, \hspace{0.5cm}
+\bigg(-\frac{2\P^j}{[\P^2\!+\!\bar Q^2]^2}
+\frac{(z_2\!-\!z_1)}{[\P^2\!+\!\bar Q^2]^2}\left(\k^j
-4\frac{(\k\!\cdot\!\P)\P^j}{[\P^2\!+\!\bar Q^2]}\right)
\bigg)
m\gamma^l
\bigg]
t^{a'} \int  dv^+\,
 \mathcal{U}_A\left(+\infty,v^+;\b\right)_{a'a} g{\mathcal{F}_{j}^{\;-}}_a (v^+,\b)\; 
\nonumber\\
&\, \hspace{0.5cm}
+
t^{a'} t^{b'}
\int  dv^+ \int  dw^+ 
 \mathcal{U}_A\left(+\infty,v^+;\b\right)_{a'a} g{\mathcal{F}_{i}^{\;-}}_a (v^+,\b)\,
\mathcal{U}_A\left(+\infty,w^+;\b\right)_{b'b} g{\mathcal{F}_{j}^{\;-}}_b (w^+,\b)
\nonumber\\
&\,
\hspace{1cm} \times\,
\bigg[
\bigg(
-\frac{\left(\delta^{ij}\P^m+\delta^{im}\P^j+\delta^{jm}\P^i\right)}{[\P^2\!+\!\bar Q^2]^2}
+\frac{4\P^i\P^j\P^m}{[\P^2\!+\!\bar Q^2]^3}
\bigg)
\bigg(
(z_2\!-\!z_1)\,  \delta^{lm}+\frac{[\gamma^l,\gamma^m]}{2}
\bigg) 
\nonumber\\
&\,\hspace{1.8cm}
 +
\bigg(
-\frac{\delta^{ij}}{[\P^2\!+\!\bar Q^2]^2}
+\frac{4\P^i\P^j}{[\P^2\!+\!\bar Q^2]^3}
\bigg)
m\gamma^l
\bigg]
\Bigg\}
 v(2)
 +O\left(\frac{|\k|}{|\P|^4}\right)
\label{Ampl-Eik_T_b2b} 
\, .
\end{align}
Thanks to the expansion \eqref{O_dec_q_qbar_1plus2}, the total contribution from the decorated Wilson lines $\mathcal{U}^{(1)}_{F;i}$ and $\mathcal{U}^{(2)}_F$ to the transverse photon amplitude is found to be
\begin{align}
&i{\cal M }_{q_1 \bar q_2 \leftarrow \gamma^*_T}^{(1)+(2)}
=\, 
 \frac{e e_f}{2q^+}\,  \varepsilon_{\lambda}^l\,  \frac{i}{(2q^+)z_1 z_2}\,
  \bar u(1) \gamma^+\int_{\b}\,   e^{-i\b\cdot\k}\, 
  \nn \\
&
\times
\Bigg\{
\bigg[
\bigg(
-\frac{2\P^j\P^m}{[\P^2\!+\!\bar Q^2]}
+(z_2\!-\!z_1)\frac{[\k^j\P^m+2\P^j\k^m]}{[\P^2\!+\!\bar Q^2]}
-4(z_2\!-\!z_1)\frac{(\k\!\cdot\!\P)\P^j\P^m}{[\P^2\!+\!\bar Q^2]^2}
\bigg)
\bigg(
(z_2\!-\!z_1)\,  \delta^{lm}+\frac{[\gamma^l,\gamma^m]}{2}
\bigg) 
  \nn \\
&
\, \hspace{0.5cm}
+\bigg(-\frac{2\P^j}{[\P^2\!+\!\bar Q^2]}
+\frac{(z_2\!-\!z_1)\k^j}{[\P^2\!+\!\bar Q^2]}
-4(z_2\!-\!z_1)\frac{(\k\!\cdot\!\P)\P^j}{[\P^2\!+\!\bar Q^2]^2}
\bigg)
m\gamma^l
\bigg]
t^{a'} \int  dv^+\, v^+\, 
 \mathcal{U}_A\left(+\infty,v^+;\b\right)_{a'a} g{\mathcal{F}_{j}^{\;-}}_a (v^+,\b)\; 
\nonumber\\
&
+
t^{a'} t^{b'}
\int  dv^+ \int  dw^+ 
 \mathcal{U}_A\left(+\infty,v^+;\b\right)_{a'a} g{\mathcal{F}_{i}^{\;-}}_a (v^+,\b)\,
\mathcal{U}_A\left(+\infty,w^+;\b\right)_{b'b} g{\mathcal{F}_{j}^{\;-}}_b (w^+,\b)
\nonumber\\
&\,
\hspace{0.5cm} \times\,
\bigg[
\bigg(
-\frac{\delta^{ij}\P^m}{[\P^2\!+\!\bar Q^2]}\, \min(v^+,w^+) 
+\bigg(-\frac{(\delta^{im}\P^j+\delta^{jm}\P^i)}{[\P^2\!+\!\bar Q^2]}
+\frac{4\P^i\P^j\P^m}{[\P^2\!+\!\bar Q^2]^2}\bigg)( z_2 v^+\!+\!z_1 w^+) 
\nonumber\\
&\,\hspace{1.5cm}
+\frac{(\delta^{im}\P^j-\delta^{jm}\P^i)}{[\P^2\!+\!\bar Q^2]}( z_2 v^+\!-\!z_1 w^+) 
\bigg)
\bigg(
(z_2\!-\!z_1)\,  \delta^{lm}+\frac{[\gamma^l,\gamma^m]}{2}
\bigg) 
\nonumber\\
&\,\hspace{1cm}
+
\bigg(
-\frac{\delta^{ij}}{[\P^2\!+\!\bar Q^2]}\, \min(v^+,w^+) 
+\frac{4\P^i\P^j}{[\P^2\!+\!\bar Q^2]^2}\, ( z_2 v^+\!+\!z_1 w^+)
\bigg)
m\gamma^l
\bigg]
\Bigg\}
 v(2)
 +O\left(\frac{|\k|}{|\P|^2 W^2}\right)
\label{Ampl-12NEik_T_b2b} 
\, .
\end{align}
The total contribution from the decorated Wilson lines $\mathcal{U}^{(3)}_{F; ij}$ to the amplitude is obtained as
\begin{align}
i{\cal M }_{q_1 \bar q_2 \leftarrow \gamma^*_T}^{(3)}
 = &\,
   \frac{ e e_f}{4 (q^+)^2}\, \varepsilon_{\lambda}^l  \,
   \frac{1}{[\P^2\!+\!\bar Q^2]}\:
\bar u(1) \gamma^+ 
\bigg\{  \frac{[\gamma^i,\gamma^j]}{4z_1}\,
\bigg[
 \P^m\left((z_2\!-\!z_1)\, \delta^{lm}+\frac{[\gamma^l,\gamma^m]}{2}\right) +m\, \gamma^l
\bigg]  
\nn \\
& \hspace{5cm}
+\bigg[
 \P^m\left((z_2\!-\!z_1)\, \delta^{lm}+\frac{[\gamma^l,\gamma^m]}{2}\right) +m\, \gamma^l
\bigg] \, 
\frac{[\gamma^i,\gamma^j]}{4z_2}
 \bigg\} v(2)
  \nn \\
& \times
  \int_{\b} \, e^{-i\b\cdot\k}\, 
  \int dz^+ \,  t^{a'} \mathcal{U}_A(+\infty,z^+;\b)_{a'a} \; g \mathcal{F}_{ij}^a(z^+,\b)\:
 +O\left(\frac{|\k|}{|\P|^2 W^2}\right)
 \, ,
\label{ampl-3_T_b2b} 
\end{align}
thanks to the expansions~\eqref{O_dec_q_3} and \eqref{O_dec_qbar_3} .
Using the expansion~\eqref{O_dyn_tar}, the contribution~\eqref{Ampl-dyn_T} to the amplitude is rewritten as 
\begin{align}
i{\cal M }_{q_1 \bar q_2 \leftarrow \gamma^*_T}^{\textrm{dyn. tar.}}
=&\,
 \frac{ e e_f}{2(q^+)^2}\, \varepsilon_{\lambda}^i\,
 \bar u(1)  \gamma^+
 \Bigg\{
 \frac{2\P^i}{[\P^2\!+\!\bar Q^2]} 
 +
\frac{(z_2\!-\!z_1)Q^2}{[\P^2\!+\!\bar Q^2]^2}  
\bigg[
(z_2\!-\!z_1)\P^i+ \frac{[\gamma^i,\gamma^j]}{2}\, \P^j +m\gamma^i
\bigg]
\Bigg\}v(2)
\nonumber\\
&\, \times\,
\int_{\b}\  e^{-i\b\cdot\k}  \int dz^+ 
\,  t^{a'}
\mathcal{U}_A(+\infty,z^+;\b)_{a'a} \,
g \mathcal{F}^{+-}_a(z^+,\b)\:
+O\left(\frac{|\k|}{|\P|^2 W^2}\right)
\label{ampl-dyn_tar_T_b2b}
\, .
\end{align}
 In the transverse photon case, there are two extra contributions, that we have already combined in Eq.~\eqref{Ampl-Lplus-in_combined}. In these contributions, all the Wilson lines and field strengths already sit at the same transverse position $\z$, so that there is no expansion of the color structure to perform at $\k\ll\P$. With the color structure rewritten in Eq.~\eqref{O_-Lplus-in_combined}, one finds
\begin{align}
i{\cal M }_{q_1 \bar q_2 \leftarrow \gamma^*_T}^{L^+\textrm{ phase}}
+ i{\cal M }_{q_1 \bar q_2 \leftarrow \gamma^*_T}^{\textrm{in}} 
=  &\, 
  e e_f \, \varepsilon_{\lambda}^i \,
  \frac{i}{4z_1z_2 (q^+)^2}\;
  \bar u(1) \gamma^+\!
  \left[(z_2\!-\!z_1)\, \delta^{ij}+\frac{[\gamma^i,\gamma^j]}{2}\right]  
  v(2) \: 
 \nn \\
& \times\,
 \int_{\b}\, e^{-i\b \cdot\k}\,
\int dz^+ \, z^+ \,  
t^{a'} \, 
\mathcal{U}_A(+\infty,z^+;\b)_{a'a}\;  g {\mathcal{F}_{j}^{\;-}}_a (z^+,\b)
\label{Ampl-Lplus-in_combined_corrlim} 
\, .
\end{align}
Since the contributions \eqref{Ampl-Eik_T_b2b}, \eqref{Ampl-12NEik_T_b2b} and \eqref{Ampl-Lplus-in_combined_corrlim} are expressed in terms of the same two color structure, with one or two factors of the leading components $\mathcal{F}_{\perp}^{\;-}$ of the background field strength, they can be combined as 
\begin{align}
&i{\cal M }_{q_1 \bar q_2 \leftarrow \gamma^*_T}^{\textrm{Eik.}+(1)+(2)+L^+\textrm{ phase} +\textrm{in}}
=\, 
 \frac{e e_f}{2q^+}\,  \varepsilon_{\lambda}^l\,  
  \bar u(1) \gamma^+\int_{\b}\,   e^{-i\b\cdot\k}\, 
  \Bigg\{
  t^{a'} \int  dv^+\, 
 \mathcal{U}_A\left(+\infty,v^+;\b\right)_{a'a} g{\mathcal{F}_{j}^{\;-}}_a (v^+,\b)\; 
  \nn \\
&
\times
\Bigg[
\left(1+iv^+\frac{[\P^2\!+\!\bar Q^2]}{2z_1 z_2 q^+}\right)
\bigg(-\frac{2\P^j}{[\P^2\!+\!\bar Q^2]^2}
+\frac{(z_2\!-\!z_1)}{[\P^2\!+\!\bar Q^2]^2}
\left(\k^j
-\frac{4(\k\!\cdot\!\P)\P^j}{[\P^2\!+\!\bar Q^2]}\right)
\bigg)
m\gamma^l
  \nn \\
&
\, \hspace{0.2cm}
+
\bigg[
\left(1+iv^+\frac{[\P^2\!+\!\bar Q^2]}{2z_1 z_2 q^+}\right)
\bigg(
\frac{\delta^{jm}}{[\P^2\!+\!\bar Q^2]}
-\frac{2\P^j\P^m}{[\P^2\!+\!\bar Q^2]^2}
+\frac{(z_2\!-\!z_1)}{[\P^2\!+\!\bar Q^2]^2}\left(\k^j\P^m+\P^j\k^m+(\k\!\cdot\!\P)\delta^{jm}
-4\frac{(\k\!\cdot\!\P)\P^j\P^m}{[\P^2\!+\!\bar Q^2]}\right)
\bigg)
 \nn \\
&
\, \hspace{1cm}
+\frac{iv^+}{2z_1 z_2 q^+}\, \frac{(z_2\!-\!z_1)}{[\P^2\!+\!\bar Q^2]}\, 
\left[
\P^j\k^m-(\k\!\cdot\!\P)\delta^{jm}
\right]\,
\bigg]
\bigg(
(z_2\!-\!z_1)\,  \delta^{lm}+\frac{[\gamma^l,\gamma^m]}{2}
\bigg)
\Bigg]
\nonumber\\
&
+
t^{a'} t^{b'}
\int  dv^+ \int  dw^+ 
 \mathcal{U}_A\left(+\infty,v^+;\b\right)_{a'a} g{\mathcal{F}_{i}^{\;-}}_a (v^+,\b)\,
\mathcal{U}_A\left(+\infty,w^+;\b\right)_{b'b} g{\mathcal{F}_{j}^{\;-}}_b (w^+,\b)
\nonumber\\
&\,
\hspace{0.5cm} \times\,
\Bigg[
\Bigg(
-\frac{\delta^{ij}}{[\P^2\!+\!\bar Q^2]^2}\, 
   \left(1+i\frac{[\P^2\!+\!\bar Q^2]}{2z_1 z_2 q^+}\, \min(v^+,w^+)\right) 
+\frac{4\P^i\P^j}{[\P^2\!+\!\bar Q^2]^3}\, \left(1+i\frac{[\P^2\!+\!\bar Q^2]}{2z_1 z_2 q^+}\, ( z_2 v^+\!+\!z_1 w^+)\right)
\Bigg)
m\gamma^l
\nonumber\\
&\,\hspace{1cm}
+
\Bigg(
-\frac{\delta^{ij}\P^m}{[\P^2\!+\!\bar Q^2]^2}\, \left(1+i\frac{[\P^2\!+\!\bar Q^2]}{2z_1 z_2 q^+}\, \min(v^+,w^+)\right)
+\frac{4\P^i\P^j\P^m}{[\P^2\!+\!\bar Q^2]^3}\, \left(1+i\frac{[\P^2\!+\!\bar Q^2]}{2z_1 z_2 q^+}\, ( z_2 v^+\!+\!z_1 w^+)\right)
\nonumber\\
&\,\hspace{1.5cm}
-\frac{\delta^{jm}\P^i}{[\P^2\!+\!\bar Q^2]^2}\, \left(1+2 i z_2 v^+\, \frac{[\P^2\!+\!\bar Q^2]}{2z_1 z_2 q^+}\right)
-\frac{\delta^{im}\P^j}{[\P^2\!+\!\bar Q^2]^2}\, \left(1+2 i z_1 w^+\, \frac{[\P^2\!+\!\bar Q^2]}{2z_1 z_2 q^+}\right)
\Bigg)
\bigg(
(z_2\!-\!z_1)\,  \delta^{lm}+\frac{[\gamma^l,\gamma^m]}{2}
\bigg) 
\Bigg]
\Bigg\}
 v(2)
   \nn \\
&\,
+O\left(\frac{|\k|}{|\P|^4}\right)
+O\left(\frac{|\P|^2}{|\k|\, W^4}\right)
\label{Ampl-sumEik12+_T_b2b} 
\, .
\end{align}
Interestingly, in the second and third lines in Eq.~\eqref{Ampl-sumEik12+_T_b2b}, one recovers the same factorization between the NEik corrections (terms proportional to $v^+$ here) and the NLP corrections in the correlation limit (terms suppressed by one power of $\k/\P$) as found in the longitudinal photon case. However, in the part of the amplitude with a single factor of $\mathcal{F}_{\perp}^{\;-}$, the term in the fourth line of Eq.~\eqref{Ampl-sumEik12+_T_b2b} is breaking that factorization between NEik and NLP corrections.  

As shown in Appendix~\ref{app:rewrite_non_fact}, this particular term can be rewritten as

\begin{align}
i{\cal M }_{q_1 \bar q_2 \leftarrow \gamma^*_T}^{\textrm{non. fact.}}
\equiv&\, 
 \frac{e e_f}{2q^+}\,  \varepsilon_{\lambda}^l\,  
  \bar u(1) \gamma^+\bigg(
(z_2\!-\!z_1)\,  \delta^{lm}+\frac{[\gamma^l,\gamma^m]}{2}
\bigg) v(2)
  \int_{\b}\,   e^{-i\b\cdot\k}\, 
  t^{a'} \int  dv^+\, 
 \mathcal{U}_A\left(+\infty,v^+;\b\right)_{a'a} g{\mathcal{F}_{j}^{\;-}}_a (v^+,\b)\; 
  \nn \\
&
\times
\frac{iv^+}{2z_1 z_2 q^+}\, \frac{(z_2\!-\!z_1)}{[\P^2\!+\!\bar Q^2]}\, 
\left[
\P^j\k^m-(\k\!\cdot\!\P)\delta^{jm}
\right]\,
  \nn \\
=&\, 
   \frac{e e_f}{(2q^+)^2}\,  \varepsilon_{\lambda}^l\,
 \frac{(z_2\!-\!z_1)}{z_1 z_2}\,   
  \frac{\epsilon^{mn}\P^n}{[\P^2\!+\!\bar Q^2]}\, 
  \bar u(1) \gamma^+\bigg(
(z_2\!-\!z_1)\,  \delta^{lm}+\frac{[\gamma^l,\gamma^m]}{2}
\bigg) v(2)
  \nn \\
&
\times\, 
\epsilon^{ij}
\int_{\b}\,   e^{-i\b\cdot\k}\, 
 \bigg\{
\frac{1}{2}\, t^{a'} \int  dv^+\,\mathcal{U}_A\left(+\infty,v^+;\b\right)_{a'a}\,   g{\mathcal{F}_{ij}^a}
   \nn \\
&\hspace{-1.5cm}
 -i\,  t^{a'} t^{b'}
\int  dv^+ \int  dw^+\, \min(v^+,w^+)\, 
 \mathcal{U}_A\left(+\infty,v^+;\b\right)_{a'a} g{\mathcal{F}_{i}^{\;-}}_a (v^+,\b)\,
\mathcal{U}_A\left(+\infty,w^+;\b\right)_{b'b} g{\mathcal{F}_{j}^{\;-}}_b (w^+,\b)
 \bigg\}  
 \label{Ampl-nonFact_T_b2b_3} 
\, .
\end{align}
Hence, we have been able to trade the term violating the  factorization of NEik and NLP corrections among the single  ${\mathcal{F}_{j}^{\;-}}$ insertion contributions for the sum of a  single  ${\mathcal{F}_{ij}}$ insertion contribution and a double ${\mathcal{F}_{i}^{\;-}}{\mathcal{F}_{j}^{\;-}}$ insertion contribution.

Using the relation~\eqref{Ampl-nonFact_T_b2b_3}, the total amplitude in the transverse photon case, including Eik and NEik contributions, and LP and NLP contributions in the back-to-back dijet regime, can be expressed as
\begin{align}
i{\cal M }_{q_1 \bar q_2 \leftarrow \gamma^*_T}
=&\,
i{\cal M }_{q_1 \bar q_2 \leftarrow \gamma^*_T}\bigg|_{{\mathcal{F}^{\perp -}} }
+i{\cal M }_{q_1 \bar q_2 \leftarrow \gamma^*_T}\bigg|_{{\mathcal{F}^{\perp -}} {\mathcal{F}^{\perp -}}}
+i{\cal M }_{q_1 \bar q_2 \leftarrow \gamma^*_T}\bigg|_{{\mathcal{F}_{ij}} }
+i{\cal M }_{q_1 \bar q_2 \leftarrow \gamma^*_T}\bigg|_{{\mathcal{F}^{+ -}} }
\label{ampl_T_b2b}
\, ,
\end{align}
where the contribution
\begin{align}
i{\cal M }_{q_1 \bar q_2 \leftarrow \gamma^*_T}\bigg|_{{\mathcal{F}^{\perp -}} } 
= &\,
 \frac{e e_f}{2q^+}\,  \varepsilon_{\lambda}^l\,  
  \bar u(1) \gamma^+
  \Bigg\{
  \bigg(
(z_2\!-\!z_1)\,  \delta^{lm}+\frac{[\gamma^l,\gamma^m]}{2}
\bigg)
  \nn \\
&
\, 
\times\,
\bigg[
\frac{\delta^{jm}}{[\P^2\!+\!\bar Q^2]}
-\frac{2\P^j\P^m}{[\P^2\!+\!\bar Q^2]^2}
+\frac{(z_2\!-\!z_1)}{[\P^2\!+\!\bar Q^2]^2}\left(\k^j\P^m+\P^j\k^m+(\k\!\cdot\!\P)\delta^{jm}
-4\frac{(\k\!\cdot\!\P)\P^j\P^m}{[\P^2\!+\!\bar Q^2]}\right)
\bigg]
 \nn \\
&
+
  \bigg[-\frac{2\P^j}{[\P^2\!+\!\bar Q^2]^2}
+\frac{(z_2\!-\!z_1)}{[\P^2\!+\!\bar Q^2]^2}
\left(\k^j
-\frac{4(\k\!\cdot\!\P)\P^j}{[\P^2\!+\!\bar Q^2]}\right)
\bigg]
m\gamma^l
\Bigg\}
 v(2)
 \nn \\
&
\times\,
t^{a'}\int_{\b}\,   e^{-i\b\cdot\k}\,  \int  dv^+\, 
\left(1+iv^+\frac{[\P^2\!+\!\bar Q^2]}{2z_1 z_2 q^+}\right)\,
 \mathcal{U}_A\left(+\infty,v^+;\b\right)_{a'a} g{\mathcal{F}_{j}^{\;-}}_a (v^+,\b)
    \nn \\
&\,
+O\left(\frac{|\k|}{|\P|^4}\right)
+O\left(\frac{|\P|^2}{|\k|\, W^4}\right)
\label{ampl-Fperpmin_T_b2b}
\end{align}
now features the factorization between NEik and NLP corrections. Taking the terms obtained in Eq.~\eqref{Ampl-nonFact_T_b2b_3}  into account, the second and third contributions in Eq.~\eqref{ampl_T_b2b} for the amplitude are now
\begin{align}
&i{\cal M }_{q_1 \bar q_2 \leftarrow \gamma^*_T}\bigg|_{{\mathcal{F}^{\perp -}} {\mathcal{F}^{\perp -}}}
=\, 
 \frac{e e_f}{2q^+}\,  \varepsilon_{\lambda}^l\,  
  t^{a'} t^{b'}
  \int_{\b}\,   e^{-i\b\cdot\k}\, 
\int  dv^+ \int  dw^+ 
 \mathcal{U}_A\left(+\infty,v^+;\b\right)_{a'a} g{\mathcal{F}_{i}^{\;-}}_a (v^+,\b)\,
\mathcal{U}_A\left(+\infty,w^+;\b\right)_{b'b} g{\mathcal{F}_{j}^{\;-}}_b (w^+,\b)
\nonumber\\
&
 \times\,
  \bar u(1) \gamma^+
  \Bigg\{
\Bigg[
-\frac{\delta^{ij}}{[\P^2\!+\!\bar Q^2]^2}\, 
   \left(1+i\frac{[\P^2\!+\!\bar Q^2]}{2z_1 z_2 q^+}\, \min(v^+,w^+)\right) 
+\frac{4\P^i\P^j}{[\P^2\!+\!\bar Q^2]^3}\, \left(1+i\frac{[\P^2\!+\!\bar Q^2]}{2z_1 z_2 q^+}\, ( z_2 v^+\!+\!z_1 w^+)\right)
\Bigg]
m\gamma^l
\nonumber\\
&\,\hspace{1cm}
+
\Bigg[
-\frac{\delta^{ij}\P^m}{[\P^2\!+\!\bar Q^2]^2}\, \left(1+i\frac{[\P^2\!+\!\bar Q^2]}{2z_1 z_2 q^+}\, \min(v^+,w^+)\right)
+\frac{4\P^i\P^j\P^m}{[\P^2\!+\!\bar Q^2]^3}\, \left(1+i\frac{[\P^2\!+\!\bar Q^2]}{2z_1 z_2 q^+}\, ( z_2 v^+\!+\!z_1 w^+)\right)
\nonumber\\
&\,\hspace{1.5cm}
-\frac{\delta^{jm}\P^i}{[\P^2\!+\!\bar Q^2]^2}\, \left(1+2 i z_2 v^+\, \frac{[\P^2\!+\!\bar Q^2]}{2z_1 z_2 q^+}\right)
-\frac{\delta^{im}\P^j}{[\P^2\!+\!\bar Q^2]^2}\, \left(1+2 i z_1 w^+\, \frac{[\P^2\!+\!\bar Q^2]}{2z_1 z_2 q^+}\right)
\nn\\
&\,\hspace{1.5cm}
-i\frac{(z_2\!-\!z_1)}{2z_1 z_2 q^+}\, \frac{\big[\delta^{im}\P^j\!-\! \delta^{jm}\P^i\big]}{[\P^2\!+\!\bar Q^2]}\, \min(v^+,w^+)
\Bigg]
\bigg(
(z_2\!-\!z_1)\,  \delta^{lm}+\frac{[\gamma^l,\gamma^m]}{2}
\bigg) 
\Bigg\}
 v(2)
+O\left(\frac{|\k|}{|\P|^4}\right)
+O\left(\frac{|\P|}{W^4}\right)
   \nn \\
&=
 \frac{e e_f}{2q^+}\,  \varepsilon_{\lambda}^l\,  
  t^{a'} t^{b'}
  \int_{\b}\,   e^{-i\b\cdot\k}\, 
\int  dv^+ \int  dw^+ 
 \mathcal{U}_A\left(+\infty,v^+;\b\right)_{a'a} g{\mathcal{F}_{i}^{\;-}}_a (v^+,\b)\,
\mathcal{U}_A\left(+\infty,w^+;\b\right)_{b'b} g{\mathcal{F}_{j}^{\;-}}_b (w^+,\b)
\nonumber\\
&
 \times\,
  \bar u(1) \gamma^+
  \Bigg\{
\Bigg[
-\frac{\delta^{ij}}{[\P^2\!+\!\bar Q^2]^2}\, 
   \left(1+i\frac{[\P^2\!+\!\bar Q^2]}{2z_1 z_2 q^+}\, \min(v^+,w^+)\right) 
+\frac{4\P^i\P^j}{[\P^2\!+\!\bar Q^2]^3}\, \left(1+i\frac{[\P^2\!+\!\bar Q^2]}{2z_1 z_2 q^+}\, ( z_2 v^+\!+\!z_1 w^+)\right)
\Bigg]
m\gamma^l
\nonumber\\
&\,\hspace{1cm}
+
\Bigg[
-\frac{\delta^{ij}\P^m}{[\P^2\!+\!\bar Q^2]^2}\, \left(1+i\frac{[\P^2\!+\!\bar Q^2]}{2z_1 z_2 q^+}\, \min(v^+,w^+)\right)
+\frac{4\P^i\P^j\P^m}{[\P^2\!+\!\bar Q^2]^3}\, \left(1+i\frac{[\P^2\!+\!\bar Q^2]}{2z_1 z_2 q^+}\, ( z_2 v^+\!+\!z_1 w^+)\right)
\nonumber\\
&\,\hspace{1.5cm}
-\frac{\delta^{jm}\P^i}{[\P^2\!+\!\bar Q^2]^2}\, \left(1+ i \frac{[\P^2\!+\!\bar Q^2]}{2z_1 z_2 q^+}\, \big(v^+\!-\! (z_2\!-\!z_1)(w^+\!-\!v^+)\theta(v^+\!-\!w^+)\big)\right)
\nn\\
&\,\hspace{1.5cm}
-\frac{\delta^{im}\P^j}{[\P^2\!+\!\bar Q^2]^2}\, \left(1+ i \frac{[\P^2\!+\!\bar Q^2]}{2z_1 z_2 q^+}\, \big(w^+\!+\! (z_2\!-\!z_1)(v^+\!-\!w^+)\theta(w^+\!-\!v^+)\big)\right)
\Bigg]
\bigg(
(z_2\!-\!z_1)\,  \delta^{lm}+\frac{[\gamma^l,\gamma^m]}{2}
\bigg) 
\Bigg\}
 v(2)
    \nn \\
&\,
+O\left(\frac{|\k|}{|\P|^4}\right)
+O\left(\frac{|\P|}{W^4}\right)
\label{Ampl-Fperpmin_Fperpmin_T_b2b} 
\end{align}
and
\begin{align}
i{\cal M }_{q_1 \bar q_2 \leftarrow \gamma^*_T}\bigg|_{{\mathcal{F}_{ij}} }
\equiv = &\, 
i{\cal M }_{q_1 \bar q_2 \leftarrow \gamma^*_T}^{(3)}
+i{\cal M }_{q_1 \bar q_2 \leftarrow \gamma^*_T}^{\textrm{non. fact.}}\bigg|_{{\mathcal{F}_{ij}} }
\nn\\
 = &\,
   \frac{ e e_f}{4 (q^+)^2}\,
   \frac{ \varepsilon_{\lambda}^l }{[\P^2\!+\!\bar Q^2]}\:
\bar u(1) \gamma^+ 
\bigg\{  \frac{[\gamma^i,\gamma^j]}{4z_1}\,
\bigg[
 \P^m\left((z_2\!-\!z_1)\, \delta^{lm}+\frac{[\gamma^l,\gamma^m]}{2}\right) +m\, \gamma^l
\bigg]  
\nn \\
& 
+\bigg[
 \P^m\left((z_2\!-\!z_1)\, \delta^{lm}+\frac{[\gamma^l,\gamma^m]}{2}\right) +m\, \gamma^l
\bigg] \, 
\frac{[\gamma^i,\gamma^j]}{4z_2}
\nn \\
& 
 +\frac{(z_2\!-\!z_1)}{2z_1z_2}\, \epsilon^{ij}\, \epsilon^{mn}\P^n\, \left((z_2\!-\!z_1)\, \delta^{lm}+\frac{[\gamma^l,\gamma^m]}{2}\right) 
 \bigg\} v(2)
  \nn \\
& \times
  t^{a'}\int_{\b} \, e^{-i\b\cdot\k}\, 
  \int dv^+ \,   \mathcal{U}_A(+\infty,v^+;\b)_{a'a} \; g \mathcal{F}_{ij}^a(v^+,\b)\:
  +O\left(\frac{|\k|}{ W^2 |\P|^2}\right)
+O\left(\frac{|\P|}{W^4}\right)
 \, .
\label{ampl-Fij_T_b2b} 
\end{align}
Finally the contribution of ${\mathcal{F}^{+ -}}$ is unchanged, so that
\begin{align}
i{\cal M }_{q_1 \bar q_2 \leftarrow \gamma^*_T}\bigg|_{{\mathcal{F}^{+ -}} } 
\equiv &\,
 i{\cal M }_{q_1 \bar q_2 \leftarrow \gamma^*_T}^{\textrm{dyn. tar.}}
 \nn\\
= &\,
 \frac{ e e_f}{2(q^+)^2}\, \varepsilon_{\lambda}^l\,
 \bar u(1)  \gamma^+
 \Bigg\{
 \frac{2\P^l}{[\P^2\!+\!\bar Q^2]} 
 +
\frac{(z_2\!-\!z_1)Q^2}{[\P^2\!+\!\bar Q^2]^2}  
\bigg[
\P^m\left((z_2\!-\!z_1)\, \delta^{lm}+\frac{[\gamma^l,\gamma^m]}{2}\right) +m\gamma^l
\bigg]
\Bigg\}v(2)
\nonumber\\
&\, \times\,
\int_{\b}\  e^{-i\b\cdot\k}\, t^{a'}  \int dv^+ 
\,  
\mathcal{U}_A(+\infty,v^+;\b)_{a'a} \,
g \mathcal{F}^{+-}_a(v^+,\b)\:
+O\left(\frac{|\k|}{W^2\, |\P|^2}\right)
+O\left(\frac{|\P|}{W^4}\right)
\label{ampl-Fplusmin_T_b2b}
\, .
\end{align}


\subsection{Cross section via transverse photon}


The transverse photon amplitude~\eqref{ampl_T_b2b} can be alternatively split into two types of terms according to their Dirac structure: light-front helicity flip terms, containing one factor of $m \gamma^l$, and light-front helicity conserving terms. The interference between these two types of terms vanish by Dirac algebra. 
Then, when squaring the contribution \eqref{ampl-Fperpmin_T_b2b} to the amplitude, the non-flip and the flip terms can be squared separately thanks to the relations
%
\begin{align}
&\, 
\frac{1}{2}\sum_{\lambda}\sum_{h_1,h_2}\, 
\bigg[ \varepsilon_{\lambda}^{l'}\,\bar u(1) \gamma^+ \left((z_2\!-\!z_1)\, \delta^{l'm'}+\frac{[\gamma^{l'},\gamma^{m'}]}{2}\right)v(2)
\bigg]^{\dag}
  \varepsilon_{\lambda}^l\,\bar u(1) \gamma^+ \left((z_2\!-\!z_1)\, \delta^{lm}+\frac{[\gamma^l,\gamma^m]}{2}\right)v(2)
\nn\\
=&\, 8\, (k^+)^2\, z_1\, z_2\,  \Big[z_1^2+z_2^2\Big]\,  \delta^{m'm}
\, ,
\label{Dirac_struct_T_non_flip}
\end{align}
and
%
\begin{align}
\frac{1}{2}\sum_{\lambda}\sum_{h_1,h_2}\, 
\bigg[ \varepsilon_{\lambda}^{l'}\, m\,  \bar u(1) \gamma^+ \gamma^{l'} v(2)
\bigg]^{\dag}
  \varepsilon_{\lambda}^l\,  m\, \bar u(1) \gamma^+ \gamma^l v(2)
=&\, 8\, (k^+)^2\, z_1\, z_2\, m^2
\, .
\label{Dirac_struct_T_flip}
\end{align}
In such a way, one finds the contribution
%
\begin{align}
 \frac{d\sigma_{\gamma^{*}_T\rightarrow q_1\bar q_2}}{d {\rm P.S.}}\Bigg|_{{\mathcal{F}^{\perp -}} {\mathcal{F}^{\perp -}}}
=&\, 2q^+ 2\pi\,  \delta(q^+\!-\!k^+) \, 
\,   e^2 e_f^2\, z_1 z_2
\Bigg\{
\frac{[z_1^2+z_2^2]}{[\P^2\!+\!\bar Q^2]^2}\, 
\left[1+\frac{2(z_2\!-\!z_1)(\k\!\cdot\!\P)}{[\P^2\!+\!\bar Q^2]}\right]\delta^{ij}
\nonumber\\
 &
 +\frac{\big[[z_1^2+z_2^2]\bar Q^2-m^2\big]}{[\P^2\!+\!\bar Q^2]^4}\, 
 \Bigg[-4\P^i\P^j\left(1+\frac{4(z_2\!-\!z_1)(\k\!\cdot\!\P)}{[\P^2\!+\!\bar Q^2]}\right)
 +2(z_2\!-\!z_1) \big(\P^i\k^j+\k^i\P^j\big)
 \Bigg]
 \Bigg\}
\nonumber\\
 &\, \times\, 
  \int_{\b,\b'}\,   e^{i(\b'-\b)\cdot\k}\, 
   \int  dv^+ d{v'}^+\,
 \bigg[ 1-
 i\frac{[\P^2\!+\!\bar Q^2]}{2q^+z_1 z_2} \, 
 ({v'}^+\!-\!v^+)
 \bigg]
\nonumber\\
&
\hspace{-3cm} \times\, 
 \Big\langle 
 g{\mathcal{F}_{j}^{\;-}}_b ({v'}^+,\b')
 \Big[  \mathcal{U}_A\left(+\infty,{v'}^+;\b'\right)^{\dag}  \mathcal{U}_A\left(+\infty,v^+;\b\right)\Big]_{ba}
  g{\mathcal{F}_{i}^{\;-}}_a (v^+,\b)\Big\rangle
 +O\left(\frac{1}{\P^6}\right)+O\left(\frac{1}{\k^2 W^4}\right)
\label{X_sec_T_Fperpmin_Fperpmin}
\end{align}
%
to the cross section, keeping only terms which are of order Eik or NEik, and LP or NLP.
The interference between the contributions \eqref{Ampl-Fperpmin_Fperpmin_T_b2b}  and  \eqref{ampl-Fperpmin_T_b2b} to the amplitude can be calculated using the same relations~\eqref{Dirac_struct_T_non_flip} and~\eqref{Dirac_struct_T_flip}, and one finds
%
%
\begin{align}
&
\frac{d\sigma_{\gamma^{*}_T\rightarrow q_1\bar q_2}}{d {\rm P.S.}}\Bigg|_{{\mathcal{F}^{\perp -}} {\mathcal{F}^{\perp -}} {\mathcal{F}^{\perp -}}}
= 2q^+ 2\pi\,  \delta(q^+-k^+) \,   e^2 e_f^2\, 2 z_1 z_2\, 2\, {\textrm{Re}}\; {\rm tr}_F\big( t^{a'} t^{b'} t^{c'}\big)\;      
\,  
 \int_{\b,\b'}\,   e^{i(\b'-\b)\cdot\k}\, 
  \int  dv^+ dw^+ d{v'}^+
 \nonumber\\
 &\times\, 
 \Big\langle   \mathcal{U}_A\left(+\infty,{v'}^+;\b'\right)_{c'c}  g{\mathcal{F}_{l}^{\;-}}_c ({v'}^+,\b')
 \mathcal{U}_A\left(+\infty,{v}^+;\b\right)_{a'a}  g{\mathcal{F}_{i}^{\;-}}_a ({v}^+,\b)
  \mathcal{U}_A\left(+\infty,{w}^+;\b\right)_{b'b}  g{\mathcal{F}_{j}^{\;-}}_b ({w}^+,\b)
  \Big\rangle
   \nonumber\\
 &
\times
\Bigg\{
 \frac{\big[[z_1^2+z_2^2](\P^2\!-\!\bar Q^2)+2m^2\big]}{[\P^2\!+\!\bar Q^2]^4}\, \P^l
 \Bigg[
 \delta^{ij}\bigg(1-  i\frac{[\P^2\!+\!\bar Q^2]}{2q^+z_1 z_2} \, 
 \left({v'}^+\!-\!\min(v^+,w^+)\right)\bigg)
 \nn\\
 &\, \hspace{5cm}
-\frac{4\, \P^i\P^j}{[\P^2\!+\!\bar Q^2]}
\bigg(1-  i\frac{[\P^2\!+\!\bar Q^2]}{2q^+z_1 z_2} \, 
 \left({v'}^+\!-\!z_2 v^+\!-\!z_1 w^+)\right)\bigg)
 \Bigg]
 \nonumber\\
 &
-\frac{[z_1^2+z_2^2]}{[\P^2\!+\!\bar Q^2]^3}\, \P^i 
\left[\delta^{jl}-\frac{2\, \P^j\P^l}{[\P^2\!+\!\bar Q^2]}\right]
\bigg(1-  i\frac{[\P^2\!+\!\bar Q^2]}{2q^+z_1 z_2} \, 
 \left({v'}^+\!-\!v^+-(z_2\!-\!z_1)   (v^+\!-\! w^+)  \theta(v^+\!-\! w^+) \right)\bigg)
\nonumber\\
 &
-\frac{[z_1^2+z_2^2]}{[\P^2\!+\!\bar Q^2]^3}\, \P^j
\left[\delta^{il}-\frac{2\, \P^i\P^l}{[\P^2\!+\!\bar Q^2]}\right]
\bigg(1-  i\frac{[\P^2\!+\!\bar Q^2]}{2q^+z_1 z_2} \, 
 \left({v'}^+\!-\!w^++(z_2\!-\!z_1)   (w^+\!-\! v^+)  \theta(w^+\!-\! v^+) \right)\bigg) 
 \Bigg\}
 \nn\\
&\, 
+O\left(\frac{1}{\P^6}\right)
+O\left(\frac{1}{|\k| |\P|  W^4}\right)
   \, .
\label{X_sec_T_Fperpmin_Fperpmin_Fperpmin}    
\end{align}
Similarly, the interference between the contributions \eqref{ampl-Fplusmin_T_b2b}  and  \eqref{ampl-Fperpmin_T_b2b} is obtained as
\begin{align}
 \frac{d\sigma_{\gamma^{*}_T\rightarrow q_1\bar q_2}}{d {\rm P.S.}}\Bigg|_{{\mathcal{F}^{+ -}} {\mathcal{F}^{\perp -}}}
=&\,  2q^+ \, 2\pi\, \delta(q^+-k^+) 
\,    e^2 e_f^2\, \frac{z_1 z_2 (z_2\!-\!z_1)}{q^+} \; \frac{(-1)\P^j}{[\P^2\!+\!\bar Q^2]^4}\,
 \nonumber\\
 &\times
 \bigg\{
\big[\P^2\!+\!\bar Q^2+(z_1^2+z_2^2)Q^2\big] [\P^2\!-\!\bar Q^2]
+2m^2 Q^2
\bigg\}\,
2\, {\textrm{Re}}\; \int_{\b,\b'}\,   e^{i(\b'-\b)\cdot\k}\,
 \int  dv^+ d{v'}^+
 \nonumber\\
 &\times
%
%
 \Big\langle 
 g{\mathcal{F}_{j}^{\;-}}_b ({v'}^+,\b')
 \Big[  \mathcal{U}_A\left(+\infty,{v'}^+;\b'\right)^{\dag}  \mathcal{U}_A\left(+\infty,v^+;\b\right)\Big]_{ba}
  g\mathcal{F}^{\, +-}_a (v^+,\b)\Big\rangle
  \nn\\
  &\,
  +O\left(\frac{1}{\P^4 W^2}\right)
  +O\left(\frac{1}{|\k| |\P|  W^4}\right)
  \, .
\label{X_sec_T_Fplusmin_Fperpmin}  
\end{align}
%
Finally, the interference between the contributions~\eqref{ampl-Fij_T_b2b}  and~\eqref{ampl-Fperpmin_T_b2b} is found to be 
\begin{align}
 \frac{d\sigma_{\gamma^{*}_T\rightarrow q_1\bar q_2}}{d {\rm P.S.}}\Bigg|_{{\mathcal{F}_{ij}} {\mathcal{F}^{\perp -}}}
=&\,  2q^+ \, 2\pi\, \delta(q^+-k^+) 
\,    e^2 e_f^2\, \frac{z_1 z_2 (z_2\!-\!z_1)}{q^+} \; \frac{ \epsilon^{ml}\P^m}{[\P^2\!+\!\bar Q^2]^2}\,
\frac{\epsilon^{ij}}{2}\, 
2\, {\textrm{Re}}\; \int_{\b,\b'}\,   e^{i(\b'-\b)\cdot\k}\,
 \int  dv^+ d{v'}^+
 \nonumber\\
 &\times
%
%
 \Big\langle 
 g{\mathcal{F}_{l}^{\;-}}_b ({v'}^+,\b')
 \Big[  \mathcal{U}_A\left(+\infty,{v'}^+;\b'\right)^{\dag}  \mathcal{U}_A\left(+\infty,v^+;\b\right)\Big]_{ba}
  g\mathcal{F}_{ij}^a (v^+,\b)\Big\rangle
   \nn\\
  &\,
  +O\left(\frac{1}{\P^4 W^2}\right)
  +O\left(\frac{1}{|\k| |\P|  W^4}\right)
  \, ,
\label{X_sec_T_Fij_Fperpmin}  
\end{align}
%
using in particular the identity 
\begin{align}
{\rm tr}_D \left\{\frac{\gamma^-\gamma^+}{2} \frac{[\gamma^{i},\gamma^{j}]}{2}  \frac{[\gamma^{l},\gamma^{m}]}{2}
\right\}
=&\,
-2  \epsilon^{ij}\, \epsilon^{lm}
\, ,
\end{align}
%
valid in the case of a four-dimensional spacetime. The DIS dijet cross section via transverse photon, including contributions of order Eik or NEik at high energy and LP or NLP in the back-to-back jets regime, is then
%
\begin{align}
&
\frac{d\sigma_{\gamma^{*}_T\rightarrow q_1\bar q_2}}{d {\rm P.S.}}\Bigg|_{{\rm Eik }+{\rm NEik}}
= \frac{d\sigma_{\gamma^{*}_T\rightarrow q_1\bar q_2}}{d {\rm P.S.}}\Bigg|_{{\mathcal{F}^{\perp -}} {\mathcal{F}^{\perp -}}}
+\frac{d\sigma_{\gamma^{*}_T\rightarrow q_1\bar q_2}}{d {\rm P.S.}}\Bigg|_{{\mathcal{F}^{\perp -}} {\mathcal{F}^{\perp -}} {\mathcal{F}^{\perp -}}}
+ \frac{d\sigma_{\gamma^{*}_T\rightarrow q_1\bar q_2}}{d {\rm P.S.}}\Bigg|_{{\mathcal{F}^{+ -}} {\mathcal{F}^{\perp -}}}
+ \frac{d\sigma_{\gamma^{*}_T\rightarrow q_1\bar q_2}}{d {\rm P.S.}}\Bigg|_{{\mathcal{F}_{ij}} {\mathcal{F}^{\perp -}}}
\, ,
\label{X_sec_T_gen_form}
\end{align}
with the contributions written in Eqs.~\eqref{X_sec_T_Fperpmin_Fperpmin}, \eqref{X_sec_T_Fperpmin_Fperpmin_Fperpmin},
\eqref{X_sec_T_Fplusmin_Fperpmin} and~\eqref{X_sec_T_Fij_Fperpmin}.


\section{Consistency with the TMD formalism}
\label{sec:CGC_vs_TMD}

At this stage, we have obtained the DIS dijet cross sections via longitudinal or transverse photon, Eqs.~\eqref{X_sec_L_gen_form} and~\eqref{X_sec_T_gen_form}, as linear combinations of terms involving correlators of two or three background field strength, in the back-to-back limit at the considered accuracy.
So far, the background field has been treated as a classical field, and the averaging over it, noted $\langle \cdots\rangle$, is implicitly some statistical average generalizing the CGC formalism.
 In order to make contact with the TMD formalism, one has to promote the classical background field to a quantum field, and relate the statistical average of a quantity to its expectation value in the quantum state of the target,\footnote{In general, this procedure is ambiguous, due to the lack of knowledge of the correct ordering of the fields as quantum operators. However, for fields with zero expectation values (which is the case for color fields), the ordering of the fields do not matter in correlators of two or three fields, but only only in correlators of four or more fields. Hence, there is no ambiguity related to the quantum ordering of the field strength insertions in this study. Only the gauge links could cause ambiguities when promoting the background from classical to quantum field. }
as
(see for example Refs.~\cite{Belitsky:2002sm,Dominguez:2011wm,Marquet:2016cgx,Altinoluk:2019wyu})
\ba
\langle{\cal O} \rangle &=& \lim_{P_{{tar}}'\rightarrow P_{{tar}}}
\frac{\langle P_{{tar}}'|\hat{\cal O}| P_{{tar}}\rangle}{\langle P_{{tar}}'| P_{{tar}}\rangle}
\, .
\label{def_average}
\ea
We choose target states\footnote{The target states forming a basis should include some information about the target spin, for example its helicity. However, in this study we focus on scattering on an unpolarized target. Then, in order to simplify the notations, we drop the helicity of the target in the notation for the quantum state, and the averaging over that helicity in Eq.~\eqref{def_average} is implicit. Note that the discussion in the present section could in principle be repeated for any type of spin asymmetry of the target. } normalized as
\ba
\langle P_{{tar}}'| P_{{tar}}\rangle &=& 2P_{{tar}}^-\, (2\pi)^3 \delta(P_{{tar}}^{'-}\!-\!P_{{tar}}^-)\, \delta^{(2)}(\P_{{tar}}'\!-\!\P_{{tar}})
\, .
\label{norm_states}
\ea
Once the background field is treated as a quantum operator, translations can be performed using the action of the momentum operator. Following this idea, one can rewrite the quantity appearing in Eqs.~\eqref{X_sec_L_Fperpmin_Fperpmin} and~\eqref{X_sec_T_Fperpmin_Fperpmin} as (see for example Ref.~\cite{Altinoluk:2023qfr} for a detailed derivation)
%
\begin{align}
& \int_{\b,\b'}\,   e^{i(\b'-\b)\cdot\k}\, 
   \int  dv^+ d{v'}^+\,
 \bigg[ 1-
 i\frac{[\P^2\!+\!\bar Q^2]}{2q^+z_1 z_2} \, 
 ({v'}^+\!-\!v^+)
 \bigg]
\nonumber\\
&
\times\, 
 \Big\langle 
 {\mathcal{F}_{j}^{\;-}}_b ({v'}^+,\b')
 \Big[  \mathcal{U}_A\left(+\infty,{v'}^+;\b'\right)^\dagger  \mathcal{U}_A\left(+\infty,v^+;\b\right)\Big]_{ba}
  {\mathcal{F}_{i}^{\;-}}_a (v^+,\b)\Big\rangle
\nonumber\\
=&\,
\frac{1}{2P_{{tar}}^-}
 \int_{\Delta\b}\,   e^{i\k \cdot\Delta\b}\, 
   \int  d\Delta v^+\,
    \bigg[ 1-
 i\frac{[\P^2\!+\!\bar Q^2]}{2q^+z_1 z_2} \, 
 \Delta v^+
 \bigg]
\nonumber\\
 &\, \times\, 
  \Big\langle P_{{tar}}\Big|
 {\mathcal{F}_{j}^{\;-}}_b (\Delta v^+,\Delta\b)
 \Big[  \mathcal{U}_A\left(+\infty,\Delta v^+;\Delta\b\right)^\dagger  \mathcal{U}_A\left(+\infty,0;0\right)\Big]_{ba}
  {\mathcal{F}_{i}^{\;-}}_a (0,0)\Big| P_{{tar}}\Big\rangle
  \nonumber\\
=&\,
\frac{1}{2P_{{tar}}^-}
 \int_{\Delta\b}\,   e^{i\k \cdot\Delta\b}\, 
   \int  d\Delta v^+\,
    \bigg[ 1-
 i\frac{[\P^2\!+\!\bar Q^2]}{2q^+z_1 z_2} \, 
 \Delta v^+
 \bigg]
\nonumber\\
 &\, \times\, 
  \Big\langle P_{{tar}}\Big|
 {\mathcal{F}_{j}^{\;-}}_b (0,0)
 \Big[  \mathcal{U}_A\left(+\infty,0;0\right)^\dagger  \mathcal{U}_A\left(+\infty,-\Delta v^+;-\Delta\b\right)\Big]_{ba}
  {\mathcal{F}_{i}^{\;-}}_a (-\Delta v^+,-\Delta\b)\Big| P_{{tar}}\Big\rangle
  \nonumber\\
=&\,  
\frac{(2\pi)^3}{2}\,   \bigg[ 1 +\frac{[\P^2\!+\!\bar Q^2]}{z_1 z_2(2q^+P_{{tar}}^-)} \, \partial_{{\rm x}}\bigg]  
 \bigg[ {\rm x}\, \Phi^{j -;i-}({\rm x},\k) \bigg] \bigg|_{{\rm x}=0}
\label{class_to_quant_Fperpmin_Fperpmin}
\, ,
\end{align}
using the notation\footnote{Note that with this notation, the various $\Phi$ components do not have the same magnitude in terms of the power counting introduced in Section ~\ref{sec:powcount}.}
%
\begin{align}
\Phi^{\mu\nu;\rho\sigma}({\rm x},\k) \equiv&\,
\frac{1}{{\rm x} P_{{tar}}^-}\, \frac{1}{(2\pi)^3}
 \int d^2\z\,   e^{-i\k \cdot\z}\, 
   \int  d z^+\, e^{i {\rm x} P_{{tar}}^- z^+}\, 
\nonumber\\
 &\, \times\, 
  \Big\langle P_{{tar}}\Big|
 \mathcal{F}^{\mu\nu}_b (0)
 \Big[  \mathcal{U}_A\left(+\infty,0;0\right)^\dagger  \mathcal{U}_A\left(+\infty,z^+;\z\right)\Big]_{ba}
  \mathcal{F}^{\rho\sigma}_a (z)\Big| P_{{tar}}\Big\rangle \bigg|_{z^-=0}
\label{def_Phi_correlator_Fmunu_Frhosigma}
\, ,
\end{align}
for the gluon field strength correlator appearing in the definition of gluon TMDs in an unpolarized target, with a gauge link forming a future pointing staple.

In the same way, one can rewrite the field strength correlator appearing in Eqs.~\eqref{X_sec_L_Fplusmin_Fperpmin}  and~\eqref{X_sec_T_Fplusmin_Fperpmin} as
%
\begin{align}
& 2\, {\textrm{Re}}\; \int_{\b,\b'}\,   e^{i(\b'-\b)\cdot\k}\,
 \int  dv^+ d{v'}^+
%
%
 \Big\langle 
 {\mathcal{F}_{j}^{\;-}}_b ({v'}^+,\b')
 \Big[  \mathcal{U}_A\left(+\infty,{v'}^+;\b'\right)^\dagger  \mathcal{U}_A\left(+\infty,v^+;\b\right)\Big]_{ba}
  \mathcal{F}^{\, +-}_a (v^+,\b)\Big\rangle
\nonumber\\
=&\,
\int_{\b,\b'}\,   e^{i(\b'-\b)\cdot\k}\,
 \int  dv^+ d{v'}^+
 \Big\langle 
 {\mathcal{F}_{j}^{\;-}}_b ({v'}^+,\b')
 \Big[  \mathcal{U}_A\left(+\infty,{v'}^+;\b'\right)^\dagger  \mathcal{U}_A\left(+\infty,v^+;\b\right)\Big]_{ba}
  \mathcal{F}^{\, +-}_a (v^+,\b)\Big\rangle
  \nonumber\\
 &\,
 + 
 \int_{\b,\b'}\,   e^{i(\b'-\b)\cdot\k}\,
 \int  dv^+ d{v'}^+
 \Big\langle 
 \mathcal{F}^{\, +-}_b ({v'}^+,\b')
 \Big[  \mathcal{U}_A\left(+\infty,{v'}^+;\b'\right)^\dagger  \mathcal{U}_A\left(+\infty,v^+;\b\right)\Big]_{ba}
  {\mathcal{F}_{j}^{\;-}}_a (v^+,\b)\Big\rangle
  \nonumber\\
=&\,  
\frac{(2\pi)^3}{2}\,  
 \bigg[ - {\rm x}\, \Phi^{j-;+-}({\rm x},\k)  - {\rm x}\, \Phi^{+-;j-}({\rm x},\k)\bigg] \bigg|_{{\rm x}=0}
\label{class_to_quant_Fplusmin_Fperpmin}
\, ,
\end{align}
and the field strength correlator appearing in Eq.~\eqref{X_sec_T_Fij_Fperpmin} as
%
\begin{align}
& 2\, {\textrm{Re}}\; \int_{\b,\b'}\,   e^{i(\b'-\b)\cdot\k}\,
 \int  dv^+ d{v'}^+
 \Big\langle 
 {\mathcal{F}_{l}^{\;-}}_b ({v'}^+,\b')
 \Big[  \mathcal{U}_A\left(+\infty,{v'}^+;\b'\right)^\dagger  \mathcal{U}_A\left(+\infty,v^+;\b\right)\Big]_{ba}
  \mathcal{F}_{ij}^a (v^+,\b)\Big\rangle
\nonumber\\
=&\,
\int_{\b,\b'}\,   e^{i(\b'-\b)\cdot\k}\,
 \int  dv^+ d{v'}^+
 \Big\langle 
 {\mathcal{F}_{l}^{\;-}}_b ({v'}^+,\b')
 \Big[  \mathcal{U}_A\left(+\infty,{v'}^+;\b'\right)^\dagger  \mathcal{U}_A\left(+\infty,v^+;\b\right)\Big]_{ba}
  \mathcal{F}_{ij}^a (v^+,\b)\Big\rangle
  \nonumber\\
 &\,
 + 
 \int_{\b,\b'}\,   e^{i(\b'-\b)\cdot\k}\,
 \int  dv^+ d{v'}^+
 \Big\langle 
 \mathcal{F}_{ij}^b ({v'}^+,\b')
 \Big[  \mathcal{U}_A\left(+\infty,{v'}^+;\b'\right)^\dagger  \mathcal{U}_A\left(+\infty,v^+;\b\right)\Big]_{ba}
  {\mathcal{F}_{l}^{\;-}}_a (v^+,\b)\Big\rangle
  \nonumber\\
=&\,  
\frac{(2\pi)^3}{2}\,  
 \bigg[ - {\rm x}\, \Phi^{l-;ij}({\rm x},\k)  - {\rm x}\, \Phi^{ij;l-}({\rm x},\k)\bigg] \bigg|_{{\rm x}=0}
\label{class_to_quant_Fij_Fperpmin}
\, .
\end{align}
Hence, the cross sections~\eqref{X_sec_L_gen_form} and \eqref{X_sec_T_gen_form} can be written in the general form 
%
\begin{align}
\frac{d\sigma_{\gamma^{*}_{T,L}\rightarrow q_1\bar q_2}}{d {\rm P.S.}}\Bigg|_{{\rm Eik }+{\rm NEik}}
= &\,
2q^+ \, 2\pi\, \delta(q^+\!-\!k^+) 
\,  z_1 z_2\,  e^2 e_f^2\, g^2 \; \frac{(2\pi)^3}{2}\,
\Bigg\{
{\cal C}^{ij}_{T,L}(z_1,\P,\k)
 \bigg[ 1 +\frac{[\P^2\!+\!\bar Q^2]}{z_1 z_2(2q^+P_{{tar}}^-)} \, \partial_{{\rm x}}\bigg]  
 \bigg[ {\rm x}\, \Phi^{j -;i-}({\rm x},\k) \bigg] 
 \nn\\
 &\,
 -\frac{1}{2q^+}\, {\cal C}^{j}_{T,L}(z_1,\P)\bigg[  {\rm x}\, \Phi^{j-;+-}({\rm x},\k) + {\rm x}\, \Phi^{+-;j-}({\rm x},\k)\bigg]
  \nn\\
 &\,
 -\frac{1}{2q^+}\, {\cal C}^{ijl}_{T,L}(z_1,\P)\bigg[ {\rm x}\, \Phi^{l-;ij}({\rm x},\k)  + {\rm x}\, \Phi^{ij;l-}({\rm x},\k)\bigg]
\Bigg\}\Bigg|_{{\rm x}=0}
+3\, \mathcal{F}\textrm{ terms}
\, ,
\label{X_sec_gen_form_Phi}
\end{align}
with hard factors ${\cal C}^{ij}_{T,L}$, ${\cal C}^{j}_{T,L}$ and ${\cal C}^{ijl}_{T,L}$ which can be read off from the results from sections \ref{sec:gamma_L_case} and \ref{sec:gamma_T_case}, and which are listed in Appendix~\ref{app:coeffs}.

The first line of Eq.~\eqref{X_sec_gen_form_Phi} features the leading operator structure $\Phi^{j -;i-}$, and the factors in front contains NLP (kinematical twist) corrections (in ${\cal C}^{ij}_{T,L}$) and NEik corrections (in the $[1+\dots]$ bracket) that are factorized from eachother, as we pointed out previsoulsy. The second and third line feature new operator structures that are purley NEik corrections. Finally, as mentioned before, the $3\, \mathcal{F}$ term contains eikonal NLP (genuine twist) corrections and further contributions that are both NLP and NEik. The field strength correlators of the type \eqref{def_Phi_correlator_Fmunu_Frhosigma} can now be parametrized in terms of gluon TMD distributions that are real scalar functions, -- and all of the same magnitude under our power counting rules -- as~\cite{Mulders:2000sh,Meissner:2007rx,Lorce:2013pza} 
%
\begin{align}
 \Phi^{j -;i-}({\rm x},\k) =&\,
\frac{\delta^{ij}}{2}\, f_1^g({\rm x},\k) +\left[\k^i\k^j-\frac{\k^2}{2}\, \delta^{ij}\right]\frac{1}{2 M^2}\, h_1^{\perp g}({\rm x},\k)
\nn\\
\Phi^{j-;+-}({\rm x},\k) =&\,
\frac{\k^j}{P_{{tar}}^-}\, \left[f^{\perp g}({\rm x},\k)-i  \bar{f}^{\perp g}({\rm x},\k)\right] 
\nn\\
\Phi^{+-;j-}({\rm x},\k) =&\,
\frac{\k^j}{P_{{tar}}^-}\, \left[f^{\perp g}({\rm x},\k)+i  \bar{f}^{\perp g}({\rm x},\k)\right] 
\nn\\
\Phi^{l-;ij}({\rm x},\k) =&\,
\epsilon^{ij}\, \epsilon^{ln}\, 
\frac{\k^n}{P_{{tar}}^-}\, \left[ \bar{g}^{\perp g}({\rm x},\k)+i  g^{\perp g}({\rm x},\k)\right] 
\nn\\
\Phi^{ij;l-}({\rm x},\k) =&\,
\epsilon^{ij}\, \epsilon^{ln}\, 
\frac{\k^n}{P_{{tar}}^-}\, \left[ \bar{g}^{\perp g}({\rm x},\k)-i  g^{\perp g}({\rm x},\k)\right] 
\label{param_Phi}
\, ,
\end{align}
where $M$ is the target mass.
Inserting the relations \eqref{param_Phi} into Eq.~\eqref{X_sec_gen_form_Phi} as well as the expression \eqref{phase_space_k_P_z} for the phase-space measure, and integrating over $k^+$, one obtains
%
\begin{align}
\frac{d\sigma_{\gamma^{*}_{T,L}\rightarrow q_1\bar q_2}}{dz_1\, d^2\P\, d^2\k}\Bigg|_{{\rm Eik }+{\rm NEik}}
= &\,
  \alpha_{\rm em} e_f^2\, \alpha_{s} \; 
\Bigg\{
{\cal C}^{ f_1^g}_{T,L}(z_1,\P,\k) 
 \bigg[ 1 +\frac{[\P^2\!+\!\bar Q^2]}{z_1 z_2 W^2} \, \partial_{{\rm x}}\bigg]  
 \bigg[ {\rm x}\, f_1^g({\rm x},\k) \bigg] 
 \nn\\
 &\,
 + {\cal C}^{ h_1^{\perp g}}_{T,L}(z_1,\P,\k)
 \bigg[ 1 +\frac{[\P^2\!+\!\bar Q^2]}{z_1 z_2 W^2} \, \partial_{{\rm x}}\bigg]  
 \bigg[ {\rm x}\, h_1^{\perp g}({\rm x},\k) \bigg] 
 \nn\\
 &\,
+\frac{\k\!\cdot\!\P}{W^2}\, {\cal C}^{ f^{\perp g}}_{T,L}(z_1,\P)\,  {\rm x}\, f^{\perp g}({\rm x},\k)
+\frac{\k\!\cdot\!\P}{W^2}\, {\cal C}^{  \bar{g}^{\perp g}}_{T,L}(z_1,\P)\,
{\rm x}\,  \bar{g}^{\perp g}({\rm x},\k)
\Bigg\}\Bigg|_{{\rm x}=0}
+3\, \mathcal{F}\textrm{ terms}
\, .
\label{X_sec_gen_form_scal_TMDs}
\end{align}
The contractions of the coefficients ${\cal C}^{ij}_{T,L}$, ${\cal C}^{j}_{T,L}$ and ${\cal C}^{ijl}_{T,L}$ appearing in Eq.~\eqref{X_sec_gen_form_scal_TMDs} are found to be
%
\begin{align}
 {\cal C}^{ f_1^g}_{L}(z_1,\P,\k) 
\equiv &\,
\frac{\delta^{ij}}{2}\, {\cal C}^{ij}_{L}(z_1,\P,\k) 
=
\frac{8Q^2\, z_1^2 z_2^2}{[\P^2\!+\!\bar Q^2]^4} 
\left\{\P^2 +(z_2\!-\!z_1)(\k\!\cdot\!\P) \left[-1
+\frac{4\P^2}{[\P^2\!+\!\bar Q^2]}\right]
\right\}
\nn\\
 {\cal C}^{ h_1^{\perp g}}_{L}(z_1,\P,\k) 
\equiv &\,
\frac{1}{2 M^2}\, \left[\k^i\k^j\!-\!\frac{\k^2}{2}\, \delta^{ij}\right]\, 
 {\cal C}^{ij}_{L}(z_1,\P,\k) 
 \nn\\
 =&\,
 \frac{4Q^2\, z_1^2 z_2^2}{[\P^2\!+\!\bar Q^2]^4}\, \frac{\k^2}{M^2} 
\Bigg\{\left(\frac{2(\k\!\cdot\!\P)^2}{\k^2\P^2}-1\right)\P^2 
+(z_2\!-\!z_1)(\k\!\cdot\!\P) \left[-1
+\frac{4\P^2}{[\P^2\!+\!\bar Q^2]}\left(\frac{2(\k\!\cdot\!\P)^2}{\k^2\P^2}-1\right)\right]
\Bigg\}
\nn\\
{\cal C}^{ f^{\perp g}}_{L}(z_1,\P) 
\equiv &\,
\frac{-2 \k^j}{\k\!\cdot\!\P}\, {\cal C}^{j}_{L}(z_1,\P)
=
 -32Q^2\, z_1 z_2\,
 \frac{(z_2\!-\!z_1)[\P^2\!+\!m^2]}{ [\P^2\!+\!\bar Q^2]^4}
 \nn\\
{\cal C}^{  \bar{g}^{\perp g}}_{L}(z_1,\P) 
\equiv &\,
 - \epsilon^{ij}\, \epsilon^{ln}\, 
\frac{2\k^n}{\k\!\cdot\!\P}\, {\cal C}^{ijl}_{L}(z_1,\P)
=
0
\label{scal_coeff_contract_L}
\end{align}
in the longitudinal photon case (remembering that ${\cal C}^{ijl}_{L}=0$), and
%
\begin{align}
{\cal C}^{ f_1^g}_{T}(z_1,\P,\k) 
\equiv &\,
\frac{\delta^{ij}}{2}\, {\cal C}^{ij}_{T}(z_1,\P,\k) 
\nn\\
 =&\,
\frac{(z_1^2+z_2^2)}{[\P^2\!+\!\bar Q^2]^2}
\left[1+\frac{2(z_2\!-\!z_1)(\k\!\cdot\!\P)}{[\P^2\!+\!\bar Q^2]}
\right]
- \frac{2\big[(z_1^2\!+\!z_2^2)\bar Q^2\!-\!m^2\big]}{[\P^2\!+\!\bar Q^2]^4} 
\left\{\P^2 +(z_2\!-\!z_1)(\k\!\cdot\!\P) \left[-1
+\frac{4\P^2}{[\P^2\!+\!\bar Q^2]}\right]
\right\}
\nn\\
{\cal C}^{ h_1^{\perp g}}_{T}(z_1,\P,\k) 
\equiv &\,
\frac{1}{2 M^2}\, \left[\k^i\k^j\!-\!\frac{\k^2}{2}\, \delta^{ij}\right]\, 
 {\cal C}^{ij}_{T}(z_1,\P,\k) 
 \nn\\
 =&\, 
  -\frac{\big[(z_1^2\!+\!z_2^2)\bar Q^2\!-\!m^2\big]}{[\P^2\!+\!\bar Q^2]^4}\, \frac{\k^2}{M^2} 
\Bigg\{\left(\frac{2(\k\!\cdot\!\P)^2}{\k^2\P^2}-1\right)\P^2 
+(z_2\!-\!z_1)(\k\!\cdot\!\P) \left[-1
+\frac{4\P^2}{[\P^2\!+\!\bar Q^2]}\left(\frac{2(\k\!\cdot\!\P)^2}{\k^2\P^2}-1\right)\right]
\Bigg\}
\nn\\
{\cal C}^{ f^{\perp g}}_{T}(z_1,\P) 
\equiv &\,
\frac{-2 \k^j}{\k\!\cdot\!\P}\, {\cal C}^{j}_{T}(z_1,\P)
 =
 \frac{4(z_2\!-\!z_1)}{ [\P^2\!+\!\bar Q^2]^4}
 \bigg\{
\big[\P^2\!+\!\bar Q^2+(z_1^2+z_2^2)Q^2\big] [\P^2\!-\!\bar Q^2]
+2m^2 Q^2
\bigg\}
 \nn\\
 {\cal C}^{  \bar{g}^{\perp g}}_{T}(z_1,\P) 
\equiv &\,
 - \epsilon^{ij}\, \epsilon^{ln}\, 
\frac{2\k^n}{\k\!\cdot\!\P}\, {\cal C}^{ijl}_{T}(z_1,\P)
 =
 \frac{4(z_2\!-\!z_1)}{ [\P^2\!+\!\bar Q^2]^2}
\label{scal_coeff_contract_T}
\end{align}
in the transverse photon case. 

The cross section~\eqref{X_sec_gen_form_scal_TMDs} depends on the angle $\theta_{\k\P}$ between $\k$ and $\P$ via the scalar product $(\k\!\cdot\!\P)$. One observes that the LP terms are even in $(\k\!\cdot\!\P)$, and thus contribute only to even harmonics in that angular dependence, whereas the NLP terms are odd in  $(\k\!\cdot\!\P)$, and thus contribute only to odd harmonics.
More precisely, the LP terms in $ {\cal C}^{ f_1^g}_{L,T}$ lead to angular independent contributions, whereas the LP terms in  ${\cal C}^{ h_1^{\perp g}}_{L,T}$ have a $\cos(2\theta_{\k\P})$ modulation.
Then, the terms associated with ${\cal C}^{ f^{\perp g}}_{L,T}$ and  ${\cal C}^{  \bar{g}^{\perp g}}_{T}$, or with the NLP corrections to   $ {\cal C}^{ f_1^g}_{L,T}$ have $\cos(\theta_{\k\P})$ modulation.
Finally the  NLP corrections to    ${\cal C}^{ h_1^{\perp g}}_{L,T}$ lead to linear combinations of $\cos(\theta_{\k\P})$ and $\cos(3\theta_{\k\P})$.
In general, one can expect any even (resp. odd) power term in the back-to-back jets regime to contribute only to  even (resp. odd) harmonics in $\theta_{\k\P}$.

In Eq.~\eqref{X_sec_gen_form_scal_TMDs}, the first two lines correspond to the LP contributions in the correlation limit (with NLP corrections only in the hard coefficient). They contain eikonal terms, as well as NEik corrections, suppressed by $1/W^2$. The eikonal contributions corresponds to twist-two gluon TMDs at ${\rm x}=0$, due to the absence of a phase factor in our result in Eq.~\eqref{class_to_quant_Fperpmin_Fperpmin} compared to the definition \eqref{def_Phi_correlator_Fmunu_Frhosigma} for the field strength correlator. By contrast, the NEik corrections in the first two lines of Eq.~\eqref{X_sec_gen_form_scal_TMDs} can be interpreted as the first corrections in the Taylor expansion of the twist-two gluon TMDs around ${\rm x}=0$, providing information on the small value of ${\rm x}$ at which the TMDs are probed in this process. In particular, Eq.~\eqref{X_sec_gen_form_scal_TMDs} can be rewritten as
%
\begin{align}
\frac{d\sigma_{\gamma^{*}_{T,L}\rightarrow q_1\bar q_2}}{dz_1\, d^2\P\, d^2\k}\Bigg|_{{\rm Eik }+{\rm NEik}}
= &\,
  \alpha_{\rm em} e_f^2\, \alpha_{s} \; 
\Bigg\{
{\cal C}^{ f_1^g}_{T,L}(z_1,\P,\k)\:
 {\rm x}\, f_1^g({\rm x},\k) 
 + {\cal C}^{ h_1^{\perp g}}_{T,L}(z_1,\P,\k)\:
{\rm x}\, h_1^{\perp g}({\rm x},\k)  
 \nn\\
 &\,
+\frac{\k\!\cdot\!\P}{W^2}\, {\cal C}^{ f^{\perp g}}_{T,L}(z_1,\P)\:  {\rm x}\, f^{\perp g}({\rm x},\k)
+\frac{\k\!\cdot\!\P}{W^2}\,  {\cal C}^{  \bar{g}^{\perp g}}_{T,L}(z_1,\P)\:
{\rm x}\,  \bar{g}^{\perp g}({\rm x},\k)
\Bigg\}\Bigg|_{{\rm x}=\frac{[\P^2\!+\!\bar Q^2]}{z_1 z_2\, W^2} \,}
+3\, \mathcal{F}\textrm{ terms}
\label{X_sec_gen_form_scal_TMDs_x_non_zero}
\end{align}
at the same accuracy: including all eikonal and NEik terms which are also LP or NLP.
Hence, the dependence on ${\rm x}$ of the TMDs can be recovered in our high-energy formalism by resumming a subset of power corrections beyond the eikonal approximation. 
 In the second line of Eq.~\eqref{X_sec_gen_form_scal_TMDs_x_non_zero}, we have assumed that the twist three gluon TMDs are probed at the same value of ${\rm x}$. In order to check this, it would be necessary to push our calculation to NNEik accuracy, which is far beyond the scope of the present study.


\section{Conclusion and outlook}
\label{sec:Conc}

In this paper, we extend the studies of the interplay between the CGC and TMD formalisms beyond eikonal accuracy. Focusing on DIS dijet production, the connection between these two formalisms has been previously investigated both (i) at leading twist but at NLO in the coupling constant and (ii) at LO in the coupling constant but beyond leading (kinematical) twist approximation. However, both of these limits are considered at eikonal accuracy. In this paper, we are presenting the results of the first study of the back-to-back limit of the DIS dijet production cross section at LO in the coupling constant beyond the eikonal and leading twist approximations. 

Starting from the DIS dijet production at NEik accuracy computed in \cite{Altinoluk:2022jkk} both via longitudinal or transverse photon exchange, we first consider the back-to-back limit at the amplitude level. After rewriting the decorated Wilson lines, that appear at NEik amplitudes, in terms of insertions of the field strength tensor along the longitudinal direction, we perform the small dipole size expansion for the NEik amplitudes that contributes to the DIS dijet production. Then, squaring these amplitudes, the DIS dijet cross section at NEik accuracy in the back-to-back limit is obtained, both for longitudinal and transverse photons exchange.  

Our results for the back-to-back dijet production in DIS via longitudinal photon can be summarized as follows. The final result can be written as a sum of three contributions, according to the operator expectation value they contain: a twist-2 gluon TMD distribution (correlator of $\langle {\cal F}^{\perp -}{\cal F}^{\perp -}\rangle$ with a future pointing gauge staple), a twist-3 gluon TMD distribution (correlator of $\langle {\cal F}^{+-}{\cal F}^{\perp -}\rangle$ with a future pointing gauge staple) or a twist-3 three body contribution (correlator of $\langle {\cal F}^{\perp -}{\cal F}^{\perp -}{\cal F}^{\perp -}\rangle$ with a future pointing gauge staple) (see Eqs. \eqref{X_sec_L_gen_form}, \eqref{X_sec_L_Fperpmin_Fperpmin}, \eqref{X_sec_L_Fperpmin_Fperpmin_Fperpmin} and  \eqref{X_sec_L_Fplusmin_Fperpmin}). 

The twist-2 gluon TMD term receives both LP (kinematic twist 2)  and NLP (kinematic twist 3) contributions in $|\k|/|\P|$, in the back-to-back limit. In terms of the high-energy expansion, it receives both eikonal and NEik contributions. Remarkably, these two types of power corrections factorize from each other (see Eq. \eqref{X_sec_L_Fperpmin_Fperpmin}), with the NLP contributions correcting only the coefficient and the NEik contributions correcting only the operator itself. Moreover, NEik corrections are shown to be the first order correction in the Taylor expansion of the TMDs around the momentum fraction $\rm x=0$ which suggests that the full $\rm x$ dependence of the TMDs can be recovered by re-summing all order  corrections beyond the eikonal limit for that contribution. The two-body twist-3 contributions are purely NEik, while the three-body twist-3 contributions contain eikonal NLP (genuine twist 3) terms and others that are both NLP and NEik.
 
Regarding the back-to-back dijet production in DIS via transverse photon, the full result can be written as a sum of four contributions: a twist-2 gluon TMD (correlator of $\langle {\cal F}^{\perp -}{\cal F}^{\perp -}\rangle$ with a future pointing gauge staple) two types of twist-3 gluon TMDs  (one of them is a correlator of $\langle {\cal F}^{+-}{\cal F}^{\perp -}\rangle$ and the other one is a correlator of $\langle {\cal F}_{ij}{\cal F}^{\perp -}\rangle$, both with a future pointing gauge staple) and a three body contribution (correlator of $\langle {\cal F}^{\perp -}{\cal F}^{\perp -}{\cal F}^{\perp -}\rangle$ with a future pointing gauge staple) (see Eq. \eqref{X_sec_T_gen_form} with the contributions written in Eqs.~\eqref{X_sec_T_Fperpmin_Fperpmin}, \eqref{X_sec_T_Fperpmin_Fperpmin_Fperpmin},
\eqref{X_sec_T_Fplusmin_Fperpmin} and~\eqref{X_sec_T_Fij_Fperpmin}). All of the above mentioned comments for the production via longitudinal photon holds for the production via transverse photon. 

Finally, we discuss the relation between the CGC-like  target averaging of color operator and the quantum expectation value of that operator in a target state to make the connection of our results with the TMD formalism. Adopting the standard notation given in Eq. \eqref{def_Phi_correlator_Fmunu_Frhosigma} for the gluon field strength correlator appearing in the definition of various gluon TMDs with a future pointing gauge staple, we write our results in terms of  the unpolarized twist-2 gluon TMD $f_1^g({\rm x}, \k)$ and its linearly polarized partner  $h_1^{\perp g}({\rm x}, \k)$, twist-3 gluon TMDs $f^{\perp g}(\rm x, \k)$ and ${\bar g}^{\perp g}(\rm x, \k)$, and three-body contributions. The final result can be written in compact form as given in Eq. \eqref{X_sec_gen_form_scal_TMDs_x_non_zero} with the coefficient functions are given explicitly in Eq. \eqref{scal_coeff_contract_L} for the case of longitudinal and in Eq. \eqref{scal_coeff_contract_T} for the case of transverse photon.  
We have not analyzed the obtained three body contributions with the same depth, since these objects 
have not yet been classified in a systematic way in the TMD literature.


Another relevant observable to study at NEik accuracy in the back-to-back limit is the photon+jet production at forward rapidity in proton-nucleus collisions. Since the produced photon does not interact with the target, at eikonal accuracy this observable is known to probe a dipole gluon TMD in full generality, i.e. without the need for the back-to-back limit. It would be very interesting to see whether that remains the case beyond eikonal accuracy, as well as what kind of corrections one gets with this observable and to what extend they are related to the corrections found in the present study. We plan to address that other observable in the future.




\acknowledgements{
TA is supported in part by the National Science Centre (Poland) under the research Grant No. 2023/50/E/ST2/00133 (SONATA BIS 13). GB is supported in part by the National Science Centre (Poland) under the research Grant No. 2020/38/E/ST2/00122 (SONATA BIS 10). AC  is supported in part by the National Science Centre (Poland) under the research Grant No. 2021/43/D/ST2/01154 (SONATA 17).}

\appendix



\section{Integrals over $\r$ for the hard factors at amplitude level}
\label{app:int}


After performing the small $\r$ expansion of the color operators in Sec.~\ref{sec:color_struct_exp}, the integration over $\r$ can be performed at the amplitude level, using the following relations:
\begin{align} 
\int_{\r} e^{-i\r \cdot\P}\, K_0(\bar Q|\r|) 
=&\, 2\pi\, \frac{1}{[\P^2\!+\!\bar Q^2]}
\nonumber\\
\int_{\r} e^{-i\r \cdot\P}\, K_0(\bar Q|\r|)\, \r^j  
=&\, 2\pi\, \frac{(-2i)\P^j}{[\P^2\!+\!\bar Q^2]^2}
\nonumber\\
\int_{\r} e^{-i\r \cdot\P}\, K_0(\bar Q|\r|)\,  \r^i  \r^j  
=&\, 2\pi\,\bigg\{
 \frac{2\delta^{ij}}{[\P^2\!+\!\bar Q^2]^2}
 -\frac{8\P^i\P^j}{[\P^2\!+\!\bar Q^2]^3}
 \bigg\}
\nonumber\\
\int_{\r} e^{-i\r \cdot\P}\,|\r|\,  K_1(\bar Q|\r|) 
=&\, 2\pi\, \frac{2\bar Q}{[\P^2\!+\!\bar Q^2]^2}
\nonumber\\
\int_{\r} e^{-i\r \cdot\P}\, K_1(\bar Q|\r|)\, \frac{\r^j}{|\r|}  
=&\, 2\pi\, \frac{(-i)}{\bar Q}\, \frac{\P^j}{[\P^2\!+\!\bar Q^2]}
\nonumber\\
\int_{\r} e^{-i\r \cdot\P}\, K_1(\bar Q|\r|)\, \frac{\r^i\r^j}{|\r|}  
=&\,  \frac{2\pi}{\bar Q}\, \bigg\{
 \frac{\delta^{ij}}{[\P^2\!+\!\bar Q^2]}
 -\frac{2\P^i\P^j}{[\P^2\!+\!\bar Q^2]^2}
 \bigg\}
\nonumber\\
\int_{\r} e^{-i\r \cdot\P}\, K_1(\bar Q|\r|)\, \frac{\r^i\r^j\r^l}{|\r|}  
=&\, 2\pi\, \frac{i}{\bar Q}\, \bigg\{-2
 \frac{\big[\delta^{ij}\P^l+\delta^{il}\P^j+\delta^{jl}\P^i\big]}{[\P^2\!+\!\bar Q^2]^2}
 +\frac{8\P^i\P^j\P^l}{[\P^2\!+\!\bar Q^2]^3}
 \bigg\}
 \, .
 \label{r_integrals}
\end{align}
These results can be obtained by inserting the integral representations
\begin{align}
K_0(\bar Q|\r|)
 =&\, 
2\pi \int \frac{d^2\k}{(2\pi)^2} \frac{e^{i\k \cdot \r}}{\k^2 + \bar Q^2} 
\nonumber\\
i\bar Q \frac{\r^j}{|\r|} K_1(\bar Q|\r|) 
 =&\,  
2\pi \int \frac{d^2\k}{(2\pi)^2} \frac{\k^j e^{i\k \cdot \r}}{\k^2 + \bar Q^2} 
\nonumber\\
\frac{|\r|}{2\bar Q} K_1(\bar Q|\r|) 
 =&\, 
 2\pi \int \frac{d^2\k}{(2\pi)^2} \frac{e^{i\k \cdot \r}}{[\k^2 + \bar Q^2]^2} 
\end{align}
of the modified Bessel functions of the second kind, and rewriting extra $\r$ factors as derivatives with respect to $\P$.


\section{Study of the non-eikonal corrections included in the generalized eikonal cross section for longitudinal photon}
\label{app:geik-eik}

From the generalized eikonal cross section for back-to-back DIS dijet production via longitudinal photon~\eqref{Gen_Eik_X_sec_L}, it is quite simple to take the strict eikonal approximation and recover the result from Eq.~\eqref{str_Eik_X_sec_L}. One simply has to neglect the dependence on $\Delta b^-$ of the field strength insertions and Wilson lines. Then, the integration over $\Delta b^-$ can be performed, leading to the factor $2\pi\delta(q^+\!-\!k^+)$ (meaning that the momentum exchange with the target has a vanishing light-cone $+$ component), and the expression \eqref{str_Eik_X_sec_L} is then obtained.
In order to go beyond that, and calculate the difference between the generalized eikonal cross section~\eqref{Gen_Eik_X_sec_L} and the strict eikonal~\eqref{str_Eik_X_sec_L}, one has to study further terms in the gradient expansion in $\Delta b^-$ of the color operator.   

Since the dependence on $\Delta b^-$ appears only on the amplitude side in Eq.~\eqref{Gen_Eik_X_sec_L}, it is convenient to study the gradient expansion at the level of the amplitude~\eqref{Ampl-GenEik_L_b2b}. Moreover, let us focus on the contribution with one field strength insertion at the amplitude level (or two at the cross section level) for simplicity. At first order in the gradient expansion, the color structure present in the amplitude can be written as 
\begin{align}
 &\,\int_{\b}\,   e^{-i\b\cdot\k} \int  dv^+ 
t^{a'}\, \mathcal{U}_A\left(+\infty,v^+;\b,b^-\right)_{a'a} g{\mathcal{F}_{i}^{\;-}}_a (v^+,\b,b^-)
=
\int_{\b}\,   e^{-i\b\cdot\k}\int  dv^+ 
t^{a'}\, \mathcal{U}_A\left(+\infty,v^+;\b,0\right)_{a'a} g{\mathcal{F}_{i}^{\;-}}_a (v^+,\b,0)
\nn\\
&\,
+b^-\,\int_{\b}\,   e^{-i\b\cdot\k} \int  dv^+\, t^{a'}\, \left\{\partial_{b^-}\Big[ \mathcal{U}_A\left(+\infty,{v}^+;\b,b^-\right)_{a'a}  g{\mathcal{F}_{i}^{\;-}}_a ({v}^+,\b,b^-)\Big]\bigg|_{b^-=0}\right\}
+O\left(\partial_{b^-}^2\right)
\, ,
\end{align}
including both the strict eikonal term, and the first order non-static correction.
Using the relation~\eqref{rel_adj_fund} between fundamental and adjoint Wilson lines, that first order correction can be written as 
\begin{align}
&\,
\int_{\b}\,   e^{-i\b\cdot\k}\int  dv^+\,  t^{a'}\, \partial_{b^-}\Big[ \mathcal{U}_A\left(+\infty,{v}^+;\b,b^-\right)_{a'a}  g{\mathcal{F}_{i}^{\;-}}_a ({v}^+,\b,b^-)\Big]\bigg|_{b^-=0}
\nonumber\\
=&\,
\int_{\b}\,   e^{-i\b\cdot\k}\int  dv^+\, \partial_{b^-}\Big[ \mathcal{U}_F\left(+\infty,{v}^+;\b,b^-\right)  g{\mathcal{F}_{j}^{\;-}} ({v}^+,\b,b^-)  \mathcal{U}_F\left(+\infty,{v}^+;\b,b^-\right)^{\dag}  \Big]\bigg|_{b^-=0}
\nonumber\\
=&\,
\int_{\b}\,   e^{-i\b\cdot\k}\int  dv^+\bigg\{ 
\mathcal{U}_F\left(+\infty,{v}^+;\b,b^-\right)
\bigg[ 
\overleftarrow{\mathcal{D}_-}\,  g{\mathcal{F}_{i}^{\;-}} ({v}^+,\b,b^-) 
+\big[\mathcal{D}_-,g {\mathcal{F}_{i}^{\;-}} \big]({v}^+,\b,b^-)
\nn\\
&\, \hspace{2.5cm}
+g {\mathcal{F}_{i}^{\;-}}({v}^+,\b,b^-)\,  
\overrightarrow{\mathcal{D}_-}
\bigg]
 \mathcal{U}_F\left(+\infty,{v}^+;\b,b^-\right)^{\dag}
\bigg\}\bigg|_{b^-=0}
\, .\label{gradient_App_1}
\end{align}
Using the identity 
\begin{align}
\big[\mathcal{D}_{\mu}, \mathcal{D}_{\nu} \big] =&\, ig {\mathcal{F}}_{\mu\nu} 
\end{align}
as well as the Jacobi identity, one finds
\begin{align}
\big[\mathcal{D}_-,g {\mathcal{F}_{i}^{\;-}} \big]
=&\, -i \Big[\mathcal{D}_-,\big[\mathcal{D}_{i}, \mathcal{D}_+ \big]\Big]
=  i\Big[\mathcal{D}_i,\big[\mathcal{D}_{+}, \mathcal{D}_- \big]\Big]+ i\Big[\mathcal{D}_+,\big[\mathcal{D}_{-}, \mathcal{D}_i \big]\Big]
=\Big[\mathcal{D}_i,g {\mathcal{F}^{+-}}\Big]+ \Big[\mathcal{D}_+,g {\mathcal{F}_{i}^{\;+}}\Big]
\, .\label{Jacobi_App}
\end{align}
Moreover, integrating by parts the covariant derivative, one obtains
\begin{align}
&\, 
 \int  dv^+\, \mathcal{U}_F\left(+\infty,{v}^+;\b,0\right)\,
  \Big[\mathcal{D}_+,g {\mathcal{F}_{i}^{\;+}}\Big]({v}^+,\b,0)\:
   \mathcal{U}_F\left(+\infty,{v}^+;\b,0\right)^{\dag}
\nonumber\\
=&\,
 -\int  dv^+\, 
\mathcal{U}_F\left(+\infty,{v}^+;\b,0\right) 
\Big( 
\overleftarrow{\mathcal{D}_+}\,  
g {\mathcal{F}_{i}^{\;+}}({v}^+,\b,0)
+ g {\mathcal{F}_{i}^{\;+}}({v}^+,\b,0)\,  
\overrightarrow{\mathcal{D}_+}
\Big) 
\mathcal{U}_F\left(+\infty,{v}^+;\b,0\right)^{\dag}
\nonumber\\
=&\,
0
\, .\label{IBP_App}
\end{align}
so that  (with covariant derivatives acting only on the Wilson lines within the brackets and not on the phase factor)
\begin{align}
&\,
\int_{\b}\,   e^{-i\b\cdot\k} \int  dv^+\, t^{a'}\partial_{b^-}\Big[ \mathcal{U}_A\left(+\infty,{v}^+;\b,b^-\right)_{a'a}  g{\mathcal{F}_{i}^{\;-}}_a ({v}^+,\b,b^-)\Big]\bigg|_{b^-=0}
\nonumber\\
=&\,
\int_{\b}\,   e^{-i\b\cdot\k}\int  dv^+\bigg\{ 
\mathcal{U}_F\left(+\infty,{v}^+;\b,b^-\right)
\Big[ 
\overleftarrow{\mathcal{D}_-}\,  g{\mathcal{F}_{i}^{\;-}} ({v}^+,\b,b^-) 
 +\Big[\mathcal{D}_i,g {\mathcal{F}^{+-}}\Big]({v}^+,\b,b^-)
\nonumber\\
&\,
\hspace{1.5cm}
+g {\mathcal{F}_{i}^{\;-}}({v}^+,\b,b^-)\,  
\overrightarrow{\mathcal{D}_-}
\Big]
 \mathcal{U}_F\left(+\infty,{v}^+;\b,b^-\right)^{\dag}
\bigg\}\bigg|_{b^-=0}\nonumber\\
=&\,
\int_{\b}\,   e^{-i\b\cdot\k}\int  dv^+\bigg\{ 
\mathcal{U}_F\left(+\infty,{v}^+;\b,b^-\right)
\Big[ 
\overleftarrow{\mathcal{D}_-}\,  g{\mathcal{F}_{i}^{\;-}} ({v}^+,\b,b^-) 
-\overleftarrow{\mathcal{D}_i}\,  g{\mathcal{F}^{+-}} ({v}^+,\b,b^-) 
\nonumber\\
&\,
\hspace{1.5cm}
-g{\mathcal{F}^{+-}} ({v}^+,\b,b^-)\,  \overrightarrow{\mathcal{D}_i}
+g {\mathcal{F}_{i}^{\;-}}({v}^+,\b,b^-)\,  
\overrightarrow{\mathcal{D}_-}
\Big]
 \mathcal{U}_F\left(+\infty,{v}^+;\b,b^-\right)^{\dag}
\bigg\}\bigg|_{b^-=0}
\nonumber\\
&\,
+\int_{\b}\,   e^{-i\b\cdot\k}\int  dv^+ \partial_{\b^i}\bigg\{ 
\mathcal{U}_F\left(+\infty,{v}^+;\b,b^-\right) 
  g{\mathcal{F}^{+-}} ({v}^+,\b,b^-) 
 \mathcal{U}_F\left(+\infty,{v}^+;\b,b^-\right)^{\dag}
\bigg\}\bigg|_{b^-=0}
\nonumber\\
=&\,
\int_{\b}\,   e^{-i\b\cdot\k}\int  dv^+\int  dw^+\, \theta(w^+\!-\!v^+)\,
\mathcal{U}_F\left(+\infty,{w}^+;\b\right)\,
 \bigg\{ 
 -ig{\mathcal{F}^{+-}} ({w}^+,\b)\, \mathcal{U}_F\left({w}^+,{v}^+;\b\right)\,
 g{\mathcal{F}_{i}^{\;-}} ({v}^+,\b) 
 \nonumber\\
&\,
\hspace{1.5cm}
+i g{\mathcal{F}_{i}^{\;-}} ({w}^+,\b)\, \mathcal{U}_F\left({w}^+,{v}^+;\b\right)\,
  g{\mathcal{F}^{+-}}({v}^+,\b) 
\bigg\} \mathcal{U}_F\left(+\infty,{v}^+;\b\right)^{\dag}
 \nonumber\\
&\,
+\int_{\b}\,   e^{-i\b\cdot\k}\int  dv^+\int  dw^+\, \theta(w^+\!-\!v^+)\,
\mathcal{U}_F\left(+\infty,{v}^+;\b\right)\,
 \bigg\{ 
 -ig{\mathcal{F}^{+-}} ({v}^+,\b)\, \mathcal{U}_F\left({w}^+,{v}^+;\b\right)^{\dag}\,
 g{\mathcal{F}_{i}^{\;-}} ({w}^+,\b) 
 \nonumber\\
&\,
\hspace{1.5cm}
+ ig{\mathcal{F}_{i}^{\;-}} ({v}^+,\b)\, \mathcal{U}_F\left({w}^+,{v}^+;\b\right)^{\dag}\,
  g{\mathcal{F}^{+-}}({w}^+,\b) 
\bigg\} \mathcal{U}_F\left(+\infty,{w}^+;\b\right)^{\dag}
\nonumber\\
&\,
+i\k^i\int_{\b}\,   e^{-i\b\cdot\k}\int  dv^+ 
\mathcal{U}_F\left(+\infty,{v}^+;\b\right) 
  g{\mathcal{F}^{+-}} ({v}^+,\b) 
 \mathcal{U}_F\left(+\infty,{v}^+;\b\right)^{\dag}
\nonumber\\
=&\,
-i\,[t^{a'},t^{b'}]  \int_{\b}\,   e^{-i\b\cdot\k}\int  dv^+\int  dw^+\, \mathcal{U}_A\left(+\infty,{v}^+;\b\right)_{a'a}\, g{\mathcal{F}_a^{+-}} ({v}^+,\b)\, \mathcal{U}_A\left(+\infty,{w}^+;\b\right)_{b'b}\,  
 g{\mathcal{F}_{i}^{\;-}}_b ({w}^+,\b) 
 \nonumber\\
&\,
+i\k^i\, t^{a'}\int_{\b}\,   e^{-i\b\cdot\k}\int  dv^+ 
\mathcal{U}_A\left(+\infty,{v}^+;\b\right)_{a'a}\, g{\mathcal{F}_a^{+-}} ({v}^+,\b)
\end{align}

Hence, at the amplitude level, one obtains two types of terms in the difference between the generalized eikonal and the strict eikonal expressions, at the first order in the gradient expansion in $b^-$. 
At cross section level, they lead to two types of NEik contributions in the difference between Eqs.~\eqref{Gen_Eik_X_sec_L} and \eqref{str_Eik_X_sec_L}:
either terms with a color operator of the type $f^{abc} \langle{\mathcal{F}_a^{+-}} (\b) {\mathcal{F}_{i}^{\;-}}_b (\b)  {\mathcal{F}_{j}^{\;-}}_c (\b')\rangle$ (omitting Wilson lines and $+$ coordinates) or terms of the type $\k^i \langle{\mathcal{F}_a^{+-}} (\b) {\mathcal{F}_{j}^{\;-}}_a (\b')\rangle$.
Both of these types of terms are doubly power suppressed in the back-to-back regime. In the first type of term, one power suppression comes from having three field strength insertions instead of two, and another power suppression from the presence of a subleading component $\mathcal{F}^{+-}$ of the field strength. In the second type of term, one power suppression comes from the subleading component $\mathcal{F}^{+-}$, and a second one from the presence of an explicit factor of the small momentum $\k$.   
 Hence, in the difference between the generalized eikonal cross section~\eqref{Gen_Eik_X_sec_L} and the strict eikonal cross section~\eqref{str_Eik_X_sec_L}, all the contributions which are of order NEik in the high-energy limit (single gradient in $\Delta b^-$) are of order NNLP (or beyond) in the back-to-back jets regime, often called twist 4 contributions. This is beyond the accuracy of the present study, aimed at calculating NLP corrections. 
 
Since we have not used the specific form of the hard factor in this appendix, but only analysed the color structure, the derivation is valid not only for the cross section via longitudinal photon exchange, but also for the cross section via transverse photon exchange.

%


\section{Rearranging the back-to-back dijet production amplitude via transverse photon}
\label{app:rewrite_non_fact}


In this appendix, the term in the fourth line of Eq.~\eqref{Ampl-sumEik12+_T_b2b} is rewritten in terms of different color structures, for convenience.
Using the relation $\delta^{im} \delta^{jn}-\delta^{in} \delta^{jm} = \epsilon^{ij}\epsilon^{mn}$, valid in the case of 2 transverse dimensions exactly, one finds
$[\P^j\k^m-(\k\!\cdot\!\P)\delta^{jm}]=  (\k^i\epsilon^{ij})(\epsilon^{mn}\P^n)$.
Hence, that term from Eq.~\eqref{Ampl-sumEik12+_T_b2b} writes
\begin{align}
i{\cal M }_{q_1 \bar q_2 \leftarrow \gamma^*_T}^{\textrm{non. fact.}}
\equiv&\, 
 \frac{e e_f}{2q^+}\,  \varepsilon_{\lambda}^l\,  
  \bar u(1) \gamma^+\bigg(
(z_2\!-\!z_1)\,  \delta^{lm}+\frac{[\gamma^l,\gamma^m]}{2}
\bigg) v(2)
  \int_{\b}\,   e^{-i\b\cdot\k}\, 
  t^{a'} \int  dv^+\, 
 \mathcal{U}_A\left(+\infty,v^+;\b\right)_{a'a} g{\mathcal{F}_{j}^{\;-}}_a (v^+,\b)\; 
  \nn \\
&
\times
\frac{iv^+}{2z_1 z_2 q^+}\, \frac{(z_2\!-\!z_1)}{[\P^2\!+\!\bar Q^2]}\, 
\left[
\P^j\k^m-(\k\!\cdot\!\P)\delta^{jm}
\right]\,
  \nn \\
=&\, 
 \frac{e e_f}{(2q^+)^2}\,  \varepsilon_{\lambda}^l\,
 \frac{(z_2\!-\!z_1)}{z_1 z_2}\,   
  \frac{\epsilon^{mn}\P^n}{[\P^2\!+\!\bar Q^2]}\, 
  \bar u(1) \gamma^+\bigg(
(z_2\!-\!z_1)\,  \delta^{lm}+\frac{[\gamma^l,\gamma^m]}{2}
\bigg) v(2)
  \nn \\
&
\times\, 
\epsilon^{ij}
\int_{\b}\,   e^{-i\b\cdot\k}\, (i\k^i)
 \int  dv^+\, v^+\, 
 \mathcal{U}_F\left(+\infty,v^+;\b\right) g{\mathcal{F}_{j}^{\;-}} (v^+,\b) \mathcal{U}_F\left(+\infty,v^+;\b\right)^{\dag} \; 
   \nn \\
=&\, 
 \frac{e e_f}{(2q^+)^2}\,  \varepsilon_{\lambda}^l\,
 \frac{(z_2\!-\!z_1)}{z_1 z_2}\,   
  \frac{\epsilon^{mn}\P^n}{[\P^2\!+\!\bar Q^2]}\, 
  \bar u(1) \gamma^+\bigg(
(z_2\!-\!z_1)\,  \delta^{lm}+\frac{[\gamma^l,\gamma^m]}{2}
\bigg) v(2)
  \nn \\
&
\times\, 
\epsilon^{ij}
\int_{\b}\,   e^{-i\b\cdot\k}\, 
 \int  dv^+\, v^+\, \overrightarrow{\partial_{\b^i}}
 \left[\mathcal{U}_F\left(+\infty,v^+;\b\right) g{\mathcal{F}_{j}^{\;-}} (v^+,\b) \mathcal{U}_F\left(+\infty,v^+;\b\right)^{\dag}\right] \; 
   \nn \\
=&\, 
 \frac{e e_f}{(2q^+)^2}\,  \varepsilon_{\lambda}^l\,
 \frac{(z_2\!-\!z_1)}{z_1 z_2}\,   
  \frac{\epsilon^{mn}\P^n}{[\P^2\!+\!\bar Q^2]}\, 
  \bar u(1) \gamma^+\bigg(
(z_2\!-\!z_1)\,  \delta^{lm}+\frac{[\gamma^l,\gamma^m]}{2}
\bigg) v(2)
  \nn \\
&
\times\, 
\epsilon^{ij}
\int_{\b}\,   e^{-i\b\cdot\k}\, 
 \int  dv^+\, v^+\, 
 \bigg[
 \mathcal{U}_F\left(+\infty,v^+;\b\right)  \overleftarrow{\mathcal{D}_{i}}\,   g{\mathcal{F}_{j}^{\;-}} (v^+,\b) \mathcal{U}_F\left(+\infty,v^+;\b\right)^{\dag}
   \nn \\
&
 +\mathcal{U}_F\left(+\infty,v^+;\b\right)\, \left[\mathcal{D}_{i},   g{\mathcal{F}_{j}^{\;-}}\right]\,  \mathcal{U}_F\left(+\infty,v^+;\b\right)^{\dag}
 +\mathcal{U}_F\left(+\infty,v^+;\b\right) g{\mathcal{F}_{j}^{\;-}} (v^+,\b)\, \overrightarrow{\mathcal{D}_{i}}\, \mathcal{U}_F\left(+\infty,v^+;\b\right)^{\dag}
 \bigg]
 \label{Ampl-nonFact_T_b2b} 
\, .
\end{align}
Here, the covariant derivatives act only within the square brackets.  

The specificity of the contribution \eqref{Ampl-nonFact_T_b2b} comes from the possibility, using $\epsilon^{ij}$ and the Jacobi identity, to entirely remove the single ${\mathcal{F}_{j}^{\;-}}$, as
\begin{align}
 \epsilon^{ij}\left[\mathcal{D}_{i},   g{\mathcal{F}_{j}^{\;-}}\right]
 =&\,
 (-i)\epsilon^{ij}\big[\mathcal{D}_{i}, [\mathcal{D}_{j},\mathcal{D}_{+}] \big]
 =
  (-i)\frac{\epsilon^{ij}}{2}\, \Big(\big[\mathcal{D}_{i}, [\mathcal{D}_{j},\mathcal{D}_{+}] \big]-\big[\mathcal{D}_{j}, [\mathcal{D}_{i},\mathcal{D}_{+}] \big]\Big)
  \nn\\
=&\,  
 (-i)\frac{\epsilon^{ij}}{2}\, \Big(\big[\mathcal{D}_{i}, [\mathcal{D}_{j},\mathcal{D}_{+}] \big]+\big[\mathcal{D}_{j}, [\mathcal{D}_{+},\mathcal{D}_{i}] \big]\Big)
 = 
 (-i)\frac{\epsilon^{ij}}{2}\, \Big(- \big[\mathcal{D}_{+}, [\mathcal{D}_{i},\mathcal{D}_{j}] \big]\Big)
   \nn\\
=&\,  
-\frac{\epsilon^{ij}}{2}\, \left[\mathcal{D}_{+},   g{\mathcal{F}_{ij}}\right]
\, .
\end{align}
Using this relation as well as the identity Eq.~\eqref{deriv_Wilson}, the contribution \eqref{Ampl-nonFact_T_b2b} becomes
\begin{align}
i{\cal M }_{q_1 \bar q_2 \leftarrow \gamma^*_T}^{\textrm{non. fact.}}
=&\, 
 \frac{e e_f}{(2q^+)^2}\,  \varepsilon_{\lambda}^l\,
 \frac{(z_2\!-\!z_1)}{z_1 z_2}\,   
  \frac{\epsilon^{mn}\P^n}{[\P^2\!+\!\bar Q^2]}\, 
  \bar u(1) \gamma^+\bigg(
(z_2\!-\!z_1)\,  \delta^{lm}+\frac{[\gamma^l,\gamma^m]}{2}
\bigg) v(2)
  \nn \\
&
\times\, 
\epsilon^{ij}
\int_{\b}\,   e^{-i\b\cdot\k}\, 
 \int  dv^+\, v^+\, 
 \bigg[
 -\frac{1}{2}\, \mathcal{U}_F\left(+\infty,v^+;\b\right)\, \left[\mathcal{D}_{+},   g{\mathcal{F}_{ij}}\right]\,  \mathcal{U}_F\left(+\infty,v^+;\b\right)^{\dag}
   \nn \\
&
 -i  \int_{v^+}^{+\infty}  \!\!\!\!\!\!\! dw^+\, \mathcal{U}_F\left(+\infty,w^+;\b\right) g{\mathcal{F}_{i}^{\;-}} (w^+,\b)     \mathcal{U}_F\left(w^+,v^+;\b\right)   g{\mathcal{F}_{j}^{\;-}} (v^+,\b) \mathcal{U}_F\left(+\infty,v^+;\b\right)^{\dag}
   \nn \\
&
 +i  \int_{v^+}^{+\infty}  \!\!\!\!\!\!\! dw^+\, \mathcal{U}_F\left(+\infty,v^+;\b\right) g{\mathcal{F}_{j}^{\;-}} (v^+,\b)     \mathcal{U}_F\left(w^+,v^+;\b\right)^{\dag} g{\mathcal{F}_{i}^{\;-}} (w^+,\b) \mathcal{U}_F\left(+\infty,w^+;\b\right)^{\dag}
 \bigg]
\nn \\
=&\, 
   \frac{e e_f}{(2q^+)^2}\,  \varepsilon_{\lambda}^l\,
 \frac{(z_2\!-\!z_1)}{z_1 z_2}\,   
  \frac{\epsilon^{mn}\P^n}{[\P^2\!+\!\bar Q^2]}\, 
  \bar u(1) \gamma^+\bigg(
(z_2\!-\!z_1)\,  \delta^{lm}+\frac{[\gamma^l,\gamma^m]}{2}
\bigg) v(2)
  \nn \\
&
\times\, 
\epsilon^{ij}
\int_{\b}\,   e^{-i\b\cdot\k}\, 
 \int  dv^+\,
 \bigg\{
 -\frac{v^+}{2}\, \overrightarrow{\partial_{v^+}}\left[\mathcal{U}_F\left(+\infty,v^+;\b\right)\,   g{\mathcal{F}_{ij}}\,  \mathcal{U}_F\left(+\infty,v^+;\b\right)^{\dag}\right]
   \nn \\
&\hspace{0.5cm}
 -i  \int dw^+\, \mathcal{U}_F\left(+\infty,v^+;\b\right) g{\mathcal{F}_{i}^{\;-}} (v^+,\b)     \mathcal{U}_F\left(v^+,w^+;\b\right)   g{\mathcal{F}_{j}^{\;-}} (w^+,\b) \mathcal{U}_F\left(+\infty,w^+;\b\right)^{\dag}
   \nn \\
& \hspace{1.5cm}\times\,
\big(\theta(v^+\!-\!w^+)w^+ + \theta(w^+\!-\!v^+)v^+ \big) 
 \bigg\}   
 \label{Ampl-nonFact_T_b2b_2} 
\, .
\end{align}
Finally, integrating by parts the first term in $v^+$ and
rewriting the fundamental Wilson lines in terms of adjoint Wilson lines, one obtains the expression \eqref{Ampl-nonFact_T_b2b_3}.


\section{Tensor hard factors}
\label{app:coeffs}


In this appendix, we provide explicit expressions for the hard factors ${\cal C}^{ij}_{T,L}$, ${\cal C}^{j}_{T,L}$ and ${\cal C}^{ijl}_{T,L}$ with open transverse indices, which appear in the expression~\eqref{X_sec_gen_form_Phi} for the DIS dijet cross sections in the back-to-back regime. 
From Eqs.~\eqref{X_sec_L_gen_form}, \eqref{X_sec_L_Fperpmin_Fperpmin} and \eqref{X_sec_L_Fplusmin_Fperpmin}, these hard factors are found to be   
%
\begin{align}
 {\cal C}^{ij}_{L}(z_1,\P,\k) 
=&\,
4\, Q^2\,   z_1^2 z_2^2
\Bigg[
\frac{4\P^i\P^j}{[\P^2\!+\!\bar Q^2]^4}
-2(z_2\!-\!z_1)\frac{(\P^i\k^j+\k^i\P^j)}{[\P^2\!+\!\bar Q^2]^4}
 +16(z_2\!-\!z_1)\frac{(\k\!\cdot\!\P)\P^i\P^j}{[\P^2\!+\!\bar Q^2]^5}
\Bigg]
+O\left(\frac{Q^2 \k^2}{\P^8}\right)
\nn\\
 {\cal C}^{j}_{L}(z_1,\P)
 =&\,
 16\, Q^2\, z_1 z_2\,  (z_2\!-\!z_1) \; \frac{\P^j[\P^2+m^2]}{[\P^2\!+\!\bar Q^2]^4}\,
 +O\left(\frac{Q^2 |\k|}{\P^6}\right)
 \nn\\
{\cal C}^{ijl}_{L}(z_1,\P)
 =&\,
 0
\label{tensor_hard_fact_L}
\end{align}
in the longitudinal photon case.
From Eqs.~\eqref{X_sec_T_Fperpmin_Fperpmin}, \eqref{X_sec_T_Fplusmin_Fperpmin}, \eqref{X_sec_T_Fij_Fperpmin} and \eqref{X_sec_T_gen_form} by contrast, these hard factors are found to be   
%
\begin{align}
 {\cal C}^{ij}_{T}(z_1,\P,\k) 
=&\,
\frac{[z_1^2+z_2^2]}{[\P^2\!+\!\bar Q^2]^2}\, 
\left[1+\frac{2(z_2\!-\!z_1)(\k\!\cdot\!\P)}{[\P^2\!+\!\bar Q^2]}\right]\delta^{ij}
\nonumber\\
 &
 +\frac{\big[[z_1^2+z_2^2]\bar Q^2-m^2\big]}{[\P^2\!+\!\bar Q^2]^4}\, 
 \Bigg[-4\P^i\P^j\left(1+\frac{4(z_2\!-\!z_1)(\k\!\cdot\!\P)}{[\P^2\!+\!\bar Q^2]}\right)
 +2(z_2\!-\!z_1) \big(\P^i\k^j+\k^i\P^j\big)
 \Bigg]
 +O\left(\frac{\k^2}{\P^6}\right)
\nn\\
 {\cal C}^{j}_{T}(z_1,\P)
 =&\,
 -2(z_2\!-\!z_1) \; \frac{\P^j}{[\P^2\!+\!\bar Q^2]^4}\,
 \bigg\{
\big[\P^2\!+\!\bar Q^2+(z_1^2+z_2^2)Q^2\big] [\P^2\!-\!\bar Q^2]
+2m^2 Q^2
\bigg\}\,
 +O\left(\frac{|\k|}{|\P|^4}\right)
 \nn\\
{\cal C}^{ijl}_{T}(z_1,\P)
 =&\,
 2 (z_2\!-\!z_1) \; \frac{ \epsilon^{ml}\P^m}{[\P^2\!+\!\bar Q^2]^2}\,
\frac{\epsilon^{ij}}{2}\, 
 +O\left(\frac{|\k|}{|\P|^4}\right)
\label{tensor_hard_fact_T}
\end{align}
in the transverse photon case.



\bibliography{mybib_New}

\end{document}